\documentclass[fleqn,10pt]{article}
\usepackage{latexsym, graphicx, epsfig, amsmath, amssymb,amsfonts}
\usepackage{natbib,amsthm,version}
\usepackage{amsbsy,bm,multirow,enumerate}
\usepackage[titletoc,page]{appendix}
\usepackage[mathscr]{eucal}
\usepackage{mathtools}
\usepackage{color}
\usepackage{subfigure}
\usepackage[utf8]{inputenc}
\usepackage[english]{babel}
\usepackage{float}
\usepackage{caption}
\usepackage{systeme}
\usepackage[affil-it]{authblk}
\usepackage{hyperref}

\newtheorem{theorem}{Theorem}[section]

\begin{document}

\title{ 
	A consistent and conservative volume distribution algorithm and its applications to multiphase flows using Phase-Field models
	\footnote{\copyright $<$2021$>$. This manuscript version is made available under the CC-BY-NC-ND 4.0 license \url{http://creativecommons.org/licenses/by-nc-nd/4.0/}. }
	\footnote{This manuscript was accepted for publication in International Journal of Multiphase Flow, Vol 142, Ziyang Huang, Guang Lin, Arezoo M. Ardekani, A consistent and conservative volume distribution algorithm and its applications to multiphase flows using Phase-Field models, Page 103727, Copyright Elsevier (2021).}
} 
\author[1]{
	Ziyang Huang%
	\thanks{Email: \texttt{huan1020@purdue.edu}}}

\author[1,2]{
	Guang Lin%
	\thanks{Email: \texttt{guanglin@purdue.edu}; Corresponding author}}

\author[1]{
	Arezoo M. Ardekani%
	\thanks{Email: \texttt{ardekani@purdue.edu}; Corresponding author}}

\affil[1]{
	School of Mechanical Engineering, Purdue University, West Lafayette, IN 47907, USA}
\affil[2]{
	Department of Mathematics, Purdue University, West Lafayette, IN 47907, USA}

\date{(June 11, 2021)}

\maketitle


\begin{abstract}
In the present study, the multiphase volume distribution problem, where there can be an arbitrary number of phases, is addressed using a consistent and conservative volume distribution algorithm. The proposed algorithm satisfies the summation constraint, the conservation constraint, and the \textit{consistency of reduction}.
The first application of the volume distribution algorithm is to determine the Lagrange multipliers in multiphase Phase-Field models that enforce the mass conservation, and a multiphase conservative Allen-Cahn model that satisfies the \textit{consistency of reduction} is developed. A corresponding consistent and conservative numerical scheme is developed for the model. The multiphase conservative Allen-Cahn model has a better ability than the multiphase Cahn-Hilliard model to preserve under-resolved structures.
The second application is to develop a numerical procedure, called the boundedness mapping, to map the order parameters, obtained numerically from a multiphase model, into their physical interval, and at the same time to preserve the physical properties of the order parameters. Along with the consistent and conservative schemes for the multiphase Phase-Field models, the numerical solutions of the order parameters are reduction consistent, conservative, and bounded, which are theoretically analyzed and numerically validated. 
Then, the multiphase Phase-Field models are coupled with the momentum equation by satisfying the \textit{consistency of mass conservation} and the \textit{consistency of mass and momentum transport}, thanks to the consistent formulation. It is demonstrated that the proposed model and scheme converge to the sharp-interface solution and are capable of capturing the complicated multiphase dynamics even when there is a large density and/or viscosity ratio.
\end{abstract}

\vspace{0.05cm}

\section{Introduction}\label{Sec Introduction}
Multiphase flows are ubiquitous and have wide-spread applications. For example, the oil spill accident in 2010 \citep{united2011scene} has gained lots of attention worldwide due to its dramatic damage to the environment. In order to predict the spread of the oil and to provide remediation strategies, a model that is capable of capturing interactions of the water, oil, and air is needed. Other examples include the enhanced oil recovery, where $\mathrm{CO}_2$ is injected along with the water into oil reservoirs \citep{aramideh2019unstable,wang2017assessing,alvarado2010enhanced}, and dynamics of compound drops \citep{kan1998hydrodynamics,gao2011spreading,zhu2020impact}, where the drop is composed of different fluids. Lots of efforts have been focused on modeling and simulating two-phase flows, and the one-fluid formulation \cite{Tryggvasonetal2011,ProsperettiTryggvason2007}, where the motion of the fluids is governed by a single equation of their mixture, is one of the most popular ones. Under this framework, many successful numerical models or methods have been developed to specify locations of interfaces, e.g., 
the front-tracking method \cite{UnverdiTryggvason1992,Tryggvasonetal2001}, 
the level-set method \cite{OsherSethian1988,Sussmanetal1994,SethianSmereka2003,Gibouetal2018}, 
the conservative level-set method \cite{OlssonKreiss2005,Olssonetal2007,ChiodiDesjardins2017}, 
the volume-of-fluid (VOF) method \cite{HirtNichols1981,ScardovelliZaleski1999,OwkesDesjardins2017}, 
the THINC method \cite{Xiaoetal2005,Iietal2012,XieXiao2017,Qianetal2018}, and the Phase-Field (or Diffuse-Interface) method \cite{Andersonetal1998,Jacqmin1999,Shen2011,Huangetal2020}.
A recent review of various interface-capturing methods is available in \cite{Mirjalilietal2017}.
The surface tension can be modeled by the smoothed surface stress method \cite{Gueyffieretal1999}, the continuous surface force (CSF) \cite{Brackbilletal1992}, the ghost fluid method (GFM) \cite{Fedkiwetal1999,Lalanneetal2015}, the conservative and well-balanced surface tension model \cite{Abu-Al-Saud2018}, and the Phase-Field method derived from the energy balance or the least-action principle \cite{Jacqmin1999,Yueetal2004}, and the surface tension model is incorporated into the momentum equation by the balanced-force algorithm \cite{Francoisetal2006}. A recent review of various numerical models for surface tension is available in \cite{Popinet2018}. 
The physical coupling between the mass and momentum transport is enforced numerically by a consistent scheme, e.g., \cite{Rudman1998,Bussmannetal2002,ChenadecPitsch2013,OwkesDesjardins2017} for the volume-of-fluid (VOF) method, \cite{RaessiPitsch2012,Nangiaetal2019} for the level-set method, \cite{Xieetal2020} for the THINC method, and \cite{Huangetal2020,Huangetal2020CAC} for the Phase-Field methods. 
Many recent studies investigate three-phase flows, e.g., \cite{Schofieldetal2009,Schofieldetal2010,Francois2015,Losassoetal2006,Starinshaketal2002,Boyeretal2010,KimLowengrub2005,Kim2007,ZhangWang2016,Zhangetal2016,Abadi2018}, and even extend the model to general $N$-phase flows $(N \geqslant 1)$, e.g., \cite{BoyerMinjeaud2014,Kim2009,Kim2012,LeeKim2015,KimLee2017,Dong2014,Dong2015,Dong2017,Dong2018,YangDong2018}. Most of the three-phase and $N$-phase models belong to the category of the Phase-Field model due to its simplicity and effectiveness. In the Phase-Field model, a set of order parameters, which are commonly related to the volume fractions of the phases, are introduced to indicate the locations of different phases. The sharp interfaces are replaced by small but finite interfacial regions, inside which there are thermodynamical compression and diffusion of the model to preserve the thickness of the interfacial regions.

In the present work, we consider the multiphase volume distribution problem, where there can be an arbitrary number of phases. In some previous two-phase studies, this is also called the mass distribution or the mass redistribution. Since the problem is not related to the densities of individual phases, it is more precise to call it the volume distribution.
Given the volume changes of individual phases going to be distributed to the domain and a set of order parameters representing the locations of different phases, we need to specify the volume distribution functions of individual phases at every location of the domain. It should be noted that the domain is fixed. Therefore, before and after the volume distribution, the volume of the domain does not change.
This is a kind of inverse problem and its solution is not necessarily unique. However, the admissible solution should not produce any fictitious phases, local voids, or overfilling. In addition to that, the integrals of the volume distribution functions of the solution over the domain should be the given volume changes correspondingly. These goals are achieved by satisfying the proposed summation and conservation constraints for volume distribution and the \textit{consistency of reduction}, which will be discussed in detail in the present study. We call the volume distribution is consistent and conservative if it satisfies all the aforementioned physical constraints.
It is relatively straightforward to solve the volume distribution problem and to satisfy the physical constraints in two-phase cases, while it becomes non-trivial for general multiphase cases. 
In a two-phase case, increasing the volume of one of the two phases corresponds to a decrease of the same amount of the volume of another phase from the summation constraint. As a result, only one of the phases is necessarily considered and the volume distribution is solved phase-wise. In a general multiphase case, there can be more than two phases at a specific location. When any one of them changes its volume, the others have to respond to that simultaneously to satisfy the summation constraint. In other words, all the phases have to be considered at the same time.
In the two-phase case, the phases inside interfacial regions are fixed. However, the number of phases inside a specific interfacial region is varied from at least two to at most $N$, and there are lots of different possible combinations of the phases inside that region in the multiphase case. This also casts difficulty to satisfy the \textit{consistency of reduction} in the general multiphase setup.
To the best of our knowledge, this problem has never been addressed in the previous studies. In the present study, the consistent and conservative volume distribution algorithm is proposed to solve the problem.

The first application, motivated us to address the multiphase volume distribution problem, is to design the physical and general Lagrange multipliers that enforce the mass conservation for a variety of multiphase Phase-Field models. As a specific example, a multiphase conservative Allen-Cahn model that satisfies the \textit{consistency of reduction} is developed in the present work, with the help of the proposed consistent and conservative volume distribution algorithm. Almost all the Phase-Field models for multiphase flows are Cahn-Hilliard type \citep{CahnHilliard1958} since it has a conservative form. Therefore, the mass conservation of each phase is satisfied. 
However, by adding a Lagrange multiplier, we can obtain the so-called conservative Allen-Cahn model, which satisfies the mass conservation as well. 
By appropriately designing the Lagrange multiplier, Brassel and Bretin \cite{BrasselBretin2011} proposed a two-phase conservative Allen-Cahn model that is applicable for two-phase flow modeling. The two-phase conservative Allen-Cahn model is a 2nd-order partial differential equation, while the Cahn-Hilliard models are usually 4th order. Therefore the conservative Allen-Cahn model is easier to solve. In addition, it enjoys the maximum principle so that its solution has an upper and lower bound. Both the analysis \cite{BrasselBretin2011} and numerical comparison \cite{LeeKim2016} suggest that the two-phase conservative Allen-Cahn model has a better ability than the Cahn-Hilliard models to preserve the under-resolved structure. Our previous analysis \cite{Huangetal2020CAC} also shows that the two-phase conservative Allen-Cahn model satisfies the \textit{consistency of reduction}.
Therefore, it is attractive to develop a multiphase conservative Allen-Cahn model for multiphase flows. Such a model is developed by Kim and Lee \cite{KimLee2017}. Although they numerically show that their model is able to preserve small structures and their solution is inside the physical interval, the model violates the \textit{consistency of reduction}, and as a result produces fictitious phases. This will be analyzed and numerically demonstrated in the present work.
The \textit{consistency of reduction} is of great importance for a multiphase model since it avoids generating fictitious phases \citep{BoyerMinjeaud2014,LeeKim2015,Dong2017,Dong2018,Abadi2018,Huangetal2020N} and recovers the sing-phase dynamics inside individual bulk-phase regions \citep{Huangetal2020N}. It has a significant effect on the flow dynamics, especially when the density ratio or viscosity ratio in the problem is large. For example, in a water-oil-air system, the maximum density and viscosity ratios are of the order of $1000$. Violating the \textit{consistency of reduction} can unphysically generate the oil at the interface of a water-air bubble. The bubble becomes much heavier and more viscous than it should be even though only a small amount of the oil is generated. Therefore, the rising motion of the bubble is slowed down due to violating the \textit{consistency of reduction}. This behavior has been demonstrated in \cite{BoyerMinjeaud2014}. 
To the best of our knowledge, the proposed multiphase conservative Allen-Cahn model is the first model of this kind satisfying the \textit{consistency of reduction}. A corresponding consistent and conservative scheme is developed, which preserves all the physical properties of the model on the discrete level.

The second application related to the multiphase volume distribution problem is to map the order parameters into their physical interval. In multiphase flows, the order parameters are not only the indicators of different phases but also used to compute the density and viscosity of the fluid mixture. This computation is based on the assumption that the order parameters have a physical bound. For example, the order parameters should be in $[0,1]$ if they are the volume fractions. If some of the order parameters are beyond their physical interval, there is no physical interpretation for them, and the density of the fluid mixture, for example, can be smaller than the minimum density of the phases, and can even be negative, resulting in an ill-posed momentum equation. Problems having large density and/or viscosity ratios are less tolerant to the out-of-bound order parameters, and a small out-of-bound error can become problematic in a computation.
The out-of-bound order parameters can be generated due to the defect of the model. For example, the widely-used two-phase Cahn-Hilliard model \cite{Shen2011}, with constant mobility and the Ginzburg-Landau double-well potential, admits an out-of-bound solution \cite{Yueetal2007,TierraGuillen-Gonzalez2015,Chenetal2019,Franketal2020}. Fortunately, both the asymptotic analysis \cite{Magalettietal2013,Abelsetal2012} and the scaling analysis \cite{Yueetal2007} suggest that the out-of-bound issue is controlled by the interface thickness, which is normally as small as the grid size.
Another source of the out-of-bound order parameters is from numerical errors, even though the model has the maximum principle. However, designing a bound-preserving scheme is not a trivial task, especially when the model is non-linear and complicated. In addition, the bound-preserving scheme usually casts an additional constraint on the time step, e.g., \cite{Mirjalilietal2020} for the two-phase conservative Phase-Field model \cite{ChiuLin2011}, and \cite{Huangetal2020CAC} for the two-phase conservative Allen-Cahn model \cite{BrasselBretin2011}. The out-of-bound issue from the numerical error is the truncation error of the scheme, which again is related to the grid size. 
Since the out-of-bound error from either the defect of the model or the numerical error is related to the grid size, which is usually small, a more common practice is to clip the out-of-bound solution, e.g., in \cite{ChiuLin2011,DongShen2012,Dong2018}. However, a volume distribution algorithm has to be supplemented following the clipping operation. Otherwise, the mass conservation is destroyed \cite{Huangetal2019}. Chiu and Lin \cite{ChiuLin2011} evenly distributed the volume, which is lost from the clipping operation, to the interfacial regions, and this algorithm is applied in \cite{Zhangetal2019}. Huang et al. \cite{Huangetal2020CAC} distributed the volume based on a weight function that is the same as the one in the two-phase conservative Allen-Cahn model \cite{BrasselBretin2011}. 
The clipping operation is also commonly used in the volume-of-fluid (VOF) or THINC methods, e.g., \cite{Baraldietal2014,Fusteretal2018,Qianetal2018}, and a volume distribution algorithm is required to achieve mass conservation \cite{Baraldietal2014}.
As discussed, the multiphase volume distribution problem is far more challenging than the two-phase one and, thus, both the clipping operation and the volume distribution have to be carefully designed. Otherwise, fictitious phases, local voids, or overfilling can be artificially produced. To the best of our knowledge, the general solution of this problem has not been proposed in previous studies.
Thanks to the consistent and conservative volume distribution algorithm, we develop a numerical procedure for multiphase problems, called the boundedness mapping, which maps the order parameters, obtained numerically from a multiphase model, into their physical interval, and at the same time, the physical properties of the order parameters, i.e., their summation constraint, mass conservation, and \textit{consistency of reduction}, are preserved. 

To simulate multiphase flows, the Phase-Field models need to be coupled with the momentum equation appropriately. By satisfying the \textit{consistency of mass conservation} and the \textit{consistency of mass and momentum transport}, the resulting momentum equation is compatible with the actual mass conservation equation of the Phase-Field model and the kinetic energy conservation, and is Galilean invariant \citep{Huangetal2020N}. Otherwise, unphysical velocity and pressure fluctuations, as well as interface deformation, appear, which can result in numerical instability, especially in problems with large density ratios. These have been analyzed and demonstrated in our previous studies for two- and multi-phase flows \cite{Huangetal2020,Huangetal2020N,Huangetal2020CAC}. 
Another challenge is cast when implementing the Phase-Field models that are not written in a conservative form for multiphase flows. For example, the multiphase conservative Allen-Cahn model has a Lagrange multiplier to enforce the mass conservation. As a result, it is not in a conservative form and the consistency analysis proposed in \cite{Huangetal2020} is not valid. The same issue will also appear for the Phase-Field models that are in a conservative form, e.g., the Cahn-Hilliard models, when some operations are performed to correct their out-of-bound solutions. Those operations contribute to the violation of the consistency conditions. Actually, this issue was neither addressed nor explicitly discussed in many previous studies for two-phase flows using the volume-of-fluid (VOF) or Phase-Field methods, where the clipping operation and volume distribution were performed, although the schemes in those studies were claimed to be consistent. 
Recently in \cite{Huangetal2020CAC}, the two-phase conservative Allen-Cahn model \citep{BrasselBretin2011} is applied to study two-phase flows, and the operations to correct the numerical solution of the Phase-Field model are represented as a discrete Lagrange multiplier. To satisfy the consistency conditions, the consistent formulation is proposed to deal with all the Lagrange multipliers, continuous or discrete, in the two-phase conservative Allen-Cahn model. In the present study, the consistent formulation is applied to both the generic Phase-Field model for multiphase flows and the multiphase boundedness mapping, in order to enforce the \textit{consistency of reduction}, the \textit{consistency of mass conservation}, and the \textit{consistency of mass and momentum transport}, on both the continuous and discrete levels.
Combining the consistent and conservative scheme for the proposed multiphase conservative Allen-Cahn model and the proposed boundedness mapping, the overall scheme for the order parameters honors the summation constraint for the order parameters, the mass conservation of individual phases, and the boundedness of the order parameters, and is reduction consistent, on the discrete level. Comparison studies are performed with the multiphase Cahn-Hilliard model in \cite{Dong2018,Huangetal2020N}, which is solved by the mass conservative and reduction consistent scheme in \citep{Huangetal2020N} and the proposed boundedness mapping is again applied.
With the help of the consistent formulation, the overall schemes for the two Phase-Field models, respectively, are physically connected with the momentum conservative scheme in \citep{Huangetal2020,Huangetal2020CAC,Huangetal2020N} for the momentum equation, and the \textit{consistency of mass conservation} and the \textit{consistency of mass and momentum transport} are satisfied on the discrete level. 
We demonstrate that the proposed model and scheme are capable of capturing the complicated multiphase dynamics even when there is a large density or viscosity ratio.

The rest of the paper is organized as follows.
In Section \ref{Sec Governing equations and discretizations}, the multiphase problem and three consistency conditions for multiphase flows are defined, followed by the introduction of the governing equations. The formulations in this section are presented in a way that are generally valid for a variety of Phase-Field models for multiphase flows.
In Section \ref{Sec Volume distribution and its application}, the consistent and conservative volume distribution algorithm is elaborated, followed by its applications to develop the multiphase conservative Allen-Cahn model that satisfies the \textit{consistency of reduction}, and to develop the boundedness mapping to map the order parameters into their physical interval. 
In Section \ref{Sec Discretizations}, the consistent and conservative scheme for the multiphase conservative Allen-Cahn model is developed, and the implementation of the boundedness mapping and the consistent formulation on the discrete level are described. 
In Section \ref{Sec Results}, various numerical tests are performed to validate the properties of the models and their schemes, and to demonstrate their capability of solving complicated multiphase flows.
In Section \ref{Sec Conclusions}, the present study is concluded.

\section{Problem definition and governing equations}\label{Sec Governing equations and discretizations}
In this section, the problem of interest and several important consistency conditions are firstly defined in Section \ref{Sec Problem definition}, after which the governing equations are introduced in Section \ref{Sec Governing equations}. The formulations proposed in this section are generic for various Phase-Field models of multiphase incompressible flows.

\subsection{Problem definition}\label{Sec Problem definition}
In the present work, we confine our study to multiphase incompressible flows where the number of phases is $N$ ($N \geqslant 1$). All the phases are immiscible with each other and have their own constant densities $\{\rho_p\}_{p=1}^N$ and viscosities $\{\mu_p\}_{p=1}^N$. Each pair of phases has a surface tension $\sigma_{p,q}$ $(1 \leqslant p,q \leqslant N)$. Locations of the phases are labeled by a set of order parameters $\{\phi_p\}_{p=1}^N$, which specifically is the volume fraction contrasts. The physical range of $\{\phi_p\}_{p=1}^N$ is in $[-1,1]$. The order parameters are not independent and their summation should satisfy the summation constraint for the order parameters, i.e.,
\begin{equation}\label{Eq Summation constraint phi}
\sum_{p=1}^N \phi_p=2-N.
\end{equation}
Eq.(\ref{Eq Summation constraint phi}) is equivalent to the summation of the volume fractions of the phases being unity if the volume fractions $\{C_p\}_{p=1}^N$ are defined as
\begin{equation}\label{Eq Volume fraction}
C_p=\frac{1+\phi_p}{2}, \quad 1 \leqslant p \leqslant N.
\end{equation}
Therefore, the volume fraction of Phase $p$ is $1$ where $\phi_p=1$, while it is $0$ where $\phi_p=-1$. 
Without considering any sources of the phases in or at the boundary of the domain, the mass (or volume) conservation of individual phases requires that
\begin{equation}\label{Eq Conservation phi}
\frac{d}{dt}\int_{\Omega} \phi_p d\Omega =0, \quad 1\leqslant p \leqslant N,
\end{equation}
where $\Omega$ denotes the domain considered, and Eq.(\ref{Eq Conservation phi}) is also called the conservation constraint for the order parameters.

The density and viscosity of the fluid mixture are
\begin{equation}\label{Eq Density}
\rho=\sum_{p=1}^N \rho_p C_p=\sum_{p=1}^N \rho_p \frac{1+\phi_p}{2},
\end{equation}
\begin{equation}\label{Eq Vicosity}
\mu=\sum_{p=1}^N \mu_p C_p=\sum_{p=1}^N \mu_p \frac{1+\phi_p}{2}.
\end{equation}

The flow is incompressible so its velocity $\mathbf{u}$ is divergence-free, i.e.,
\begin{equation}\label{Eq Divergence-free}
\nabla \cdot \mathbf{u}=0.
\end{equation}
Such a divergence-free velocity is also referred to as the volume-averaged velocity \citep{Abelsetal2012,Dong2018}. A series of theoretical analyses and discussions related to the volume-averaged velocity and the models based on that is performed by Brenner, e.g., in \citep{Brenner2004,Brenner2005Kinematics,Brenner2005NS,Brenner2006}. 

As discussed in \cite{Huangetal2020,Huangetal2020CAC,Huangetal2020N}, the following three consistency conditions are of great importance for a multiphase flow model. The definitions of the consistency conditions are
\begin{itemize}
	\item the \textit{consistency of reduction}: A $N$-phase system should be able to recover the corresponding $M$-phase system ($1 \leqslant M \leqslant N-1$) when ($N-M$) phases are absent. 
	\item the \textit{consistency of mass conservation}: The mass conservation equation should be consistent with the transport equation defined from the Phase-Field equation and the density of the fluid mixture. The mass flux in the mass conservation equation should lead to a zero mass source. 
	\item the \textit{consistency of mass and momentum transport}: The momentum flux in the momentum equation should be a tensor product between the mass flux and the flow velocity, where the mass flux should be identical to the one in the mass conservation equation.
\end{itemize}

\begin{figure}[!t]
	\centering
	\includegraphics[scale=.4]{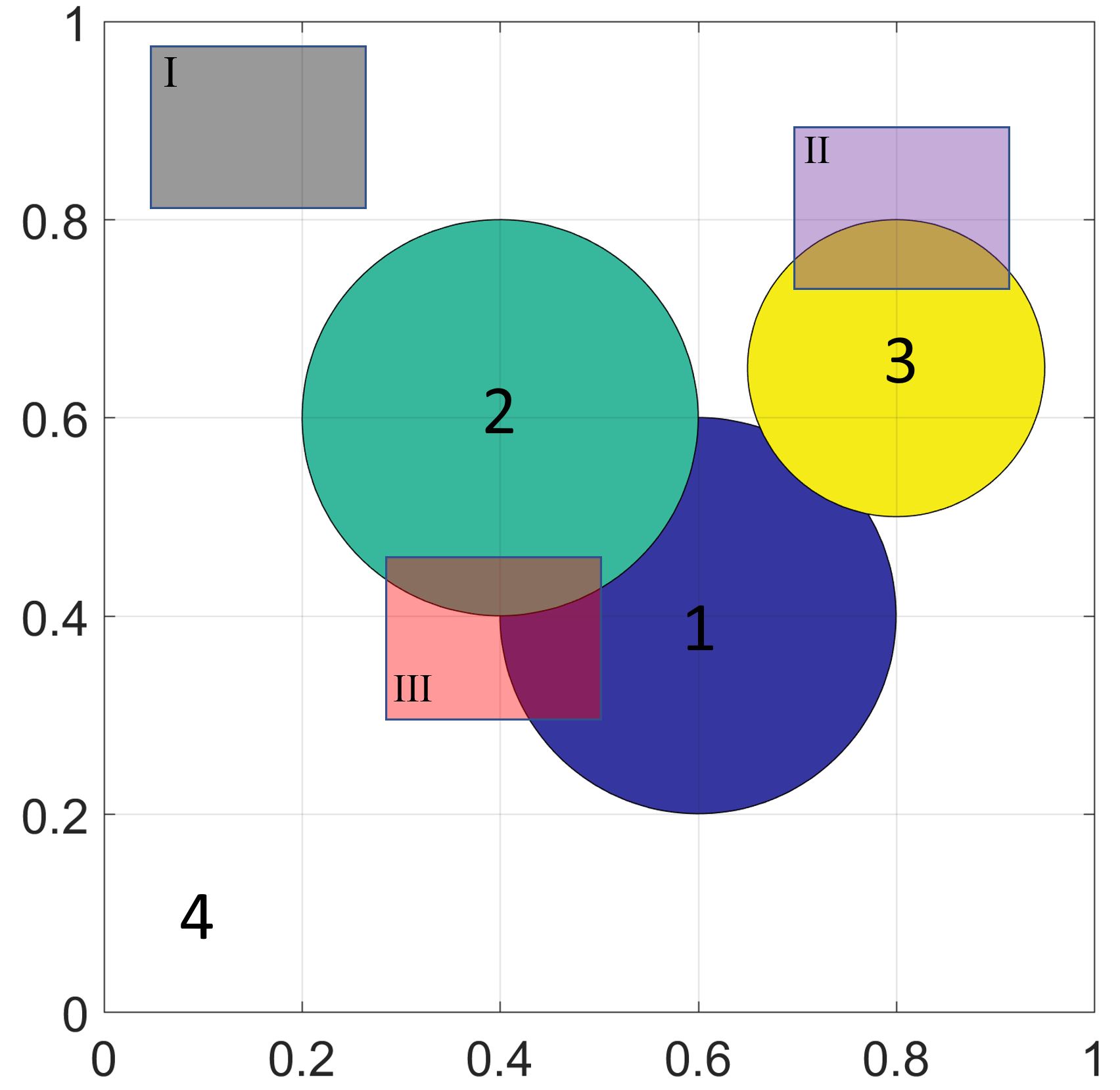}
	\caption{Schematic of the \textit{consistency of reduction}. Blue: Phase 1. Green: Phase 2. Yellow: Phase 3. White: Phase 4. Regions I, II, III are examples of local regions that only include single-, two-, and three-phase dynamics, respectively, although the entire domain is governed by a 4-phase model. \label{Fig RC-Schematic}}
\end{figure}

When considering the \textit{consistency of reduction}, the understanding of absent phases needs to be clarified. In the previous studies, the absence of phases is considered globally, for example, the absence of Phase $p$ means that Phase $p$ dose not appear anywhere, i.e., $\phi_p \equiv -1$. In the present work, we consider the absence of phases locally. As an example, the absence of Phase $p$ means where $\phi_p=-1$ and all its spatial derivatives in the Phase-Field model are zero. Such a consideration in the present study has not only theoretical but also practical values. 
It is obvious that if the \textit{consistency of reduction} is true when the absent phases are considered locally, it will be true as well when the absent phases are considered globally, because $\phi_p \equiv -1$ implies all the spatial derivatives of $\phi_p$ are zero. Therefore, the local consideration in the present study won't contradict with previously developed theories about the \textit{consistency of reduction}, e.g., in \citep{BoyerMinjeaud2014,Dong2018,Huangetal2020N}.
From practical point of view, absent phases won't be initiated, because that will increase the cost of computation and the demand of storage. It should be noted that any $N$-phase problems are locally a $M$-phase problem $(M \leqslant N)$, and, in most cases, $M$ is much less than $N$. Therefore, it is more important that the \textit{consistency of reduction} is still true when phases are absent locally, instead of globally. Fig.\ref{Fig RC-Schematic} is a schematic showing interaction of 4 phases described by a 4-phase model in the entire domain. Regions I, II, III are examples of local regions that only include single-, two-, and three-phase dynamics, respectively. In Region I, the single-phase dynamics of Phase 4 needs to be recovered by the 4-phase model, and Phases 1, 2, and 3 should be absent. In Region II, the two-phase dynamics of Phases 3 and 4 needs to be recovered by the 4-phase model, and Phases 1 and 2 should be absent. In Region III, the three-phase dynamics of Phases 1, 2, and 4 needs to be recovered by the 4-phase model, and Phase 3 should be absent. 

The \textit{consistency of mass conservation} and the \textit{consistency of mass and momentum transport} are general principles to physically connect the Phase-Field model to the hydrodynamics. The effect of these two consistency conditions has been summarized in Section \ref{Sec Introduction} and detail analyses are available in \citep{Huangetal2020,Huangetal2020CAC,Huangetal2020N}. 

\subsection{Governing equations}\label{Sec Governing equations}
The governing equations consist of the multiphase Phase-Field model, which is presented in a general form, in Section \ref{Sec Phase-Field} to locate different phases and the momentum equation in Section \ref{Sec Momentum equation} to describe the fluid motion. These two components are physically connected by considering the \textit{consistency of mass conservation} and the \textit{consistency of mass and momentum transport}, with the help of the consistent formulation in Section \ref{Sec Consistent formulation}. They conserve not only the mass of each phase but also the momentum of the multiphase flow (with a proper model for the interfacial tensions).

\subsubsection{The multiphase Phase-Field model}\label{Sec Phase-Field}
The order parameters are governed by a multiphase Phase-Field model, and the Phase-Field model is either the Cahn-Hilliard type or the conservative Allen-Cahn type. In the present work, we consider a Phase-Field model to be the Cahn-Hilliard type if it is written in a conservative form. On the other hand, if a Phase-Field model includes Lagrange multipliers to enforce the mass conservation Eq.(\ref{Eq Conservation phi}), it belongs to the conservative Allen-Cahn type. A physically admissible Phase-Field model not only ensures the summation constraint Eq.(\ref{Eq Summation constraint phi}) and the mass conservation Eq.(\ref{Eq Conservation phi}) but also the \textit{consistency of reduction}. 

Without loss of generality, the multiphase Phase-Field model is written as
\begin{equation}\label{Eq Phase-Field}
\frac{\partial \phi_p}{\partial t}
+
\nabla \cdot (\mathbf{u} \phi_p)
=
\nabla \cdot \mathbf{J}_p
+
L_p^R
+
L_p^c,
\quad
1 \leqslant p \leqslant N,
\end{equation}
and the convection term is written in its conservative form, thanks to the divergence-free velocity Eq.(\ref{Eq Divergence-free}). Here, $\{\mathbf{J}_p\}_{p=1}^N$ are the diffusion fluxes and $\{L_p^R\}_{p=1}^N$ are the reaction terms of the Phase-Field model, and $\{L_p^c\}_{p=1}^N$ are the Lagrange multipliers supplemented to enforce the mass conservation Eq.(\ref{Eq Conservation phi}). 
Unless otherwise specified, the normal components of the diffusion fluxes vanish at the boundary of domain $\Omega$.
Therefore, the Cahn-Hilliard type Phase-Field model does not include the reaction terms $\{L_p^R\}_{p=1}^N$ nor the Lagrange multipliers $\{L_p^c\}_{p=1}^N$, but the conservative Allen-Cahn type Phase-Field mode does. 

In the present study, detail expressions of the diffusion fluxes $\{\mathbf{J}_p\}_{p=1}^N$ and the reaction terms $\{L_p^R\}_{p=1}^N$ do not matter, because the major focus is on determining physical $\{L_p^c\}_{p=1}^N$ with any given admissible $\{\mathbf{J}_p\}_{p=1}^N$ and $\{L_p^R\}_{p=1}^N$. The admissibility has the following two requirements for $\{\mathbf{J}_p\}_{p=1}^N$ and $\{L_p^R\}_{p=1}^N$. First, both $\{\mathbf{J}_p\}_{p=1}^N$ and $\{L_p^R\}_{p=1}^N$ are reduction consistent in such a way that their values of the absent phases vanish, see the definition of reduction consistent functions in \citep{Dong2018}. Second, both the summations of $\{\mathbf{J}_p\}_{p=1}^N$ and $\{L_p^R\}_{p=1}^N$ over $p$ are zero. Then several constraints are cast on $\{L_p^c\}_{p=1}^N$ so that the Phase-Field model Eq.(\ref{Eq Phase-Field}) is physical.

To satisfy the summation constraint for the order parameters Eq.(\ref{Eq Summation constraint phi}), both sides of the Phase-Field model Eq.(\ref{Eq Phase-Field}) should become zero after summing them over $p$, and therefore the summation of $\{L_p^c\}_{p=1}^N$ is zero, i.e.,
\begin{equation}\label{Eq Summation constraint L_p^c}
\sum_{p=1}^N L_p^c=0.
\end{equation}

To satisfy the mass conservation Eq.(\ref{Eq Conservation phi}), the integral of the right-hand side of the Phase-Field model Eq.(\ref{Eq Phase-Field}) should be zero, and we have
\begin{equation}\label{Eq Conservation L_p^c}
\int_{\Omega} (L_p^R+L_p^c) d\Omega=0,
\quad \mathrm{or} \quad
\int_{\Omega}L_p^c d\Omega=-\int_{\Omega} L_p^R d\Omega=S_p^c,
\quad 1 \leqslant p \leqslant N.
\end{equation}
It should be noted that $\sum_{p=1}^N S_p^c=0$ because the summation of admissible $\{L_p^R\}_{p=1}^N$ is zero.

Finally, $\{L_p^c\}_{p=1}^N$ should also satisfy the \textit{consistency of reduction}, and have their values of the absent phases vanished, i.e.,
\begin{equation}\label{Eq Consistency L_p^c}
L_p^c|_{\phi_p=-1}=0, \quad 1 \leqslant p \leqslant N.
\end{equation}
We only need $\phi_p=-1$ to denote the absence of Phase $p$ in Eq.(\ref{Eq Consistency L_p^c}) since $\{L_p^c\}_{p=1}^N$ do not include any derivatives of the order parameters.

The consistent and conservative volume distribution algorithm is developed in the present study to determine $\{L_p^c\}_{p=1}^N$ that satisfy all the physical constraints in Eq.(\ref{Eq Summation constraint L_p^c}), Eq.(\ref{Eq Conservation L_p^c}), and Eq.(\ref{Eq Consistency L_p^c}), and details are provided in Section \ref{Sec Volume distribution}. The rest of the governing equations introduced in this section again do not rely on the explicit forms of $\{\mathbf{J}_p\}_{p=1}^N$, $\{L_p^R\}_{p=1}^N$, or $\{L_p^c\}_{p=1}^N$.

\subsubsection{The mass conservation and consistent formulation}\label{Sec Consistent formulation}
Combining the definition of the density of the fluid mixture Eq.(\ref{Eq Density}), the Phase-Field model Eq.(\ref{Eq Phase-Field}), and the divergence-free velocity Eq.(\ref{Eq Divergence-free}), the transport of the density of the fluid mixture is governed by
\begin{equation}\label{Eq Mass Phase-Field}
\frac{\partial \rho}{\partial t}
+
\nabla \cdot \mathbf{m}^*
=
S_m^*,
\end{equation}
where $\mathbf{m}^*$ is the mass flux and $S_m^*$ is the mass source. Depending on how the terms are arranged, the definition of the mass flux is different and the mass source is determined correspondingly. For example, the corresponding mass source is $S_m^*=\sum_{p=1}^N \frac{\rho_p}{2} \left(\nabla \cdot \mathbf{J}_p + L_p^R + L_p^c \right)$ when the mass flux is defined as $\mathbf{m}^*=\rho \mathbf{u}$. The \textit{consistency of mass conservation} is applied to determine the consistent mass flux $\mathbf{m}$ such that its corresponding mass source $S_m$ is zero. Due to the presence of $\{L_p^R\}_{p=1}^N$ and $\{L_p^c\}_{p=1}^N$, there is always a non-zero mass source, no matter how the terms are arranged, and, as a result, the \textit{ consistency of mass conservation} is unable to be satisfied.

To address this issue, we follow the consistent formulation proposed in \citep{Huangetal2020CAC} and apply it to individual phases. Specifically, a set of auxiliary variables $\{Q_p\}_{p=1}^N$ is introduced, whose governing equations are
\begin{equation}\label{Eq Consistent formulation}
\nabla \cdot ( W(\phi_p) \nabla Q_p )=L_p^Q,  
\quad
L_p^Q = L_p^R+L_p^c,
\quad
1 \leqslant p \leqslant N,
\end{equation}
where 
\begin{equation}\label{Eq Weight}
W(\phi)=1-\phi^2,
\end{equation}
is the weight function satisfying the \textit{consistency of reduction}, see \cite{Huangetal2020CAC}. Due to the mass conservation, $\left( \int_{\Omega} L_p^Q d\Omega \right)$ is zero for all $p$, see Eq.(\ref{Eq Conservation L_p^c}). Therefore, Eq.(\ref{Eq Consistent formulation}) along with the homogeneous Neumann or with the periodic boundary condition is compatible with its source $L_p^Q$, and it is solvable.
For convenience, we use $W_p$ to denote $W(\phi_p)$ in the rest of the paper. As a result, the Phase-Field model Eq.(\ref{Eq Phase-Field}) can be reformulated into a conservative form, i.e.,
\begin{equation}\label{Eq Phase-Field conservation}
\frac{\partial \phi_p}{\partial t}
+
\nabla \cdot \mathbf{m}_{\phi_p}
=0,
\quad
1 \leqslant p \leqslant N,
\end{equation}
where $\{\mathbf{m}_{\phi_p}\}_{p=1}^N$ are the Phase-Field fluxes, and specifically they are
\begin{equation}\label{Eq Phase-Field flux}
\mathbf{m}_{\phi_p}
=
\mathbf{u} \phi_p 
- 
\mathbf{J}_p
- 
W_p \nabla Q_p,
\quad
1 \leqslant p \leqslant N.
\end{equation}
Once the Phase-Field model is written as Eq.(\ref{Eq Phase-Field conservation}), the consistent mass flux is immediately obtained using the general formulation in \citep{Huangetal2020N}, which reads
\begin{equation}\label{Eq Mass flux}
\mathbf{m}=\sum_{p=1}^N \frac{\rho_p}{2} (\mathbf{u}+\mathbf{m}_{\phi_p}).
\end{equation}
Plugging the consistent mass flux Eq.(\ref{Eq Mass flux}) into Eq.(\ref{Eq Mass Phase-Field}), one can easily show that the density of the fluid mixture Eq.(\ref{Eq Density}) and the consistent mass flux Eq.(\ref{Eq Mass flux}) satisfy the following mass conservation equation
\begin{equation}\label{Eq Mass}
\frac{\partial \rho}{\partial t}
+
\nabla \cdot \mathbf{m}
=
0,
\end{equation}
with the help of Eq.(\ref{Eq Divergence-free}) and Eq.(\ref{Eq Phase-Field conservation}). Therefore the \textit{consistency of mass conservation} is achieved. It should be noted that Eq.(\ref{Eq Mass}) is the actual mass conservation equation of the Phase-Field model.

\subsubsection{The momentum equation}\label{Sec Momentum equation}
The motion of the fluid phases is governed by the momentum equation
\begin{equation}\label{Eq Momentum}
\frac{\partial (\rho \mathbf{u}) }{\partial t}
+
\nabla \cdot ( \mathbf{m} \otimes \mathbf{u} ) 
=
-\nabla P
+
\nabla \cdot \left( \mu (\nabla \mathbf{u} + \nabla \mathbf{u}^T ) \right)
+
\rho \mathbf{g}
+
\mathbf{f}_s,
\end{equation}
where $P$ is the pressure to enforce the divergence-free condition Eq.(\ref{Eq Divergence-free}), $\mathbf{g}$ is the gravity, and $\mathbf{f}_s$ is the surface force due to the interfacial tensions between the phases. It should be noted that the consistent mass flux $\mathbf{m}$ defined in Eq.(\ref{Eq Mass flux}) appears in the inertia term of the momentum equation Eq.(\ref{Eq Momentum}), after applying the \textit{consistency of mass and momentum transport}, so that the momentum equation Eq.(\ref{Eq Momentum}) is consistent with the actual mass conservation equation of the Phase-Field model, i.e., Eq.(\ref{Eq Mass}). As long as the \textit{consistency of mass conservation} and the \textit{consistency of mass and momentum transport} are satisfied, the kinetic energy conservation as well as Galilean invariance can be derived from the resulting momentum equation. The related simplified analysis for two-phase flows are available in \citep{Huangetal2020} and the more formal and complete analysis for an arbitrary number of phases is given in \citep{Huangetal2020N} along with the physical interpretations of the consistency conditions and their formulations. Thus, details of those analyses are not repeated here. The momentum equation Eq.(\ref{Eq Momentum}), along with the consistent mass flux Eq.(\ref{Eq Mass flux}), satisfies the \textit{consistency of reduction} if it is the case for the Phase-Field model Eq.(\ref{Eq Phase-Field}), see the analyses in \cite{Huangetal2020N,Huangetal2020CAC}. Eq.(\ref{Eq Momentum}) has the same form as the momentum equation in \citep{Huangetal2020,Huangetal2020CAC,Huangetal2020N} derived from the same consistency conditions, and is equivalent to the one in \citep{Abelsetal2012,Dong2018}, as well as in \citep{Brenner2006} from GENERIC theory \citep{Ottinger2005}.

The reduction-consistent and momentum-conservative surface force in \citep{Dong2018,Huangetal2020N} is used to model the effect of interfacial tensions and it reads
\begin{equation}\label{Eq Surface force}
\mathbf{f}_s=\frac{1}{2} \sum_{p=1}^N \xi_p \nabla \phi_p.
\end{equation}
Therefore, the momentum equation Eq.(\ref{Eq Momentum}) conserves the momentum of the multiphase flow even the effect of interfacial tensions is included, with the surface force in Eq.(\ref{Eq Surface force}).
Here
\begin{equation}\label{Eq Chemical Potential}
\xi_p
=
\sum_{q=1}^N \lambda_{p,q} \left( 
\frac{1}{\eta^2} \left( g'_1(\phi_p)-g'_2(\phi_p+\phi_q) \right)
+
\nabla^2 \phi_q 
\right),
\end{equation}
is the chemical potential of Phase $p$, where $\lambda_{p,q}=\frac{3}{2\sqrt{2}}\sigma_{p,q}\eta$ is the mixing energy density of Phases $p$ and $q$, $\eta$ is the thickness of the interface, $g_1(\phi)=\frac{1}{4} (1-\phi^2)^2$ and $g_2(\phi)=\frac{1}{4} \phi^2 (\phi+2)^2$ are the potential functions, and $g'_1(\phi)$ and $g'_2(\phi)$ are their derivatives with respect to $\phi$.
This numerical model for interfacial tensions has been demonstrated in \citep{Dong2018,Huangetal2020N,HowardTartakovsky2020} and further discussed in \citep{Huetal2020}.
An alternative option is the generalized continuous surface tension force in \cite{Kim2009,Aiharaetal2019}. However, it is unclear whether this surface tension force is reduction-consistent or momentum-conservative.

\textit{\textbf{Remark}:
\begin{itemize}
    \item 
    One can achieve the \textit{consistency of mass and momentum transport} without performing the consistent formulation in Section \ref{Sec Consistent formulation}. Given $\mathbf{m}^*$ and $S_m^*$ that satisfy Eq.(\ref{Eq Mass Phase-Field}), the \textit{consistency of mass and momentum transport} is achieved by applying $\mathbf{m}^*$ in the inertia term and adding a momentum source $\mathbf{u} S_m^*$ on the right-hand side of the momentum equation. One can again show that such a momentum equation is Galilean invariant. However, the momentum conservation is unfortunately destroyed and, as a result, the momentum equation is inconsistent with the kinetic energy conservation. Therefore, it is critical to apply the consistent formulation so that both the \textit{consistency of mass conservation} and the \textit{consistency of mass and momentum transport} are satisfied simultaneously.
    \item
    Another category of multiphase flow models considers the non-divergence-free ``mass-averaged'' velocity, and examples include \citep{LowengrubTruskinovsky1998,Shenetal2013,Guoetal2014,GuoLin2015,Guoetal2017,Shenetal2020} for two-phase flows, \citep{KimLowengrub2005} for three-phase flows, \citep{LiWang2014,Odenetal2010} for $N$-phase flows. These models are also called the ``quasi-incompressible'' models. Lowengrub and colleagues developed both the two- and three-phase flow models \citep{LowengrubTruskinovsky1998,KimLowengrub2005} of this kind. Recent numerical implementations of this kind of model are restricted to two-phase flows \citep{Guoetal2014,GuoLin2015,Guoetal2017,Shenetal2020}. The consistency of reduction, whose importance has been realized in the studies developing the volume-averaged velocity models, like in \citep{BoyerLapuerta2006,BoyerMinjeaud2014,LeeKim2015,Dong2017,Dong2018,Huangetal2020N}, has not been explicitly analyzed or discussed based on the mass-averaged velocity models. Such analyses and discussions, however, are outside the scope of the present study. Primary comparisons between the models of volume- and mass-averaged velocities, respectively, were performed in \citep{Shenetal2013}, and little difference was observed in two-phase flow applications. This attributes to the fact that the inequality of the two averaged velocities in multiphase flow problems are confined in the small interfacial regions. Of course, further careful qualitative and quantitative comparisons are deserved but this is not the issue to be addressed in the present study.
    \item
    The mass Eq.(\ref{Eq Mass}) and momentum Eq.(\ref{Eq Momentum}) equations are generally valid for both the models using the volume- and mass-averaged velocities, respectively, although the former case is considered in the present study. As illustrated in \citep{Huangetal2020N}, whether the velocity is the volume- or mass-averaged velocity only depends on the condition provided by the Phase-Field model, and we briefly present the theoretical result in \citep{Huangetal2020N} here. To simplify the notation, we consider the volume fractions $\{C_p\}_{p=1}^N$ and denote $\{L_p^{RHS}\}_{p=1}^N$ as the right-hand side of the volume fraction equation, i.e.,
    \begin{equation}\nonumber
    \frac{\partial C_p}{\partial t}
    +
    \nabla \cdot (\mathbf{u} C_p)
    =
    L_p^{RHS},
    \quad 1 \leqslant p \leqslant N,
    \end{equation}
    derived from Eq.(\ref{Eq Phase-Field}) and Eq.(\ref{Eq Volume fraction}). If $\sum_{p=1}^N L_p^{RHS}=0$ is required, one obtains the divergence-free condition Eq.(\ref{Eq Divergence-free}) and the consistent mass flux equivalent to Eq.(\ref{Eq Mass flux}). On the other hand, if $\sum_{p=1}^N \rho_p L_p^{RHS}=0$ is required, one obtains $\nabla \cdot \mathbf{u}=\sum_{p=1}^N L_p^{RHS}$, after summing the volume fraction equation over $p$ and noticing $\sum_{p=1}^N C_p=1$, and $\mathbf{m}=\rho \mathbf{u}$. After involving the mixture theory, $\sum_{p=1}^N L_p^{RHS}=0$ relates $\mathbf{u}$ to the volume-averaged velocity, while $\sum_{p=1}^N \rho_p L_p^{RHS}=0$ implies the mass-averaged one.
    \item
    The algorithm of determining $\{L_p^c\}_{p=1}^N$ in Eq.(\ref{Eq Phase-Field}) will be introduced in the following section, and it can also be used to obtain $\{L_p^{mc}\}_{p=1}^N$, corresponding to $\{L_p^c\}_{p=1}^N$, in the mass-averaged velocity models based on $\{C_p\}_{p=1}^N$. Similar to the constraints in Eq.(\ref{Eq Conservation L_p^c}) and Eq.(\ref{Eq Consistency L_p^c}), $(\int_\Omega L_p^{mc} d\Omega=S_p^{mc})$ and $(L_p^{mc}|_{C_p=0}=0)$ should be followed by $\{L_p^{mc}\}_{p=1}^N$, but the summation constraint becomes $(\sum_{p=1}^N \rho_p L_p^{mc}=0)$, different from Eq.(\ref{Eq Summation constraint L_p^c}). Note that $\{S_p^{mc}\}_{p=1}^N$ are known, like their correspondences $\{S_p^c\}_{p=1}^N$ in Eq.(\ref{Eq Conservation L_p^c}), and that $C_p=0$ is the same as $\phi_p=-1$, see Eq.(\ref{Eq Volume fraction}). We can first determine $\{(\rho_p L_p^{mc})\}_{p=1}^N$ that satisfy
    \begin{equation}\nonumber
        \sum_{q=1}^N (\rho_q L_q^{mc})=0,\quad
        \int_{\Omega} (\rho_p L_p^{mc}) d\Omega= \rho_p S_p^{mc},\quad
        (\rho_p L_p^{mc})|_{C_p=0}=0, \quad 1 \leqslant p \leqslant N,
    \end{equation}
    following the consistent and conservative volume distribution algorithm proposed in the present study, because the above constraints for $\{(\rho_p L_p^{mc})\}_{p=1}^N$ are equivalent to those in Eq.(\ref{Eq Summation constraint L_p^c}), Eq.(\ref{Eq Conservation L_p^c}), and Eq.(\ref{Eq Consistency L_p^c}) for $\{L_p^c\}_{p=1}^N$. Finally, $\{L_p^{mc}\}_{p=1}^N$ are obtained from $\{(\rho_p L_p^{mc})/\rho_p\}_{p=1}^N$. 
\end{itemize}
}

\section{The consistent and conservative volume distribution algorithm and its applications}\label{Sec Volume distribution and its application}
In this section, the consistent and conservative volume distribution algorithm is described in detail in Section \ref{Sec Volume distribution}. Then, its two applications, one on the continuous level and the other on the discrete level, to Phase-Field models are introduced in Section \ref{Sec L_p^c} and Section \ref{Sec Boundedness mapping}, respectively. In Section \ref{Sec L_p^c}, a multiphase conservative Allen-Cahn model that satisfies the \textit{consistency of reduction} is developed. In Section \ref{Sec Boundedness mapping}, a numerical procedure, called the boundedness mapping, is developed to physically map the numerically obtained order parameters into their physical interval.

\subsection{The consistent and conservative volume distribution algorithm}\label{Sec Volume distribution}
The purpose of the consistent and conservative volume distribution algorithm is to specify the volume distribution functions of individual phases, denoted as $\{L_p\}_{p=1}^N$, in a consistent and conservative manner.

\textit{\textbf{Problem statement:}} Given a set of order parameters $\{\phi_p\}_{p=1}^N$ that satisfy their summation constraint Eq.(\ref{Eq Summation constraint phi}), i.e., $\sum_{p=1}^N \phi_p=(2-N)$, and a set of scalars $\{S_p\}_{p=1}^N$ that satisfy $\sum_{p=1}^N S_p=0$, determine a set of spatial functions $\{L_p\}_{p=1}^N$ such that
\begin{equation}\label{Eq Constraints L}
\sum_{q=1}^N L_q=0,
\quad
\int_{\Omega} L_p d\Omega = S_p, 
\quad
L_p|_{\phi_p= -1}=0,
\quad 1 \leqslant p \leqslant N.
\end{equation}
Here, $\{S_p\}_{p=1}^N$ are related to the volume changes of individual phases, and the admissible set has a zero summation over $p$. In other words, the net volume added to the domain is zero. Therefore, the present algorithm keeps the domain volume fixed but adjusts the phase volumes based on the given values.

\begin{figure}[!t]
	\centering
	\includegraphics[scale=.4]{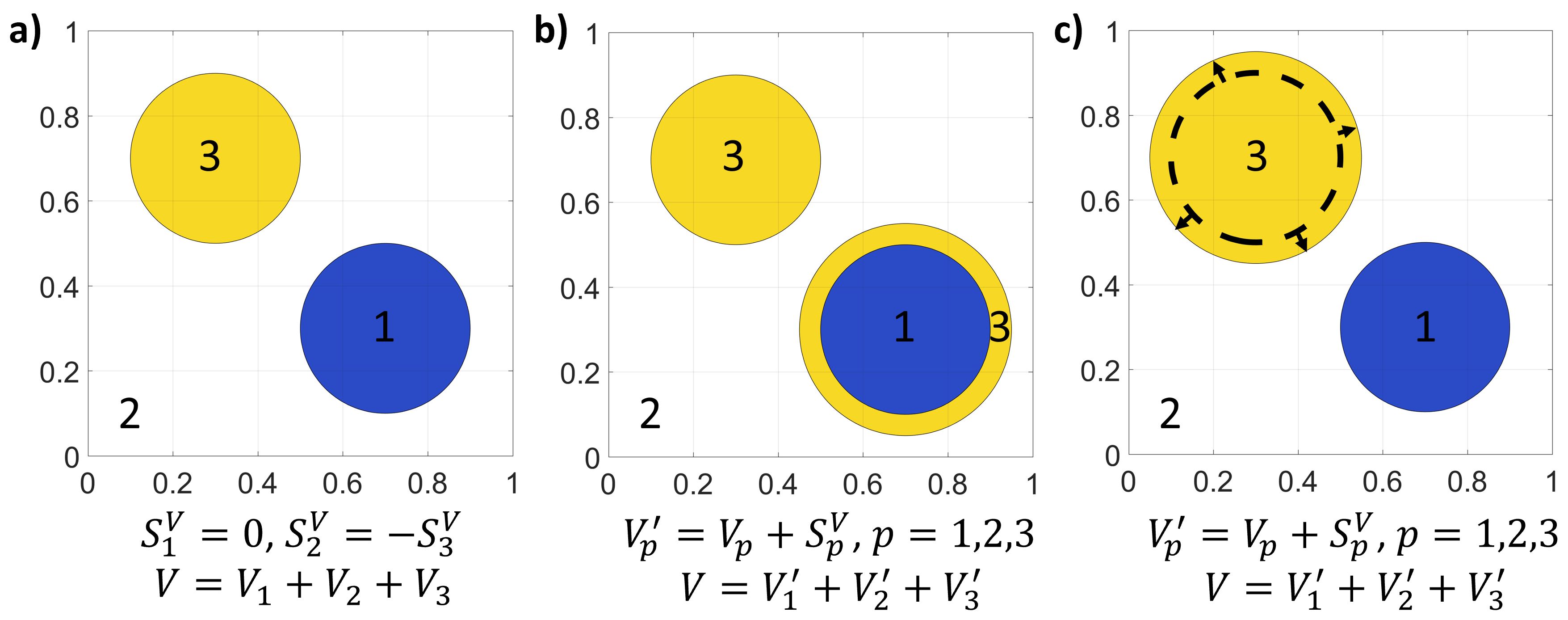}
	\caption{Schematic of the volume distribution problem. a) Configuration of the phases before the volume distribution. b) Example of an inadmissible solution of the volume distribution problem. c) Example of an admissible solution of the volume distribution problem. Blue: Phase 1. White: Phase 2. Yellow: Phase 3. $\{S_p^V\}_{p=1}^3$ are the given volume changes of the phases to be distributed to the domain. $\{V_p\}_{p=1}^3$ are the volumes of the phases before the volume distribution. $\{V'_p\}_{p=1}^3$ are the volumes of the phases after the volume distribution. $V$ is the total volume of the domain which does not change after the volume distribution. The volume changes are magnified for illustration purpose.\label{Fig VolumeDistribution-Schematic}}
\end{figure}

Given $\{S_p\}_{p=1}^N$ only, there can be multiple choices of $\{L_p\}_{p=1}^N$, while not all of them are admissible. Fig.\ref{Fig VolumeDistribution-Schematic} is a schematic showing two possible solutions, in Fig.\ref{Fig VolumeDistribution-Schematic} b) and c), of the volume distribution problem illustrated in Fig.\ref{Fig VolumeDistribution-Schematic} a). If Phases 1, 2, and 3 represent the air, water, and oil, respectively, Fig.\ref{Fig VolumeDistribution-Schematic} b) shows an oil ring suddenly produced, surrounding the air, after the volume distribution. This is unphysical and will significantly change the dynamics, since the density and viscosity around the air are greatly changed. In Fig.\ref{Fig VolumeDistribution-Schematic} c), the added volume of Phase 3 is placed around the region where Phase 3 was originally located. Therefore, there is no sudden appearance of Phase 3 near the interface of Phases 1 and 2 after the volume distribution. Moreover, one will obtain the same solution in Fig.\ref{Fig VolumeDistribution-Schematic} c) even though Phase 1 is absent, while this is not the case in Fig.\ref{Fig VolumeDistribution-Schematic} b). Therefore, the solution in Fig.\ref{Fig VolumeDistribution-Schematic} c) is admissible. It should be noted that the volume changes in Fig.\ref{Fig VolumeDistribution-Schematic} are magnified for illustration purpose. 
As a result, in addition to $\{S_p\}_{p=1}^N$, three physical constraints are proposed, which need to be strictly satisfied by $\{L_p\}_{p=1}^N$, and they are formulated in Eq.(\ref{Eq Constraints L}). 
The first constraint is called the summation constraint for volume distribution, which states that the summation of the order parameters after the volume distribution does not change, i.e., $\sum_{p=1}^N (\phi_p+L_p)=(2-N)$, so that the summation of the volume fractions of the phases is always unity, see Eq.(\ref{Eq Summation constraint phi}) and Eq.(\ref{Eq Volume fraction}). In other words, no local void or overfill can be generated by the volume distribution.
The second constraint in Eq.(\ref{Eq Constraints L}) is called the conservation constraint for volume distribution, which requires the total amounts of $\{L_p\}_{p=1}^N$ equal to the given values $\{S_p\}_{p=1}^N$. Otherwise, the volume distribution problem is not solved successfully. It should be noted that the summation and conservation constraints for volume distribution are consistent with each other due to $\sum_{p=1}^N S_p=0$.
The last constraint is related to the \textit{consistency of reduction}. If Phase $p$ is labeled absent by $\phi_p$ at a specific location, then there should not be any volume of Phase $p$ being distributed to that location. This constraint avoids producing any fictitious phases after the volume distribution at that location. As a result, solutions like Fig.\ref{Fig VolumeDistribution-Schematic} b) will not be produced, because Phase 3 (yellow) is absent near Phase 1 (blue) circle before the volume distribution (see Fig.\ref{Fig VolumeDistribution-Schematic} a)).
Combining the first and last constraints in Eq.(\ref{Eq Constraints L}), one can easily obtain $L_p|_{\phi_p= 1}=0$. In other words, the volume distribution only happens at the interfacial regions, while it is deactivated inside the bulk-phase regions. 

Specifying a set of $\{L_p\}_{p=1}^N$ that satisfy all the aforementioned constraints is not a trivial task. A successful algorithm is developed in \cite{BrasselBretin2011,Huangetal2020CAC} for two-phase flows. However, it is far more challenging in a general multiphase case.
When there are only two phases, the constraints in Eq.(\ref{Eq Constraints L}) for $L_1$ turn into $\int_\Omega L_1 d\Omega=S_1$ and $L_1|_{\phi_1=\pm1}=0$. Once $L_1$ is determined, $L_2=-L_1$ is directly obtained from the first constraint in Eq.(\ref{Eq Constraints L}), and $\{L_p\}_{p=1}^2$ satisfy all the constraints in Eq.(\ref{Eq Constraints L}). As a result, only phase-wise formulations are needed.
On the other hand, if we consider a three-phase example, even though $L_1$ is specified, one still can not determine $L_2$ and $L_3$ directly or uniquely. If $L_1$ and $L_2$ are specified independently from phase-wise formulations, there is no guarantee that $L_3=-(L_1+L_2)$ from the first constraint in Eq.(\ref{Eq Constraints L}) satisfies the rest of the constraints.
Due to the complexity of including multiple phases, the volume distribution is unable to be performed phase-wise. Instead, a coupled formulation is proposed for the volume distribution functions:
\begin{equation}\label{Eq L_p}
L_p=\sum_{q=1}^N W_{p,q} B_q, \quad 1 \leqslant p \leqslant N,
\end{equation}
where $W_{p,q}$ is the weight function for volume distribution and should be non-zero only in the interfacial regions including Phases $p$ and $q$ from the \textit{consistency of reduction}.
Eq.(\ref{Eq L_p}) can be conceptually understood as following. $B_q$, which is related to the total volume of Phase $q$ to be distributed in the domain, can only be distributed to the interfacial regions including Phase $q$. Therefore, the volume distributed to the interfacial regions including both Phases $p$ and $q$ from $B_q$ is $W_{p,q}B_q$. Then $L_p$ is obtained by summing all the contributions from the phases.  The constraints in Eq.(\ref{Eq Constraints L}) for $\{L_p\}_{p=1}^N$ turn into the following for $\{W_{p,q}\}_{p,q=1}^N$, i.e.,
\begin{equation}\label{Eq Constraints W_{p.q}}
\sum_{r=1}^N W_{r,q} =0,
\quad
\sum_{r=1}^N \left(\int_{\Omega} W_{p,r}  d\Omega \right) B_r= S_p,
\quad
W_{p,q}|_{\phi_p=-1}=0,
\quad
1 \leqslant p,q \leqslant N.
\end{equation}

Based on the \textit{consistency of reduction}, i.e., the third constraint in Eq.(\ref{Eq Constraints W_{p.q}}), we construct the weight function for volume distribution to be
\begin{equation}\label{Eq W_{p,q}}
W_{p,q}=\left\{
\begin{array}{ll}
-(1+\phi_p)(1+\phi_q), p \neq q,\\
(1+\phi_p)(1-\phi_q),  p=q,
\end{array}
\right.
\quad 1 \leqslant p,q \leqslant N.
\end{equation}
It should be noted that $W_{p,q}$ indicates the interfacial regions including both Phases $p$ and $q$ because it is non-zero only where $-1 < \phi_p,\phi_q < 1$. Moreover, the summation constraint, i.e., the first constraint in Eq.(\ref{Eq Constraints W_{p.q}}), is also satisfied by $W_{p,q}$ in Eq.(\ref{Eq W_{p,q}}) given $\sum_{p=1}^N \phi_p=(2-N)$. The remaining step is to satisfy the conservation constraint, i.e., the second constraint in Eq.(\ref{Eq Constraints W_{p.q}}). This is achieved by solving the linear system for $\{B_p\}_{p=1}^N$, i.e.,
\begin{equation}\label{Eq B}
[A_{p,q}]_{N\times N}[B_q]_{N\times1}=[S_p]_{N\times1},
\quad
A_{p,q}=\int_{\Omega} W_{p,q} d\Omega.
\end{equation}
The coefficient matrix of the linear system Eq.(\ref{Eq B}), i.e., $[A_{p,q}]_{N\times N}$, includes the integrals of the weight function $W_{p,q}$ $(1\leqslant p,q \leqslant N)$ over the domain. Since $W_{p,q}$ is symmetric, $[A_{p,q}]_{N\times N}$ is symmetric as well. It should be noted that all the diagonal elements of $[A_{p,q}]_{N\times N}$ are positive while all the off-diagonal ones are negative, and additionally that $\sum_{q=1}^N A_{p,q}$ $(1\leqslant p \leqslant N)$ is zero, from the definition of $W_{p,q}$ in Eq.(\ref{Eq W_{p,q}}).  This implies that $\sum_{q=1,q \neq p}^N |A_{p,q}|=|A_{p,p}|$ $(1 \leqslant p \leqslant N)$, which shows that the coefficient matrix $[A_{p,q}]_{N\times N}$ in Eq.(\ref{Eq B}) is not only symmetry but also diagonally dominant.
Another important observation of Eq.(\ref{Eq B}) is that the rank of $[A_{p,q}]_{N\times N}$ is at most $(N-1)$. After summing Eq.(\ref{Eq B}) over $p$, an equation of ``$0=0$'' is obtained because both $\sum_{p=1}^N W_{p,q}$ and $\sum_{p=1}^N S_p$ are zero. As a result, the linear system in Eq.(\ref{Eq B}) has multiple solutions, and our implementations show that solving Eq.(\ref{Eq B}) following the scaling argument below is critical to specify the admissible solution and for the success of the algorithm, especially when $\{S_p\}_{p=1}^N$ are close to the round-off error.
If $[|A_{p,q}|]_{N\times N}$ is of $O(1)$, then $[|B_p|]_{N\times 1}$ should share the same order of magnitude as $[|S_p|]_{N\times 1}$. A robust way to employ this scaling argument is to let $B_{q^*}$ equal to $\max|S_p|$, where $q^*$ is chosen in such a way that the minimum absolute value other than zero of $[A_{p,q}]_{N\times N}$ is in column $[A_{p,q^*}]_{N\times 1}$. Consequently, Eq.(\ref{Eq B}) has a unique solution that honors the scaling argument. 
A special case is when there is a phase, e.g., Phase $p$, doesn't have any interfacial regions in the whole domain. Equivalently, Phase $p$ is either globally absent, i.e., $\phi_p\equiv-1$, or filling the entire domain, i.e., $\phi_p\equiv1$. As a result, both the $p$th row and column in $[A_{p,q}]_{N\times N}$ are zero, and we set $B_p=0$ when this happens.   
In order to obtain $[|A_{p,q}|]_{N\times N} \sim O(1)$, the coefficient matrix $[A_{p,q}]_{N\times N}$ is rescaled by its maximum absolute value, i.e., $\max|A_{p,q}|$, if that value is not zero. Specifically, after obtaining $[A_{p,q}]_{N\times N}$ from the integrals of $\{W_{p,q}\}_{p,q=1}^N$, see Eq.(\ref{Eq B}), $[A_{p,q}]_{N\times N}$ is replaced by $[A_{p,q}]_{N\times N}/\max|A_{p,q}|$, and correspondingly $W_{p,q}$ $(1 \leqslant p,q \leqslant N)$ is replaced by $W_{p,q}/\max|A_{p,q}|$. As a result, the final coefficient matrix $[A_{p,q}]_{N\times N}$ is always of $O(1)$, independent of the domain size. 
Once $\{B_p\}_{p=1}^N$ are solved from Eq.(\ref{Eq B}), the volume distribution functions for individual phases $\{L_p\}_{p=1}^N$ are obtained from Eq.(\ref{Eq L_p}).

The volume distribution problem defined at the beginning of this section is solved by the following algorithm:
\begin{itemize}
\item \textit{Applying the weight function $W_{p,q}$ defined in Eq.(\ref{Eq W_{p,q}}) and solving $\{B_p\}_{p=1}^N$ from the $N\times N$ symmetry and diagonally dominant linear system in Eq.(\ref{Eq B}), $\{L_p\}_{p=1}^N$ are determined from Eq.(\ref{Eq L_p}), which satisfy all the physical constraints in Eq.(\ref{Eq Constraints L}).}
\end{itemize}
The algorithm is consistent and conservative in the sense that the resulting $\{L_p\}_{p=1}^N$ satisfy the \textit{consistency of reduction} (see Theorem \ref{Theorem Volume consistency of reduction} and the proof is in Appendix) and their integrals over the domain are equal to the given values $\{S_p\}_{p=1}^N$ (see Eq.(\ref{Eq B})).
\begin{theorem}\label{Theorem Volume consistency of reduction}
The proposed consistent and conservative volume distribution algorithm is reduction consistent such that the values of $\{L_p\}_{p=1}^N$ for the absent phases are zero and the formulation of $\{L_p\}_{p=1}^N$ for the present phases reduces to the corresponding one excluding the absent phases.
\end{theorem}

Lastly, we consider the two-phase case, and have $\phi_1+\phi_2=0$ and $S_1+S_2=0$ as the admissible inputs. From Eq.(\ref{Eq B}), we obtain the following equations
\begin{equation}\nonumber
\left\{
\begin{array}{ll}
S_1=B_1 \int_{\Omega} W_{1,1} d\Omega + B_2 \int_{\Omega} W_{1,2} d\Omega 
=( B_1-B_2 )\int_{\Omega} (1-\phi_1^2) d\Omega
=( B_1-B_2 )\int_{\Omega} W_1 d\Omega \\
S_2=B_1 \int_{\Omega} W_{2,1} d\Omega +B_2 \int_{\Omega} W_{2,2} d\Omega 
=-(B_1  - B_2 ) \int_{\Omega} (1-\phi_2^2) d\Omega
=-(B_1  - B_2 ) \int_{\Omega} W_2 d\Omega
\end{array}
\right.,
\end{equation}
and it should be noted that the above two equations are identical due to $S_1=-S_2$, $\phi_1=-\phi_2$, and $W_1=W_2$. Recall that $W_p=W(\phi_p)$ defined in Eq.(\ref{Eq Weight}).
Finally from Eq.(\ref{Eq L_p}), we obtain
\begin{equation}\label{Eq Volume distribution two-phase}
\left\{
\begin{array}{ll}
L_1=B_1 W_{1,1} + B_2 W_{1,2}
=( B_1  - B_2 ) (1-\phi_1^2)
=\frac{W_1}{\int_{\Omega} W_1 d\Omega} S_1,\\
L_2=B_1 W_{2,1} +B_2 W_{2,2}
=-(B_1  - B_2 ) (1-\phi_2^2)
=\frac{W_2}{\int_{\Omega} W_2 d\Omega} S_2.
\end{array}
\right.
\end{equation}
Therefore, the volume distribution algorithm becomes phase-wise in two-phase cases, and Eq.(\ref{Eq Volume distribution two-phase}) is identical to those in \citep{BrasselBretin2011,Huangetal2020CAC}.

In summary, a consistent and conservative volume distribution algorithm for multiple phases is developed. The complexity for multiphase cases originates in simultaneously satisfying the three physical constraints for volume distribution, which are the summation constraint, the conservation constraint, and the \textit{consistency of reduction} in Eq.(\ref{Eq Constraints L}) and Theorem \ref{Theorem Volume consistency of reduction}. Therefore, the volume distribution is unable to be performed phase-wise, like the two-phase case, but has to consider the contributions from different phases altogether.  

\textit{\textbf{Remark:}
In numerical implementations, the integrals in the volume distribution algorithm can be approximated by a quadrature rule.
}

\subsection{The reduction consistent multiphase conservative Allen-Cahn model}\label{Sec L_p^c}
The difficulty of specifying $\{L_p^c\}_{p=1}^N$ in Section \ref{Sec Phase-Field} that satisfy the physical constraints in Eq.(\ref{Eq Summation constraint L_p^c}), Eq.(\ref{Eq Conservation L_p^c}), and Eq.(\ref{Eq Consistency L_p^c}) is addressed in the present work using the consistent and conservative volume distribution algorithm developed in Section \ref{Sec Volume distribution}, by noticing that the constraints for $\{L_p^c\}_{p=1}^N$ are the same kind as those in Eq.(\ref{Eq Constraints L}) for $\{L_p\}_{p=1}^N$. To provide a specific example, we develop a reduction consistent multiphase conservative Allen-Cahn model. 

The diffusion fluxes and reaction terms of the multiphase conservative Allen-Cahn model considered in the present work are
\begin{equation}\label{Eq Allen-Cahn parts}
\mathbf{J}_p = M_0 \lambda_0 \nabla  \phi_p,
\quad
L_p^R=-\frac{M_0 \lambda_0}{\eta^2} \left(  g'_1(\phi_p)  -  \frac{1+\phi_p}{2} L^s \right),
\quad
L^s=\sum_{p=1}^N g'_1(\phi_p),
\end{equation}
where $M_0$ is the mobility, $\lambda_0$ is the maximum among $\lambda_{p,q}$, i.e., $\lambda_0=\max{\lambda_{p,q}}$, and $L^s$ is the Lagrange multiplier to enforce $\sum_{p=1}^N L_p^R=0$. It is obvious that both $\{\mathbf{J}_p\}_{p=1}^N$ and $\{L_p^R\}_{p=1}^N$ in Eq.(\ref{Eq Allen-Cahn parts}) are admissible, see Section \ref{Sec Phase-Field}. Correspondingly, we have
\begin{equation}\label{Eq Allen-Cahn S_p^c}
S_p^c= \int_{\Omega} \frac{M_0 \lambda_0}{\eta^2} \left(  g'_1(\phi_p)  -  \frac{1+\phi_p}{2} L^s \right)  d\Omega,
\quad
1 \leqslant p \leqslant N,
\end{equation}
and the summation of $\{S_p^c\}_{p=1}^N$ over $p$ is zero.
Plugging Eq.(\ref{Eq Allen-Cahn parts}) into the Phase-Field model Eq.(\ref{Eq Phase-Field}), the reduction consistent multiphase conservative Allen-Cahn model is
\begin{equation}\label{Eq Allen-Cahn}
\frac{\partial \phi_p}{\partial t}
+
\nabla \cdot ( \mathbf{u} \phi_p )
=
M_0 \lambda_0 \nabla^2 \phi_p
-\frac{M_0 \lambda_0}{\eta^2} \left(  g'_1(\phi_p)  -  \frac{1+\phi_p}{2} L^s \right)
+L_p^c,
\quad
1 \leqslant p \leqslant N,
\end{equation}
where $\{L_p^c\}_{p=1}^N$ are obtained from the consistent and conservative volume distribution algorithm in Section \ref{Sec Volume distribution}, using $\{\phi_p\}_{p=1}^N$ and $\{S_p^c\}_{p=1}^N$ in Eq.(\ref{Eq Allen-Cahn S_p^c}) as the inputs, and they satisfy the physical constraints in Eq.(\ref{Eq Summation constraint L_p^c}), Eq.(\ref{Eq Conservation L_p^c}), and Eq.(\ref{Eq Consistency L_p^c}). The proposed multiphase conservative Allen-Cahn model Eq.(\ref{Eq Allen-Cahn}) has the following properties and the proofs are available in Appendix.

\begin{theorem}\label{Theorem CAC summation}
The multiphse conservative Allen-Cahn model Eq.(\ref{Eq Allen-Cahn}) admits the summation constraint for the order parameters Eq.(\ref{Eq Summation constraint phi}), i.e., 
\begin{equation}\nonumber
\sum_{p=1}^N \phi_p=(2-N).
\end{equation}
\end{theorem}

\begin{theorem}\label{Theorem CAC conservation}
The multiphse conservative Allen-Cahn model Eq.(\ref{Eq Allen-Cahn}) satisfies the conservation constraint for the order parameters Eq.(\ref{Eq Conservation phi}), i.e., 
\begin{equation}\nonumber
\frac{d}{dt} \int_{\Omega} \phi_p d\Omega=0,
\quad
1 \leqslant p \leqslant N, 
\end{equation}
with a proper boundary condition.
\end{theorem}

\begin{theorem}\label{Theorem CAC reduction}
The multiphse conservative Allen-Cahn model Eq.(\ref{Eq Allen-Cahn}) satisfies the consistency of reduction such that the order parameters solved from Eq.(\ref{Eq Allen-Cahn}) for the absent phases are $-1$ and Eq.(\ref{Eq Allen-Cahn}) for the present phases reduces to the corresponding one excluding the absent phases.
\end{theorem}

It should be noted that, when there are only two phases, we have $L^s=0$, and $\{L_p^c\}_{p=1}^2$ are
\[
L_p^c=\frac{W_p}{\int_{\Omega} W_p d\Omega} S_p^c,\quad p=1,2,
\]
from Eq.(\ref{Eq Volume distribution two-phase}). As a result, the multiphase conservative Allen-Cahn model Eq.(\ref{Eq Allen-Cahn}) proposed in the present work exactly recovers the two-phase conservative Allen-Cahn model which is proposed by Brassel
and Bretin \cite{BrasselBretin2011}, further studied in \cite{Kimetal2014,LeeKim2016}, and later on applied to two-phase flows \cite{JeongKim2017,JoshiJaiman2018,JoshiJaiman2018adapt,Huangetal2020CAC}. 

Kim and Lee \cite{KimLee2017} developed a similar conservative Allen-Cahn model for multiphase flows. The only difference from the present work is that they defined $\{L_p^c\}_{p=1}^N$ as
\begin{equation}\label{Eq L_p^c KimLee}
L_p^c=\frac{ \sum_{q=1}^N W_q }{ \sum_{q=1}^N \int_{\Omega} W_q d\Omega} S_p^c,
\quad 1 \leqslant p \leqslant N.
\end{equation}
Recall that $W_p=W(\phi_p)$ is defined in Eq.(\ref{Eq Weight}). Although $\{L_p^c\}_{p=1}^N$ defined in Eq.(\ref{Eq L_p^c KimLee}) satisfy Eq.(\ref{Eq Summation constraint L_p^c}) and Eq.(\ref{Eq Conservation L_p^c}) so that $\sum_{q=1}^N \phi_q=(2-N)$ and $\frac{d}{dt} \int_{\Omega} \phi_p d\Omega=0$ $(1\leqslant p \leqslant N)$, they are not reduction consistent due to violating Eq.(\ref{Eq Consistency L_p^c}). As a result, fictitious phases can be generated by their model. Consider a three-phase example at the location where Phases 1 and 2 form an interfacial region, i.e., $-1< \phi_1,\phi_2<1$, and Phase 3 is absent around this region, i.e., $\phi_3=-1$ and $|\nabla \phi_3|=\nabla^2\phi_3=0$, like near the blue circle in Fig.\ref{Fig VolumeDistribution-Schematic} a). In Eq.(\ref{Eq L_p^c KimLee}), $\sum_{m=1}^3 W_m$ is positive and non-zero, and $S_3^c$ is not necessarily zero because (i) Phase 3 can appear somewhere away from the considered interfacial region, see the yellow circle in Fig.\ref{Fig VolumeDistribution-Schematic} a) as an example, and (ii) $S_3^c$ defined in Eq.(\ref{Eq Allen-Cahn S_p^c}) is an integral over the entire domain. As a result, around the considered interfacial region, we have $\frac{\partial \phi_3}{\partial t}=L_3^c$ with $L_3^c$ defined in Eq.(\ref{Eq L_p^c KimLee}) from \citep{KimLee2017}. If $S_3^c$ is again positive, then from Eq.(\ref{Eq L_p^c KimLee}), $L_3^c$ is positive, which leads to Phase 3 being generated around the interfacial region of Phases 1 and 2, like the yellow ring surrounding the blue circle in Fig.\ref{Fig VolumeDistribution-Schematic} b). On the other hand, $\phi_3$ will be less than $-1$ if $S_3^c$ is negative. Neither of the results is physical. One can only expect $S_3^c$ to be zero if either $\phi_3 \equiv 1$ or $\phi_3 \equiv -1$ in the entire domain. However, these two cases are meaningless in practice since they restrict the problem to be single- or two-phase. The effect of producing fictitious phases from the multiphase conservative Allen-Cahn model in \citep{KimLee2017} is demonstrated in Section \ref{Sec Fictitious phases} and the results are shown in Fig.\ref{Fig Fictitious}. On the other hand, no fictitious phase is produced by the proposed model, thanks to satisfying the \textit{consistency of reduction}, i.e., Theorem \ref{Theorem CAC reduction}, and see also Section \ref{Sec Fictitious phases} and Fig.\ref{Fig Fictitious}.

In summary, with the help of the proposed consistent and conservative volume distribution algorithm in Section \ref{Sec Volume distribution} to specify $\{L_p^c\}_{p=1}^N$, we develop a multiphase conservative Allen-Cahn model Eq.(\ref{Eq Allen-Cahn}) that satisfies not only the summation constraint for the order parameters Eq.(\ref{Eq Summation constraint phi}) and the mass conservation Eq.(\ref{Eq Conservation phi}) but also the \textit{consistency of reduction}, see Theorem \ref{Theorem CAC summation}, Theorem \ref{Theorem CAC conservation}, and Theorem \ref{Theorem CAC reduction}. A corresponding consistent and conservative numerical scheme is developed in Section \ref{Sec Scheme CAC} for the proposed multiphase conservative Allen-Cahn model Eq.(\ref{Eq Allen-Cahn}), which preserves the physical properties of the model on the discrete level.

\subsection{The boundedness mapping}\label{Sec Boundedness mapping}
The consistent and conservative volume distribution algorithm described in Section \ref{Sec Volume distribution} is used to develop a numerical procedure, called the boundedness mapping, to tackle out-of-bound order parameters that commonly appear in numerical practice, while the physical properties of the order parameters remain intact.

\textit{\textbf{Problem statement:}} Given a set of spatially discretized order parameters $\{\phi_p\}_{p=1}^N$ that satisfies the summation constraint Eq.(\ref{Eq Summation constraint phi}), i.e., $\sum_{p=1}^N \phi_p=(2-N)$, and a set of scalars $\{S_{\phi_p}\}_{p=1}^N$ that are the total amounts of individual order parameters in the domain, determine a mapping from $\{\phi_p\}_{p=1}^N$ to $\{\phi_p^b\}_{p=1}^N$, such that
\begin{equation}\label{Eq Constraints phi^b}
\sum_{q=1}^N \phi_q^b=2-N,
\quad
\sum_{n_C} [\phi_p^b \Delta \Omega]_{n_C}=S_{\phi_p},
\quad
\phi_p^b|_{\phi_p \leqslant -1}=-1,
\quad
\phi_p^b \in [-1,1],
\quad
1 \leqslant p \leqslant N.
\end{equation}
The first two constraints in Eq.(\ref{Eq Constraints phi^b}) corresponds to the summation and conservation constraints of the order parameters, i.e., Eq.(\ref{Eq Summation constraint phi}) and Eq.(\ref{Eq Conservation phi}), respectively. It should be noted that admissible $\{S_{\phi_p}\}_{p=1}^N$ satisfies $\sum_{p=1}^N S_{\phi_p}=(2-N) |\Omega|$, where $|\Omega|=\sum_{n_C} [\Delta \Omega]_{n_C}$ is the volume of the entire domain, so that the first two constraints in Eq.(\ref{Eq Constraints phi^b}) are consistent. $\{\phi_p^b\}_{p=1}^N$ are reduction consistent with $\{\phi_p\}_{p=1}^N$ in the sense that locations labeled as Phase $p$ absent by $\phi_p$ are also labeled as Phase $p$ absent by $\phi_p^b$ $(1 \leqslant p \leqslant N)$. This is formulated in the third constraint in Eq.(\ref{Eq Constraints phi^b}). In addition, the order parameters after the mapping stay in their physical interval, which is the last constraint in Eq.(\ref{Eq Constraints phi^b}). 

The boundedness mapping includes the clipping step, the rescaling step, and the conservation step, which are preformed sequentially.

The clipping step is
\begin{equation}\label{Eq Boundedness Clipping}
\phi_p^{b*}=\left\{
\begin{array}{ll}
1, \phi_p \geqslant 1,\\
-1,\phi_p \leqslant -1,\\
\phi_p, \mathrm{else},
\end{array}
\right.
\quad
1 \leqslant p \leqslant N.
\end{equation}

The rescaling step is
\begin{equation}\label{Eq Boundedness rescaling}
C_p^{b*}=\frac{1+\phi_p^{b*}}{2},
\quad
C_p^{b**}=\frac{C_p^{b*}}{\sum_{q=1}^N C_q^{b*}},
\quad
\phi_p^{b**}=2C_p^{b**}-1,
\quad
1 \leqslant p \leqslant N.
\end{equation}

The conservation step is
\begin{equation}\label{Eq Boundedness conservation}
\phi_p^b=\phi_p^{b**}+\mathcal{L}_p^b,
\quad 1 \leqslant p \leqslant N,
\end{equation}
where, from Eq.(\ref{Eq Constraints phi^b}), $\{\mathcal{L}_p^b\}_{p=1}^N$ have the following constraints
\begin{equation}\label{Eq Boundedness Constraints}
\sum_{q=1}^N \mathcal{L}_q^b=0,
\quad
\sum_{n_C} [\mathcal{L}_p^b \Delta\Omega]_{n_C}
=
S_{\phi_p}-\sum_{n_C} [\phi_p^{b**} \Delta \Omega]_{n_C}
=S_p^b,
\quad
\mathcal{L}_p^b|_{\phi_p^{b**}=-1}=0,
\quad 
1 \leqslant p \leqslant N.
\end{equation}
Notice that the constraints in Eq.(\ref{Eq Boundedness Constraints}) for $\{\mathcal{L}_p^b\}_{p=1}^N$ are the same kind as those in Eq.(\ref{Eq Constraints L}) for $\{L_p\}_{p=1}^N$ in Section \ref{Sec Volume distribution} after using the mid-point rule to approximate the integrals. Therefore, $\{\mathcal{L}_p^b\}_{p=1}^N$ are determined from the consistent and conservative volume distribution algorithm in Section \ref{Sec Volume distribution} with inputs $\{\phi_p^{b**}\}_{p=1}^N$ and $\{S_p^b\}_{p=1}^N$, and the resulting $\{\mathcal{L}_p^b\}_{p=1}^N$ satisfies Eq.(\ref{Eq Boundedness Constraints}). It should be noted that the inputs $\{\phi_p^{b**}\}_{p=1}^N$ and $\{S_p^b\}_{p=1}^N$ are admissible because $\sum_{p=1}^N \phi_p^{b**}=(2-N)$ from the rescaling step Eq.(\ref{Eq Boundedness rescaling}) and $\sum_{p=1}^N {S_p^b}=0$.

The clipping step Eq.(\ref{Eq Boundedness Clipping}) removes the out-of-bound error from the input order parameters, and the rescaling step Eq.(\ref{Eq Boundedness rescaling}) enforces the summation constraint for the order parameters. The intermediate results $\{\phi_p^{b**}\}_{p=1}^N$ satisfy the first and last constraints in Eq.(\ref{Eq Boundedness Constraints}). The conservation step Eq.(\ref{Eq Boundedness conservation}) is supplemented so that the final results $\{\phi_p^b\}_{p=1}^N$ additionally satisfy the second constraint in Eq.(\ref{Eq Boundedness Constraints}), i.e., the amounts of the order parameters in the entire domain match the given values.
After performing the boundedness mapping, i.e., the three steps above, it is obvious that the first two constraints for $\{\phi_p^b\}_{p=1}^N$ in Eq.(\ref{Eq Constraints phi^b}) are enforced. 
The third constraint in Eq.(\ref{Eq Constraints phi^b}) is also true. Given $\phi_p\leqslant-1$, from the clipping step Eq.(\ref{Eq Boundedness Clipping}), we obtain $\phi_p^{b*}=-1$. Therefore, both $C_p^{b*}$ and $C_p^{b**}$ are zero and $\phi_p^{b**}$ is again $-1$ after the rescaling step Eq.(\ref{Eq Boundedness rescaling}). Thanks to Eq.(\ref{Eq Boundedness Constraints}), we finally have $\mathcal{L}_{p}^b=0$ and obtain $\phi_p^{b}=\phi_p^{b**}=-1$ from the conservation step Eq.(\ref{Eq Boundedness conservation}). Therefore, the third constraint for $\{\phi_p^b\}_{p=1}^N$ in Eq.(\ref{Eq Constraints phi^b}) is satisfied.
Although the last constraint in Eq.(\ref{Eq Constraints phi^b}), i.e., $\{\phi_p^b\}_{p=1}^N \in [-1,1]$, is not explicitly enforced in the conservation step Eq.(\ref{Eq Boundedness conservation}), it should be noted that out-of-bound $\phi_p^b$ $(1\leqslant p \leqslant N)$, if there is any, most probably appears where $|\phi_p^{b**}|$ is close to one, due to $\mathcal{L}_p^b|_{\phi_p^{b**}=\pm 1}=0$ and $\phi_p^{b**} \in [-1,1]$. On the other hand, $\mathcal{L}_p^b$ is close to zero at those locations, see the formulations and analysis in Section \ref{Sec Volume distribution}. In practice, we always find the last constraint in Eq.(\ref{Eq Constraints phi^b}), i.e., $\{\phi_p^b\}_{p=1}^N \in [-1,1]$, satisfied after performing the boundedness mapping. If the out-of-bound issue appears in $\{\phi_p^b\}_{p=1}^N$, one can iteratively apply the boundedness mapping, letting $\{\phi_p^b\}_{p=1}^N$ as the new input, until the boundedness constraint, i.e., $\{\phi_p^b\}_{p=1}^N \in [-1,1]$, is achieved.
Following the analysis in Section \ref{Sec Volume distribution} and Eq.(\ref{Eq Volume distribution two-phase}), the proposed boundedness mapping exactly reduces to the one in \cite{Huangetal2020CAC} for two-phase flows, and it is also reduction consistent (see Theorem \ref{Theorem Boundedness reduction} and the proof is in Appendix).

\begin{theorem}\label{Theorem Boundedness reduction}
The boundedness mapping, including the clipping step Eq.(\ref{Eq Boundedness Clipping}), the rescaling step Eq.(\ref{Eq Boundedness rescaling}), and the conservation step Eq.(\ref{Eq Boundedness conservation}), satisfies the consistency of reduction such that the absent phases remain absent after the boundedness mapping and the formulation of the boundedness mapping for the present phases reduces to the corresponding one excluding the absent phases.
\end{theorem}

In summary, the boundedness mapping, which is a numerical procedure, is developed, with the help of the consistent and conservative volume distribution algorithm in Section \ref{Sec Volume distribution}. It includes the clipping step Eq.(\ref{Eq Boundedness Clipping}), the rescaling step Eq.(\ref{Eq Boundedness rescaling}), and the conservation step Eq.(\ref{Eq Boundedness conservation}), and is shown to be reduction consistent. Given a set of out-of-bound order parameters, the output of the mapping, i.e., $\{\phi_p^b\}_{p=1}^N$, not only are bounded by their physical interval, i.e., $[-1,1]$ in the present work, but also satisfy the summation constraint for the order parameters Eq.(\ref{Eq Summation constraint phi}), i.e., $\sum_{p=1}^N \phi_p^b=(2-N)$, match the given amounts of the order parameters in the entire domain $\{S_{\phi_p}\}_{p=1}^N$, and is reduction consistent with the input order parameters $\{\phi_p\}_{p=1}^N$ in the sense that Phase $p$ $(1\leqslant p \leqslant N)$ won't be mapped to the location where Phase $p$ is labeled absent by $\phi_p$. This mapping is directly applicable to numerical solutions of various multiphase models.

\section{Discretizations}\label{Sec Discretizations}
The discretizations of all the differential operators follow those in \cite{Huangetal2020}. In summary, we consider the collocated grid arrangement, where all the variables are defined at cell centers, and additional normal velocities are defined at cell faces. The convective operators are approximated by the 5th-order WENO scheme \cite{JiangShu1996}, while the other differential operators are discretized by the 2nd-order central difference. To distinguish the discrete operators from their corresponding continuous ones, we add $\tilde{(\cdot)}$ on top of them, e.g., $\tilde{\nabla}$ means the discrete gradient operator. The linear interpolation from the nearest neighbors is denoted by $\overline{f}$, while any other approximation of $f$ from its nodal values is denoted as $\tilde{f}$. The integral is approximated by the mid-point rule, i.e., $\int_{\Omega} f d\Omega \approx \sum_{n_C} [f \Delta \Omega]_{n_C}$, where $n_C$ is the cell index and $\Delta \Omega$ is the cell volume. The discrete divergence operator has the following property, i.e., $\sum_{n_C} [\tilde{\nabla} \cdot \mathbf{f} \Delta \Omega]_{n_C}=0$, if the domain is periodic or the normal component of $\mathbf{f}$ vanishes at the domain boundary, see the proof in \citep{Huangetal2020,Huangetal2020N}.
The time derivative is approximated by $\frac{\gamma_t f^{n+1}-\hat{f}}{\Delta t}$, where $f^{n+1}$ is the value of $f$ at time level $n+1$, $\Delta t$ is the times step, and $\hat{f}$ and $\gamma_t=\hat{1}$ are scheme dependent. Unless otherwise specified, we use the 2nd-order backward difference to approximate the time derivative, and $\gamma_t=1.5$ and $\hat{f}=2f^{n}-0.5f^{n-1}$ in this case. $f^{*,n+1}$ is an extrapolation along the time direction and it is $2f^{n}-f^{n-1}$ for the 2nd-order case.

The momentum equation Eq.(\ref{Eq Momentum}) is solved by the scheme in \cite{Huangetal2020}, and the divergence-free condition Eq.(\ref{Eq Divergence-free}) is enforced by the cell-face velocity, i.e., 
\begin{equation}\label{Eq Divergence-free discrete}
\tilde{\nabla} \cdot \mathbf{u}=0,
\end{equation} 
at all the discrete cells and time levels. This scheme has been extensively analyzed and successfully applied to two- and multi-phase problems in \cite{Huangetal2020,Huangetal2020N,Huangetal2020CAC}. Without considering the surface force Eq.(\ref{Eq Surface force}) (and the gravity), the momentum of the multiphase flow is conserved at the discrete level, i.e., $\sum_{n_C} [\rho \mathbf{u} \Delta \Omega]_{n_C}^n=\sum_{n_C} [\rho \mathbf{u} \Delta \Omega]_{n_C}^0$, in a periodic domain. The surface force Eq.(\ref{Eq Surface force}) can be discretized by either the balanced-force method, which achieves better numeral force balance, or the conservative method, which fully conserves the momentum \cite{Huangetal2020N}.
As long as the scheme for the Phase-Field model preserves its \textit{consistency of reduction} and the discrete mass flux $\mathbf{\tilde{m}}$ in the discretized momentum equation satisfies the \textit{consistency of mass conservation} on the discrete level, the scheme in \citep{Huangetal2020} for the momentum equation Eq.(\ref{Eq Momentum}) satisfies the \textit{consistency of reduction} and the \textit{consistency of mass and momentum transport} on the discrete level. Those analyses are given in detail in \cite{Huangetal2020,Huangetal2020N,Huangetal2020CAC}. 

Therefore, in the rest of this section, the major focus is on the scheme for the proposed multiphase conservative Allen-Cahn model Eq.(\ref{Eq Allen-Cahn}) in Section \ref{Sec L_p^c}, and the formulation of the discrete consistent mass flux.
In Section \ref{Sec Scheme CAC}, a semi-implicit, mass conservative, and reduction consistent scheme is developed to solve the proposed multiphase conservative Allen-Cahn model Eq.(\ref{Eq Allen-Cahn}). The resulting order parameters from the scheme satisfy their summation and conservation constraints, i.e., Eq.(\ref{Eq Summation constraint phi}) and Eq.(\ref{Eq Conservation phi}), and is reduction consistent, in the discrete sense. Then, in Section \ref{Sec Apply}, the boundedness mapping in Section \ref{Sec Boundedness mapping} is implemented so that the final solution of the order parameters is in addition bounded in their physical interval $[-1,1]$. The consistent formulation is applied to obtain the discrete consistent mass flux that preserves the \textit{consistency of mass conservation} on the discrete level.

For comparison and discussion in Section \ref{Sec Results}, we consider the reduction consistent multiphase Cahn-Hilliard model in \cite{Dong2018,Huangetal2020N}, whose diffusion fluxes and reaction terms are
\begin{equation}\label{Eq Cahn-Hilliard}
\mathbf{J}_p=\sum_{r=1}^N M_{p,r} \nabla \xi_r,
\quad
L_p^R=L_p^c=0,
\quad
M_{p,q}=\left\{
\begin{array}{ll}
-M_0 (1+\phi_p) (1+\phi_q), p \neq q,\\
M_0 (1+\phi_p) (1-\phi_q), p=q,
\end{array}
\right.
1 \leqslant p,q \leqslant N,
\end{equation}
where $\{\xi_p\}_{p=1}^N$ are the chemical potentials defined in Eq.(\ref{Eq Chemical Potential}). The Cahn-Hilliard model Eq.(\ref{Eq Cahn-Hilliard}) honors the summation and conservation constraints in Eq.(\ref{Eq Summation constraint phi}) and Eq.(\ref{Eq Conservation phi}), respectively, and the \textit{consistency of reduction}. 
The Canh-Hilliard model Eq.(\ref{Eq Cahn-Hilliard}) is solved by the mass conservative and reduction consistent scheme in \cite{Huangetal2020N}, and its numerical solution is proved to preserve those aforementioned physical properties. Again, the boundedness mapping in Section \ref{Sec Boundedness mapping} is supplemented in the same manner in Section \ref{Sec Apply} in order to bound the order parameters and to obtain the discrete consistent mass flux.

\subsection{The consistent and conservative scheme for the multiphase conservative Allen-Cahn model}\label{Sec Scheme CAC}
Given data at all the previous time levels, the semi-implicit, mass conservative, and reduction consistent scheme for the proposed multiphase conservative Allen-Cahn model Eq.(\ref{Eq Allen-Cahn}) in Section \ref{Sec L_p^c} solves for $\{\phi_p^{n+1}\}_{p=1}^N$ from the following four steps. 

\textbf{Step 1:} 
Solve the Allen-Cahn equation, i.e., Eq.(\ref{Eq Allen-Cahn}) excluding all the Lagrange multiplies $L^s$ and $L_p^c$, from
\begin{equation}\label{Eq CAC discrete step1}
\frac{\gamma_t \phi_p^{*}-\hat{\phi_p}}{\Delta t}
+
\tilde{\nabla} \cdot (\mathbf{u}^{*,n+1} \tilde{\phi}_p^{*,n+1})
=
M_0 \lambda_0 \tilde{\nabla} \cdot \tilde{\nabla} \phi_p^{*}
-
\frac{M_0 \lambda_0}{\eta^2} \tilde{g}'_1(\phi_p^{*}),
\quad 1 \leqslant p \leqslant N,
\end{equation}
where $\tilde{g}'_1(\phi_p^{*})$ is the linear approximation of $g'_1(\phi_p^{*})$ from its Taylor expansion at $\phi_p^{n}$, i.e.,
\begin{equation}\nonumber
\tilde{g}'_1(\phi_p^*)=g'_1(\phi_p^n)+g''_1(\phi_p^n)(\phi_p^*-\phi_p^n),
\quad
1 \leqslant p \leqslant N.
\end{equation}
This step is also implemented to solve the two-phase conservative Allen-Cahn model in \citep{Huangetal2020CAC}. The gradient-based phase selection procedure \citep{Huangetal2020N} is implemented to correct $\{(\mathbf{u}^{*,n+1} \tilde{\phi}_p^{*,n+1})\}_{p=1}^N$ in the convection term of Eq.(\ref{Eq CAC discrete step1}).

\textbf{Step 2:}
Compute the discrete Lagrange multiplier $\tilde{L}_s$ from
\begin{equation}\label{Eq L^s discrete}
\tilde{L}^s
=
\sum_{p=1}^N \left( \tilde{g}'_1(\phi_p^{*})
-
\eta^2 \tilde{\nabla} \cdot \tilde{\nabla} \phi_p^{*} \right).
\end{equation}
Compared to Eq.(\ref{Eq Allen-Cahn parts}), the appearance of $\tilde{\nabla} \cdot \tilde{\nabla} \phi_p^{*}$ in Eq.(\ref{Eq L^s discrete}) is due to $\sum_{p=1}^N \phi_p^{*} \neq (2-N)$. 

\textbf{Step 3:}
Compute the discrete Lagrange multipliers $\{\tilde{L}_p^c\}_{p=1}^N$ from the consistent and conservative volume distribution algorithm in Section \ref{Sec Volume distribution}, and the inputs are $\{\phi_p^n\}_{p=1}^N$ and $\{\tilde{S}_p^c\}_{p=1}^N$. Here
\begin{equation}\label{Eq Conservation L_p^c discrete}
\tilde{S}_p^c
=
\sum_{n_C} \left[ \frac{M_0 \lambda_0}{\eta^2} \left( \tilde{g}'_1(\phi_p^{*})-\frac{1+\phi_p^n}{2} \tilde{L}^s \right) \Delta \Omega \right]_{n_C},
\quad 1 \leqslant p \leqslant N.
\end{equation}
The resulting $\{\tilde{L}_p^c\}_{p=1}^N$ not only are reduction consistent, see Theorem \ref{Theorem Volume consistency of reduction}, but also satisfy Eq.(\ref{Eq Constraints L}) on the discrete level, i.e.,
\begin{equation}\label{Eq Constraints L_p^c discrete}
\sum_{q=1}^N \tilde{L}_q^c=0,
\quad
\sum_{n_C} [\tilde{L}_p^c \Delta\Omega]_{n_C}=\tilde{S}_p^c,
\quad
\tilde{L}_p^c|_{\phi_p^n=-1}=0, 
\quad
1 \leqslant p \leqslant N.
\end{equation}

\textbf{Step 4:}
Obtain the solution at time level $n+1$ from 
\begin{equation}\label{Eq CAC discrete step2}
\frac{\gamma_t \phi_p^{n+1} -\gamma_t \phi_p^{*}}{\Delta t}
=
\frac{M_0 \lambda_0}{\eta^2} \frac{1+\phi_p^n}{2} \tilde{L}^s + \tilde{L}_p^c, \quad 1 \leqslant p \leqslant N.
\end{equation}

In summary, $\{\phi_p^{n+1}\}_{p=1}^N$ obtained from the proposed scheme, i.e., the above four steps, satisfy the following properties:
\begin{theorem}\label{Theorem CAC summation discrete}
The proposed scheme for the multiphase conservative Allen-Cahn model Eq.(\ref{Eq Allen-Cahn}) satisfies the summation constraint for the order parameters Eq.(\ref{Eq Summation constraint phi}), i.e., 
\begin{equation}\nonumber
\sum_{p=1}^N \phi_p=(2-N),
\end{equation}
at every discrete location and time level. Therefore no local valid or overfilling can be produced numerically.
\end{theorem}

\begin{theorem}\label{Theorem CAC conservation discrete}
The proposed scheme for the multiphase conservative Allen-Cahn model Eq.(\ref{Eq Allen-Cahn}) satisfies the conservation constraint for the order parameters, i.e., 
\begin{equation}\nonumber
\sum_{n_C} [\phi_p^n \Delta \Omega]_{n_C}=\sum_{n_C} [\phi_p^0 \Delta \Omega]_{n_C},
\quad
1 \leqslant p \leqslant N,
\quad
n=1,2,3,\dots
\end{equation}
Therefore the mass of each phase is conserved numerically.
\end{theorem}

\begin{theorem}\label{Theorem CAC reduction discrete}
The proposed scheme for the multiphase conservative Allen-Cahn model Eq.(\ref{Eq Allen-Cahn}) satisfies the \textit{consistency of reduction} on the discrete level such that the absent phases remain absent and the present phases are updated from the formulations excluding the absent phases. Therefore no fictitious phases can be produced numerically.
\end{theorem}

It should be noted that several details of the scheme are critical to the success of preserving the physical properties of the order parameters, such as implementing the gradient-based phase selection procedure \citep{Huangetal2020N} to the convection term in \textbf{Step 1} (Eq.(\ref{Eq CAC discrete step1})), including $\tilde{\nabla} \cdot \tilde{\nabla} \phi_p^{*}$ in \textbf{Step 2} (Eq.(\ref{Eq L^s discrete})) when computing $\tilde{L}^s$, and solving all the order parameters from their governing equations instead of doing so for the first ($N-1$) order parameters and then using $\sum_{p=1}^N \phi_p=(2-N)$ (Eq.(\ref{Eq Summation constraint phi})), see the proofs of the above theorems as well as the remarks below them in Appendix.

\textit{\textbf{Remark:} It is suggested in \cite{LeeKim2015} that the prefactor $\frac{1+\phi_p}{2}$ of $L^s$ in Eq.(\ref{Eq Allen-Cahn}) is replaced by $\left(\frac{1+\phi_p}{2}\right)^{\gamma_s}$, where $\gamma_s$ is larger than $1$, in order to reduce the amount of fictitious phases. The numerical results in \cite{LeeKim2015} indicate that by increasing $\gamma_s$ by $1$, the peak of the fictitious phase is about two orders of magnitude smaller. However, as shown in Theorem \ref{Theorem CAC reduction} and Theorem \ref{Theorem CAC reduction discrete}, both the proposed conservative Allen-Cahn model Eq.(\ref{Eq Allen-Cahn}) using $\frac{1+\phi_p}{2}$ ahead of $L^s$ and its numerical scheme developed in this section are reduction consistent. Therefore, the production of any fictitious phases is eliminated, since the absent phases remain absent in the proofs. This property is demonstrated in Section \ref{Sec Results}. With the same procedure of the proofs of Theorem \ref{Theorem CAC reduction} and Theorem \ref{Theorem CAC reduction discrete}, choosing $\left(\frac{1+\phi_p}{2}\right)^{\gamma_s}$ $(\gamma_s \geqslant 1)$ ahead of $L^s$ in Eq.(\ref{Eq Allen-Cahn}) does not change the consistency of reduction of the proposed conservative Allen-Cahn model and its numerical scheme. The reason for generating fictitious phases in \cite{LeeKim2015} is that the scheme in \cite{LeeKim2015} does not satisfy the \textit{consistency of reduction}. 
}

\subsection{Implementation of the boundedness mapping and consistent formulation}\label{Sec Apply}
It should be noted that $\{\phi_p^{n+1}\}_{p=1}^N$, obtained from the scheme in Section \ref{Sec Scheme CAC}, are possibly outside their physical interval, i.e., $[-1,1]$ in the present study. The boundedness mapping in Section \ref{Sec Boundedness mapping} is implemented to finalize the order parameters at the new time level. Specifically, input $\{\phi_p^{n+1}\}_{p=1}^N$ and $\{S_{\phi_p^{n+1}}\}_{p=1}^N$ into the mapping and obtain $\{\phi_p^b\}_{p=1}^N$ that satisfy Eq.(\ref{Eq Constraints phi^b}). Here, $\{\phi_p^{n+1}\}_{p=1}^N$ are the solution of the scheme in Section \ref{Sec Scheme CAC}, and $\{S_{\phi_p^{n+1}}\}_{p=1}^N$ are
\begin{equation}\label{Eq Sum phi_p}
S_{\phi_p^{n+1}}= \sum_{n_C} [\phi_p^{n+1} \Delta \Omega]_{n_C},
\quad
1 \leqslant p \leqslant N,
\end{equation}
because the scheme in Section \ref{Sec Scheme CAC} satisfies the conservation constraint, see Theorem \ref{Theorem CAC conservation discrete}. Both $\{\phi_p^{n+1}\}_{p=1}^N$ and $\{S_{\phi_p^{n+1}}\}_{p=1}^N$ are the admissible inputs of the boundedness mapping algorithm in Section \ref{Sec Boundedness mapping} due to $\sum_{p=1}^N \phi_p^{n+1}=(2-N)$, see Theorem \ref{Theorem CAC summation discrete}. 

In summary, the order parameters at the new time level $\{\phi_p^b\}_{p=1}^N$, which is mapped from $\{\phi_p^{n+1}\}_{p=1}^N$, have the following properties:
\begin{itemize}
    \item 
    They satisfy the summation constraint, i.e., $\sum_{p=1}^N {\phi_p^b}=\sum_{p=1}^N {\phi_p^{n+1}}=(2-N)$, see Theorem \ref{Theorem CAC summation discrete} and the first constraint for $\{\phi_p^b\}_{p=1}^N$ in Eq.(\ref{Eq Constraints phi^b}), and therefore no local valid or overfilling can be produced numerically.
    \item 
    They satisfy the conservation constraint, i.e., $\sum_{n_C} [\phi_p^b \Delta \Omega]_{n_C}=\sum_{n_C} [\phi_p^{n+1} \Delta \Omega]_{n_C}=\sum_{n_C} [\phi_p^0 \Delta \Omega]_{n_C}$ $(1 \leqslant p \leqslant N)$, see Theorem \ref{Theorem CAC conservation discrete} and the second constraint for $\{\phi_p^b\}_{p=1}^N$ in Eq.(\ref{Eq Constraints phi^b}), and therefore the mass of each phase is conserved numerically.
    \item
    They satisfy the \textit{consistency of reduction}, see Theorem \ref{Theorem CAC reduction discrete} and Theorem \ref{Theorem Boundedness reduction}, and therefore no fictitious phases can be produced numerically.
    \item
    They satisfy the boundedness constraint, i.e., $\{\phi_p^b\}_{p=1}^N \in [-1,1]$, see the last constraint for $\{\phi_p^b\}_{p=1}^N$ in Eq.(\ref{Eq Constraints phi^b}).
\end{itemize}

\textit{\textbf{Remark:}
\begin{itemize}
    \item 
    From Theorem \ref{Theorem CAC summation discrete} and Theorem \ref{Theorem CAC reduction discrete}, one can deduce that $\{\phi_p^{n+1}\}_{p=1}^N$ are either $1$ or $-1$ inside bulk-phase regions. As a result, the out-of-bound error only possibly appears at interfacial regions. Therefore, the clipping step Eq.(\ref{Eq Boundedness Clipping}) of the boundedness mapping in Section \ref{Sec Boundedness mapping} is only effective in interfacial regions but does not modify any existing bulk-phase regions labeled by $\{\phi_p^{n+1}\}_{p=1}^N$.
    \item
    The out-of-bound error in $\{\phi_p^{n+1}\}_{p=1}^N$ is normally small. The largest out-of-bound error observed in the present study is usually of $O(10^{-5})$ in one time step even in problems including strong interactions among phases. In all the results reported in the present study, we only need to perform the boundedness mapping in Section \ref{Sec Boundedness mapping} (Eq.(\ref{Eq Boundedness Clipping}), Eq.(\ref{Eq Boundedness rescaling}), and Eq.(\ref{Eq Boundedness conservation})) once in each time step, and the resulting $\{\phi_p^b\}_{p=1}^N$ already satisfy all the constraints in Eq.(\ref{Eq Constraints phi^b}).
\end{itemize}
}

The next task is to determine the discrete consistent mass flux $\tilde{\mathbf{m}}$. We define the discrete Lagrange multipliers of the boundedness mapping $\{\tilde{L}_p^b\}_{p=1}^N$ as
\begin{equation}\label{Eq L_p^b discrete}
\tilde{L}_p^b=\frac{\gamma_t \phi_p^{b}-\gamma_t \phi_p^{n+1}}{\Delta t}, 
\quad 1 \leqslant p \leqslant N,
\end{equation}
which quantifies the effect of the boundedness mapping on the order parameters. Notice that $\sum_{n_C}[\tilde{L}_p^b\Delta \Omega]_{n_C}$ $(1 \leqslant p \leqslant N)$ is zero due to $\sum_{n_C}[\phi_p^b\Delta \Omega]_{n_C}=\sum_{n_C}[\phi_p^{n+1}\Delta \Omega]_{n_C}$.
Combining the 4 steps in Section \ref{Sec Scheme CAC} and Eq.(\ref{Eq L_p^b discrete}), the fully-discretized equation of the order parameters, including the effect of the boundedness mapping, is
\begin{equation}\label{Eq CACB discrete}
\frac{\gamma_t \phi_p^{n+1}-\hat{\phi_p}}{\Delta t}
+
\tilde{\nabla} \cdot (\mathbf{u}^{*,n+1} \tilde{\phi}_p^{*,n+1})
=
\tilde{\nabla} \cdot \underbrace{M_0 \lambda_0 \tilde{\nabla} \phi_p^{*}}_{\tilde{\mathbf{J}}_p}
\underbrace{-
\frac{M_0 \lambda_0}{\eta^2} 
\left( \tilde{g}'_1(\phi_p^{*})
-
\frac{1+\phi_p^n}{2} \tilde{L}^s 
\right)}_{\tilde{L}_p^R}
+ 
\tilde{L}_p^c
+
\tilde{L}_p^b,
\quad 1 \leqslant p \leqslant N.
\end{equation}
We have renamed $\{\phi_p^{b}\}_{p=1}^N$ as $\{\phi_p^{n+1}\}_{p=1}^N$ on the left-hand side of Eq.(\ref{Eq CACB discrete}), and labeled the discrete diffusion fluxes $\{\tilde{\mathbf{J}}_p\}_{p=1}^N$ and the discrete reaction terms $\{\tilde{L}_p^R\}_{p=1}^N$ on the right-hand side of Eq.(\ref{Eq CACB discrete}). Then the consistent formulation proposed in \citep{Huangetal2020CAC} is performed discretely, i.e.,
\begin{equation}\label{Eq Consistent formulation discrete}
\tilde{\nabla} \cdot \left( \overline{W_p^{n+1}} \tilde{\nabla} Q_p \right)
=
\tilde{L}_p^Q, 
\quad
\tilde{L}_p^Q=\tilde{L}_p^R+\tilde{L}_p^c+\tilde{L}_p^b, 
\quad
1 \leqslant p \leqslant N,
\end{equation} 
which is the discrete counterpart of Eq.(\ref{Eq Consistent formulation}). It should be noted that $\{\tilde{L}_p^b\}_{p=1}^N$, representing the effect of the boundedness mapping, are included in $\{\tilde{L}_p^Q\}_{p=1}^N$, while they, however, do not appear in Eq.(\ref{Eq Consistent formulation}). 
In order to successfully apply the consistent formulation, as discussed in Section \ref{Sec Consistent formulation}, $\sum_{n_C} [\tilde{L}_p^Q \Delta \Omega]_{n_C}$ $(1 \leqslant p \leqslant N)$ needs to be zero, which is true, see Eq.(\ref{Eq CACB discrete}), Eq.(\ref{Eq Constraints L_p^c discrete}), Eq.(\ref{Eq Conservation L_p^c discrete}), and the analysis below Eq.(\ref{Eq L_p^b discrete}).
After solving $\{Q_p\}_{p=1}^N$ from Eq.(\ref{Eq Consistent formulation discrete}), the discrete Phase-Field fluxes are computed as
\begin{equation}\label{Eq Phase-Field flux CAC discrete}
\tilde{\mathbf{m}}_{\phi_p}=\mathbf{u}^{*,n+1} \tilde{\phi}_p^{*,n+1}-\tilde{\mathbf{J}}_p-\overline{W_p^{n+1}} \tilde{\nabla} Q_p,
\quad 1 \leqslant p \leqslant N,
\end{equation}
which is the discrete counterpart of Eq.(\ref{Eq Phase-Field flux}).
As a result, the fully-discretized equation Eq.(\ref{Eq CACB discrete}) is equivalent to
\begin{equation}\label{Eq Phase-Field conservation discrete}
\frac{\gamma_t \phi_p^{n+1}-\hat{\phi_p}}{\Delta t}
+
\tilde{\nabla}\cdot \tilde{\mathbf{m}}_{\phi_p}=0,
\quad 1 \leqslant p \leqslant N,
\end{equation}
which is the discrete counterpart of Eq.(\ref{Eq Phase-Field conservation}). Since $\mathbf{u}^{*,n+1}$ appears in the convection part of the Phase-Field fluxes, i.e., $\mathbf{u}^{*,n+1} \tilde{\phi}_p^{*,n+1}$ in Eq.(\ref{Eq Phase-Field flux CAC discrete}), considering the \textit{consistency of reduction}, see the analysis in \citep{Huangetal2020N}, and from Eq.(\ref{Eq Mass flux}), the discrete consistent mass flux is finally computed from
\begin{equation}\label{Eq Mass flux discrete}
\tilde{\mathbf{m}}=\sum_{p=1}^N \frac{\rho_p}{2} (\mathbf{u}^{*,n+1}+\tilde{\mathbf{m}}_{\phi_p}).
\end{equation}

It can be easily shown, after using Eq.(\ref{Eq Phase-Field conservation discrete}) and Eq.(\ref{Eq Divergence-free discrete}), that the density of the fluid mixture Eq.(\ref{Eq Density}) and the discrete consistent mass flux Eq.(\ref{Eq Mass flux discrete}) satisfy the following fully-discretized equation:
\begin{equation}\label{Eq Mass discrete}
\frac{\gamma_t \rho^{n+1} -\hat{\rho}}{\Delta t}
+
\tilde{\nabla} \cdot \tilde{\mathbf{m}}
=
\sum_{p=1}^N \frac{\rho_p}{2} \left(
\underbrace{
	\frac{\gamma_t - \overbrace{ \hat{1} }^{\gamma_t} }{\Delta t} }_{0}
+ \underbrace{
	\tilde{\nabla} \cdot \mathbf{u}^{*,n+1} }_{0}
+ \underbrace{
	\frac{\gamma_t \phi_p^{n+1}  - \hat{\phi}_p }{\Delta t}
	+\tilde{\nabla} \cdot \tilde{\mathbf{m}}_{\phi_p} }_{0}
\right)
=
0,
\end{equation}
which is the discrete counterpart of the mass conservation equation Eq.(\ref{Eq Mass}). This is also consistent with the analysis in \citep{Huangetal2020N}. Therefore, the discrete consistent mass flux Eq.(\ref{Eq Mass flux discrete}) satisfies the \textit{consistency of mass conservation} on the discrete level. Then we can proceed to solve the momentum equation Eq.(\ref{Eq Momentum}) with the scheme in \citep{Huangetal2020}. We do not repeat the details and analyses of the scheme for the momentum equation Eq.(\ref{Eq Momentum}) here, and interested readers can refer to \citep{Huangetal2020,Huangetal2020CAC,Huangetal2020N}.

\section{Results}\label{Sec Results}
In this section, various numerical tests are performed. The Phase-Field models and their schemes, including the applications of the consistent and conservative volume distribution algorithm in Section \ref{Sec Volume distribution}, are further validated using the benchmark numerical tests in Sections \ref{Sec Fictitious phases}-\ref{Sec Convergence}, where the dimensionless values of the parameters input to the tests are directly reported. To demonstrate the capability of the present models in multiphase flows, Section \ref{Sec Dam break} reports a multiphase problem extended from the experimental configuration in \citep{MartinMoyce1952}, where physical values of the material properties are first reported and then non-dimensionalized based on the setup in \citep{MartinMoyce1952}. All the results are presented in their dimensionless forms. For brevity, we use CH and CAC to denote the multiphase Cahn-Hilliard and conservative Allen-Cahn models, respectively. If the boundedness mapping is supplemented, those models are denoted by CHB and CACB. We use $h$ to denote the grid/cell size in this section.

\subsection{Fictitious phases}\label{Sec Fictitious phases}
To illustrate the importance of satisfying the \textit{consistency of reduction}, we compare the numerical equilibrium state of the Phase-Field models without coupling to the flow, i.e., $\mathbf{u} \equiv \mathbf{0}$. In addition to CH, CHB, CAC, and CACB, the multiphase conservative Allen-Cahn model, proposed in \cite{KimLee2017}, is supplemented, and it is called CACN in this case. CACN is solved by the same scheme in Section \ref{Sec Scheme CAC} for CAC except that $\{L_p^c\}_{p=1}^N$ follow the definition in Eq.(\ref{Eq L_p^c KimLee}) from \cite{KimLee2017}. Therefore, all the differences shown here between CAC and CACN are rooted in the different definitions of $\{L_p^c\}_{p=1}^N$. As analyzed in Section \ref{Sec L_p^c}, CACN is not reduction consistent and can generate fictitious phases. 

\begin{figure}[!t]
	\centering
	\includegraphics[scale=.4]{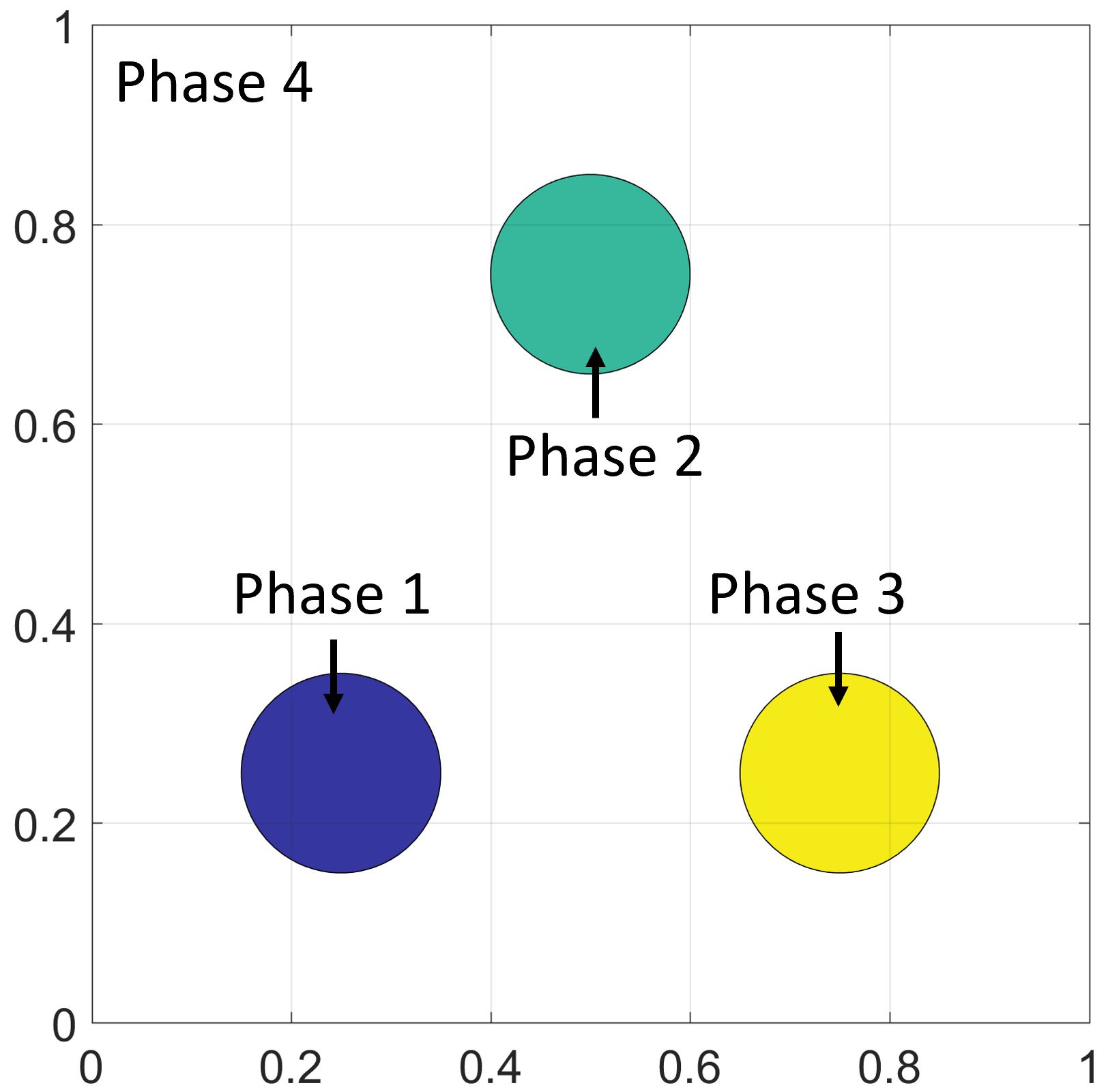}
	\caption{Schematic of the setup of the fictitious phases. Blue: Phase 1. Green: Phase 2. Yellow: Phase 3. White: Phase 4.\label{Fig Fictitious-Schematic}}
\end{figure}

Schematic of this problem is shown in Fig.\ref{Fig Fictitious-Schematic}. The domain considered is $[1\times1]$ with homogeneous Neumann boundaries. The domain is discretized by $[128\times128]$ cells and the time step is $\Delta t=10h=\frac{10}{128}$. We set $\eta=0.015$ and the off-diagonal elements of $\lambda_{p,q}$ are $\frac{3}{2\sqrt{2}}\eta$. $M_0$ is $(\eta^2/\lambda_0)$ for CAC, CACB, and CACN, and is $10^{-3} \eta^2$ for CH and CHB due to numerical stability. The circle of Phase 1 is at $(0.25,0.25)$ with a radius $0.1$. The circle of Phase 2 is at $(0.5,0.75)$ with a radius $0.1$. The circle of Phase 3 is at $(0.75,0.25)$ with a radius $0.1$. Phase 4 occupies the rest of the domain. The circles of Phases 1, 2, and 3 are separated far enough, so there are no intersections among them. Therefore, $\eta^3|\nabla \phi_1||\nabla \phi_2||\nabla \phi_3|$ should be zero. 

\begin{figure}[!t]
	\centering
	\includegraphics[scale=.4]{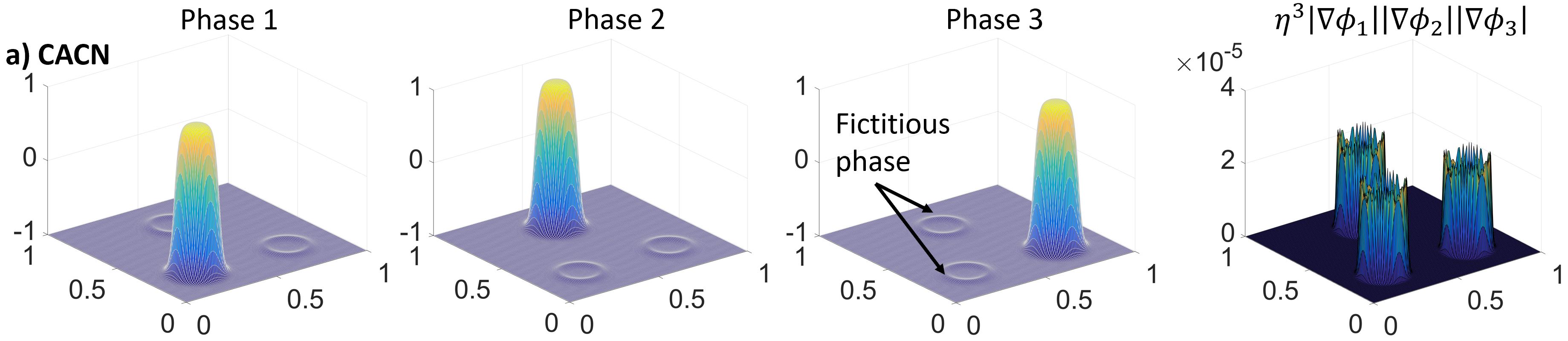}
	\includegraphics[scale=.4]{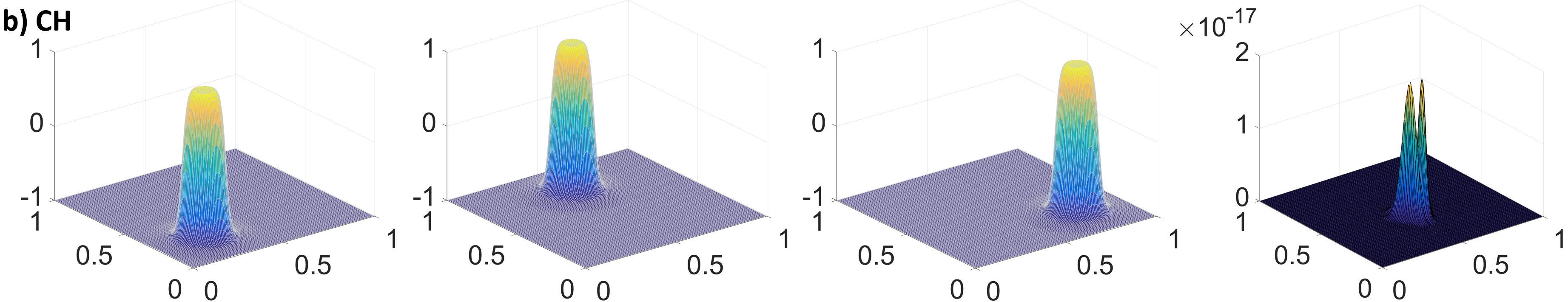}
	\includegraphics[scale=.4]{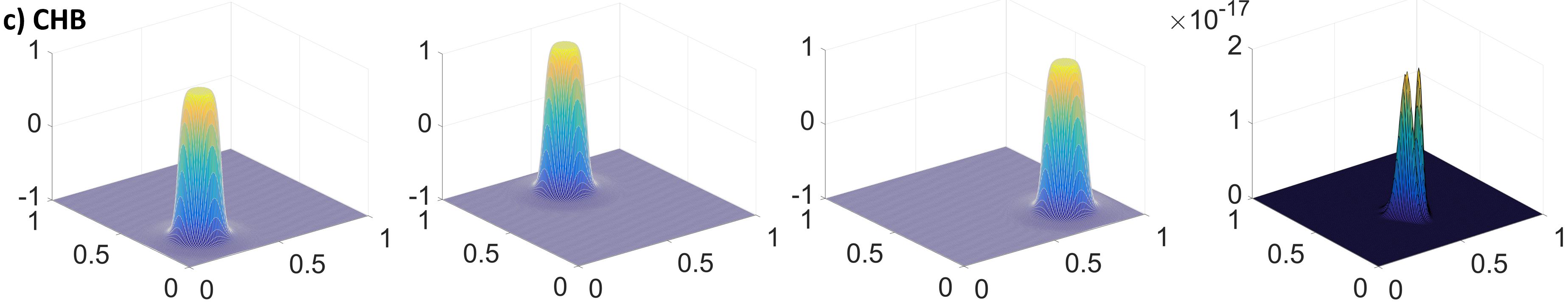}
	\includegraphics[scale=.4]{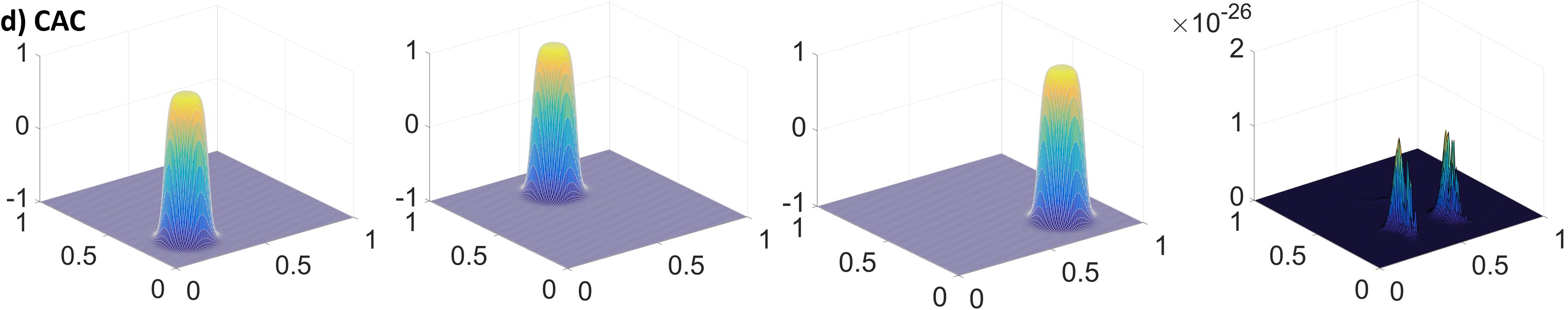}
	\includegraphics[scale=.4]{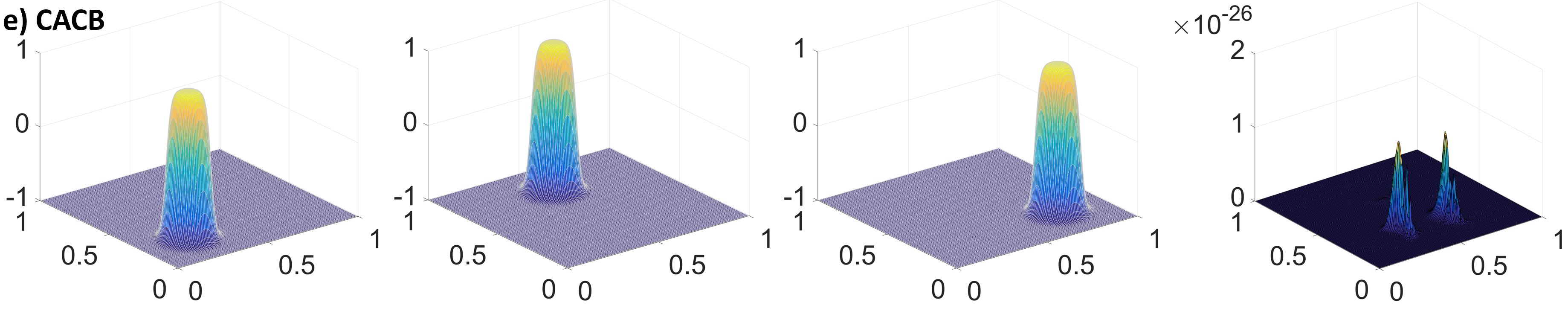}
	\caption{Results of the fictitious phases from a) CACN, b) CH, c) CHB, d) CAC, and e) CACB. The first column: Profile of phase 1. The second column: Profile of phase 2. The third column: Profile of phase 3. The fourth column: $\eta^3|\nabla \phi_1||\nabla \phi_2||\nabla \phi_3|$.\label{Fig Fictitious}}
\end{figure}

Fig.\ref{Fig Fictitious} shows the equilibrium profiles from different Phase-Field models, along with $\eta^3|\nabla \phi_1||\nabla \phi_2||\nabla \phi_3|$. It can be observed from Fig.\ref{Fig Fictitious} \textbf{a)} that some amounts of Phase 1 from CACN are generated at the interfacial regions of Phases 2 and 4, and of Phases 3 and 4. Phases  2 and 3 from CACN behave similarly to Phase 1. More clearly, $\eta^3|\nabla \phi_1||\nabla \phi_2||\nabla \phi_3|$ is non-zero at all the interfacial regions. Unphysically generating fictitious phases is because CACN violates the \textit{consistency of reduction}, as analyzed in Section \ref{Sec L_p^c}. On the other hand, the results from CH, CHB, CAC, and CACB do not generate any fictitious phases, and $\eta^3|\nabla \phi_1||\nabla \phi_2||\nabla \phi_3|$ from those models is machine zero, since the \textit{consistency of reduction} is satisfied by those models and their schemes. In addition, there is little difference in the profiles from CH, CHB, CAC, and CACB. 

\subsection{Under-resolved structures}\label{Sec Small figures}
The comparison study in \cite{LeeKim2016} shows that the two-phase CAC has a better ability to preserve under-resolved structures than CH. Here, we perform the multiphase version of that study and consider CH, CHB, CAC, and CACB. Again, the velocity is set to be zero, i.e., $\mathbf{u} \equiv \mathbf{0}$. 

The domain considered is $[1\times1]$ with homogeneous Neumann boundaries. The number of cells to discretize the domain is $[128\times128]$ and the time step is $\Delta t=10h=\frac{10}{128}$. We set $\eta=h$, $M_0=10^{-3}\eta^2$, and the off-diagonal elements of $\lambda_{p,q}$ are $\frac{3}{2\sqrt{2}}\eta$. Phase 1 is enclosed by a circle at $(0.5,0.5)$ with a diameter $0.2$. Phase 2 is enclosed by a circle at $(0.2,0.2)$ with a diameter $0.1$. Phase 3 is enclosed by a circle at $(0.8,0.8)$ with a diameter $0.05$. Phase 4 occupies the rest of the domain. 

\begin{figure}[!t]
	\centering
	\includegraphics[scale=.4]{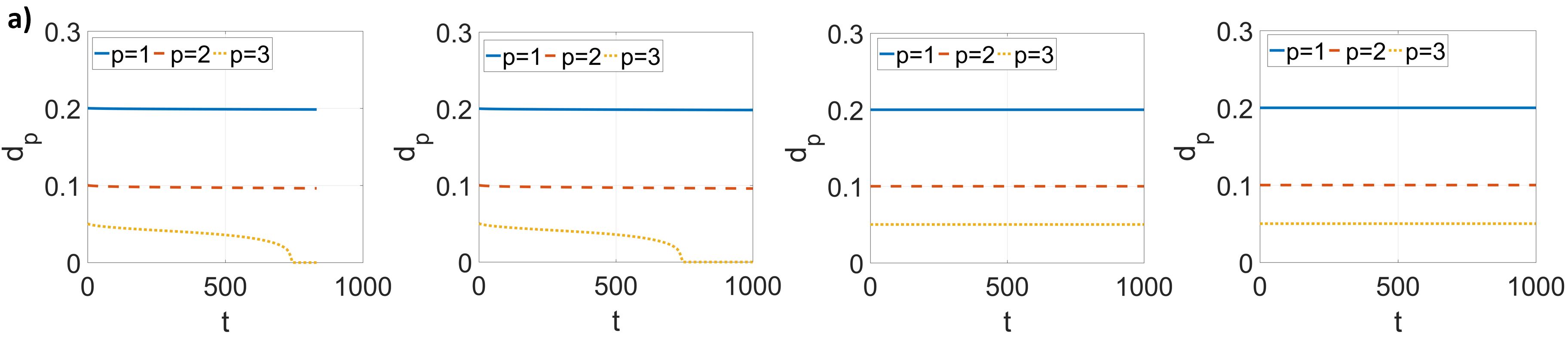}
	\includegraphics[scale=.4]{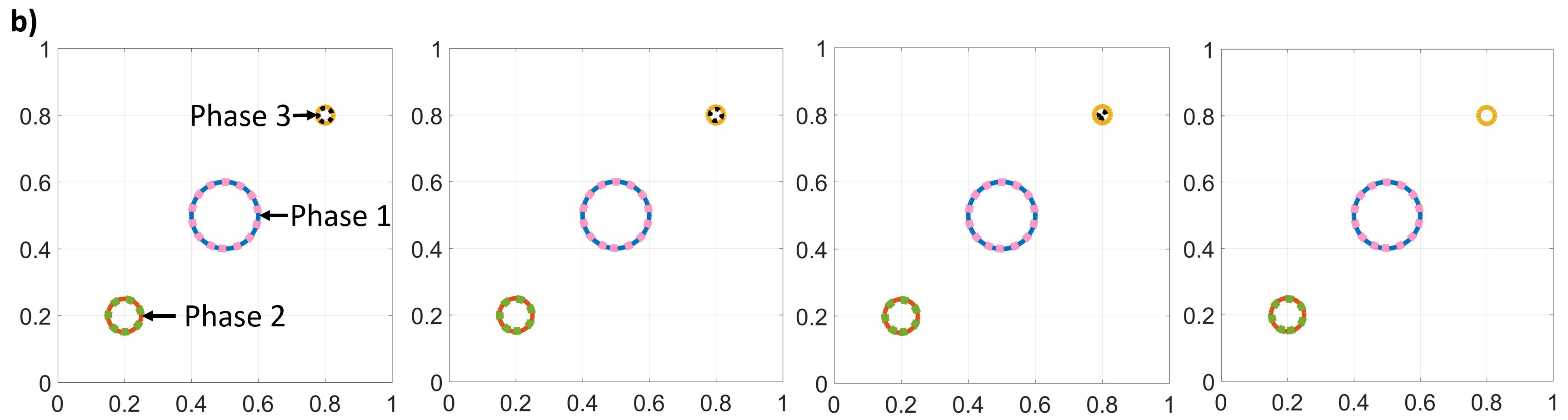}
	\includegraphics[scale=.4]{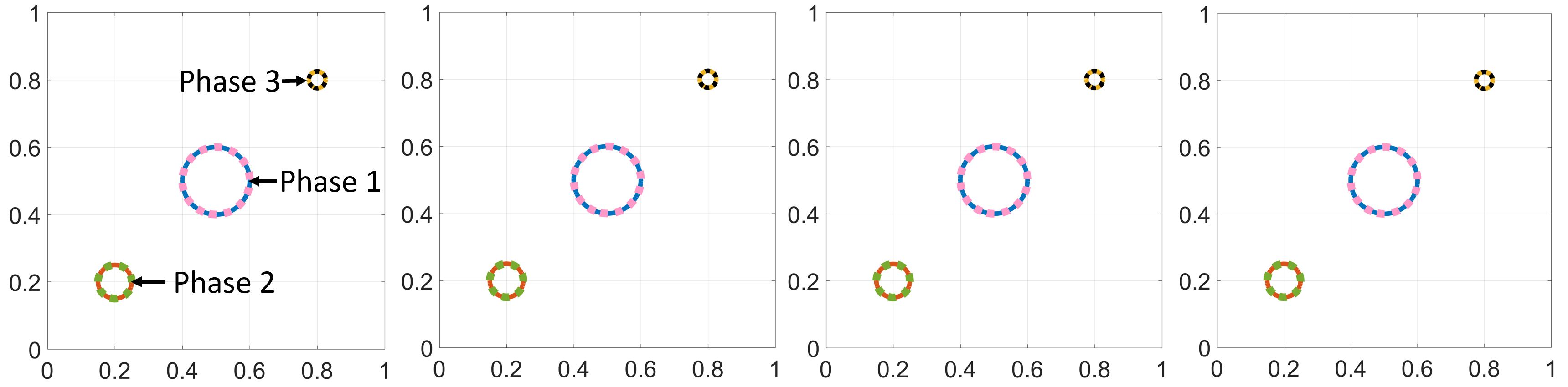}
	\includegraphics[scale=.4]{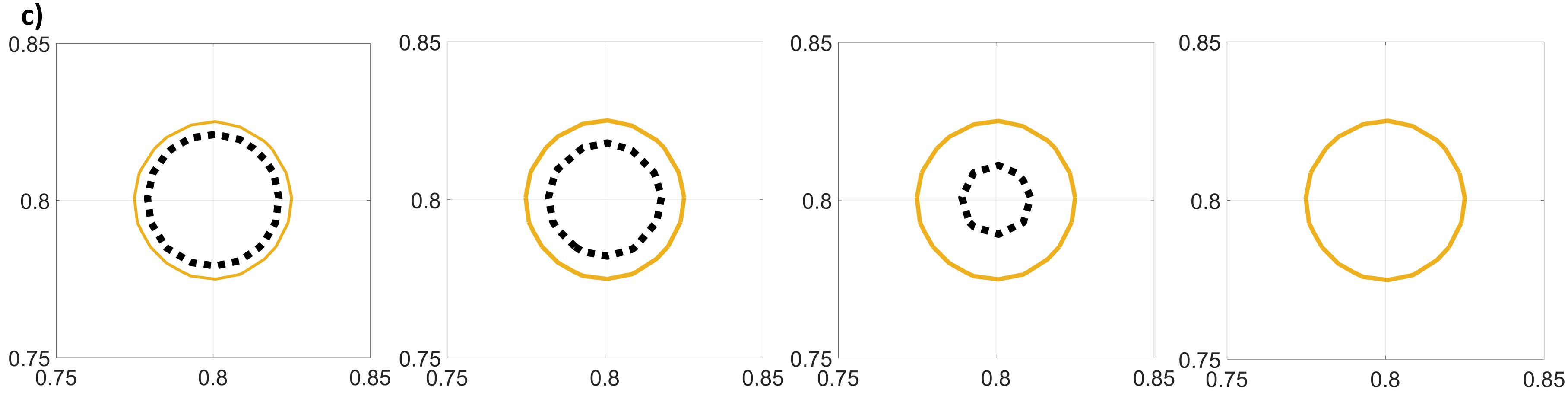}
	\includegraphics[scale=.4]{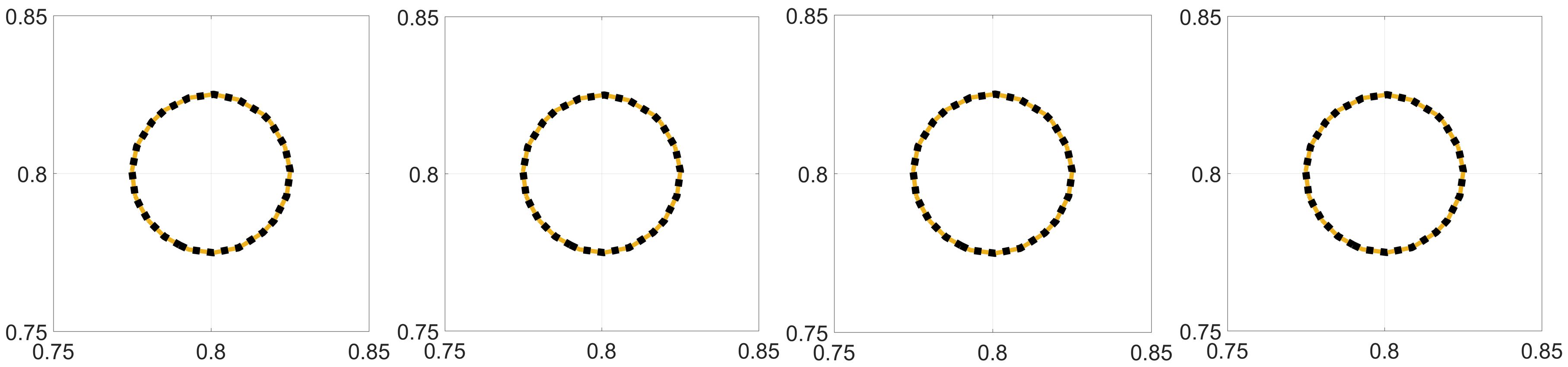}
	\caption{Evolution of the circles of the under-resolved structures. a) Evolution of the diameters of the circles from CH, CHB, CAC, and CACB (left to right). b) Evolution of the circles from CHB (top) and CACB (bottom). Solid lines: $t=0$. Dotted lines: $t=250,500,703.125,1000$ (left to right). c) Evolution of the smallest circle (Phase 3) from CHB (top) and CACB (bottom). Solid line: $t=0$. Dotted line: $t=250,500,703.125,1000$ (left to right).\label{Fig SmallFigure}}
\end{figure}

\begin{figure}[!t]
	\centering
	\includegraphics[scale=.4]{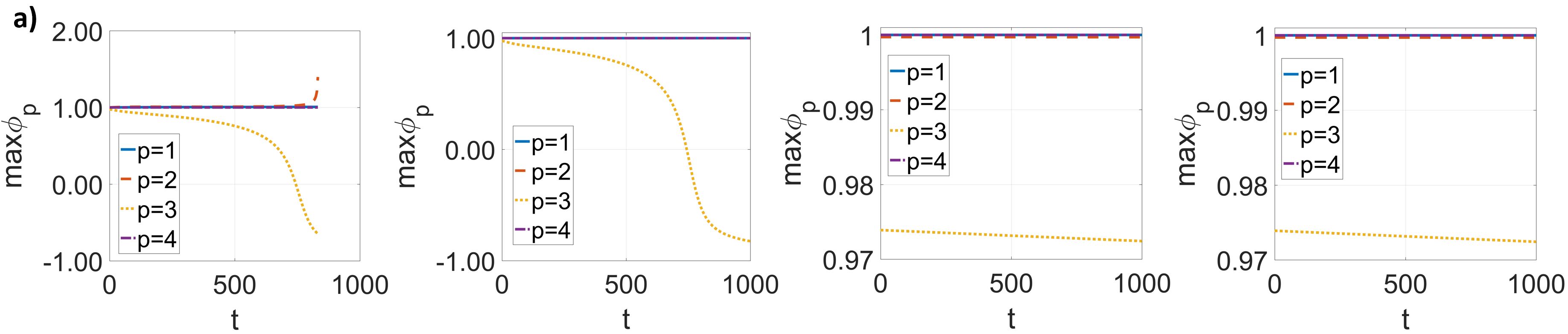}
	\includegraphics[scale=.4]{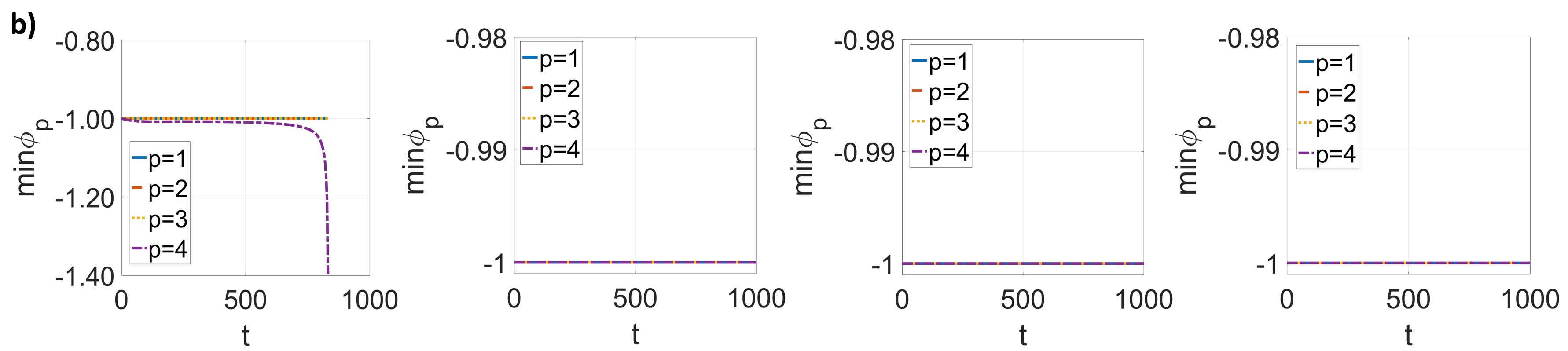}
	\includegraphics[scale=.4]{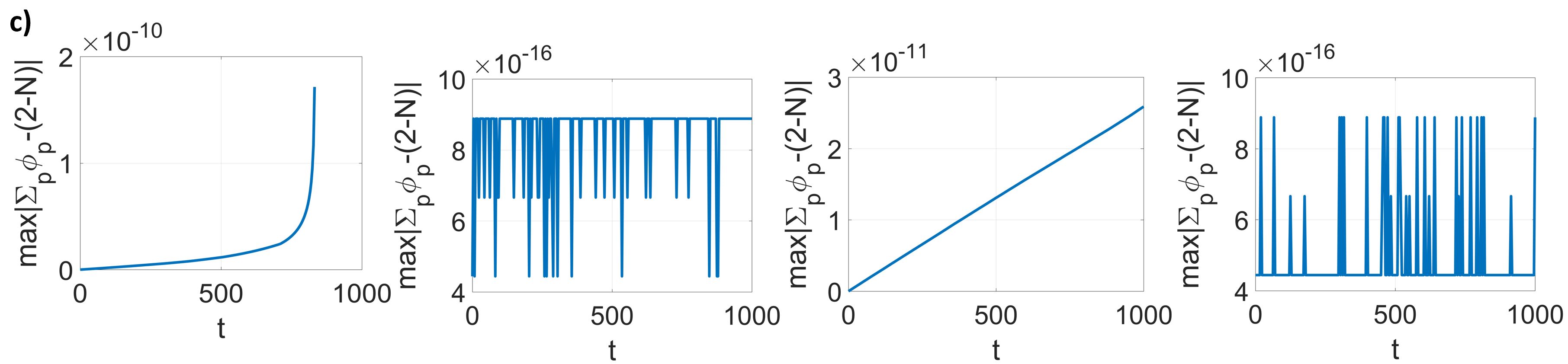}
	\caption{Results of the under-resolved structures from CH, CHB, CAC, and CACB (left to right). a) Maximum of the order parameters. b) Minimum of the order parameters. c) Error of the summation constraint for the order parameters.\label{Fig SmallFigure-Sup}}
\end{figure}

The number of grid points across the circles, i.e., $D/h$, is $25.6$ for Phase 1, $12.8$ for Phase 2, and $6.4$ for Phase 3. If the interfacial region is defined as $-0.995<\phi<0.995$, the number of grid points across the interfacial region is $n_I=2\sqrt{2}\tanh^{-1}(0.995)\eta/h \approx 8.5$. Therefore, about 17 grid points across the circle of Phase 1 are inside the bulk-phase region of Phase 1. Inside the bulk-phase region of Phase 2, there are 5 grid points across the circle of Phase 2, while there is none inside the bulk-phase region of Phase 3. We quantify the evolution of the individual circles by measuring their diameters for a long period of time, and the results are shown in Fig.\ref{Fig SmallFigure} \textbf{a)}. It is clear that the diameter of the smallest circle (Phase 3) decreases and it finally disappears at about $t=750$ from CH and CHB. The shrinkage of the other two circles from CH and CHB is negligible, considering the long period of simulation time. On the other hand, all the circles are preserved and their diameters do not change with time from CAC and CACB. Fig.\ref{Fig SmallFigure} \textbf{b)} shows the evolution of the circles and Fig.\ref{Fig SmallFigure} \textbf{c)} shows that of the smallest circle (Phase 3) from CHB and CACB, which are consistent with Fig.\ref{Fig SmallFigure} \textbf{a)}.
The maximum of $\phi_3$ less than $1$ at $t=0$ (see the last two columns of Fig.\ref{Fig SmallFigure-Sup} \textbf{a)}) represents the poor resolution of the smallest circle in the sense that there is no bulk-phase region of Phase 3 at the beginning. Nevertheless, unlike CH and CHB, both CAC and CACB preserve this under-resolved structure very well. 
As indicated in Fig.\ref{Fig SmallFigure-Sup} \textbf{a)} and \textbf{b)}, the order parameters from CHB, CAC, and CACB are in the physical interval $[-1,1]$ but those from CH are not. In this case, the out-of-bound error from CAC is in the order of the round-off error. Thus, the difference between CAC and CACB is negligible. Although the summation constraint for the order parameters is satisfied in all the Phase-Field models, see Fig.\ref{Fig SmallFigure-Sup} \textbf{c)}, it is enforced more accurately when the boundedness mapping is included (CHB and CACB). 

\begin{figure}[!t]
	\centering
	\includegraphics[scale=.4]{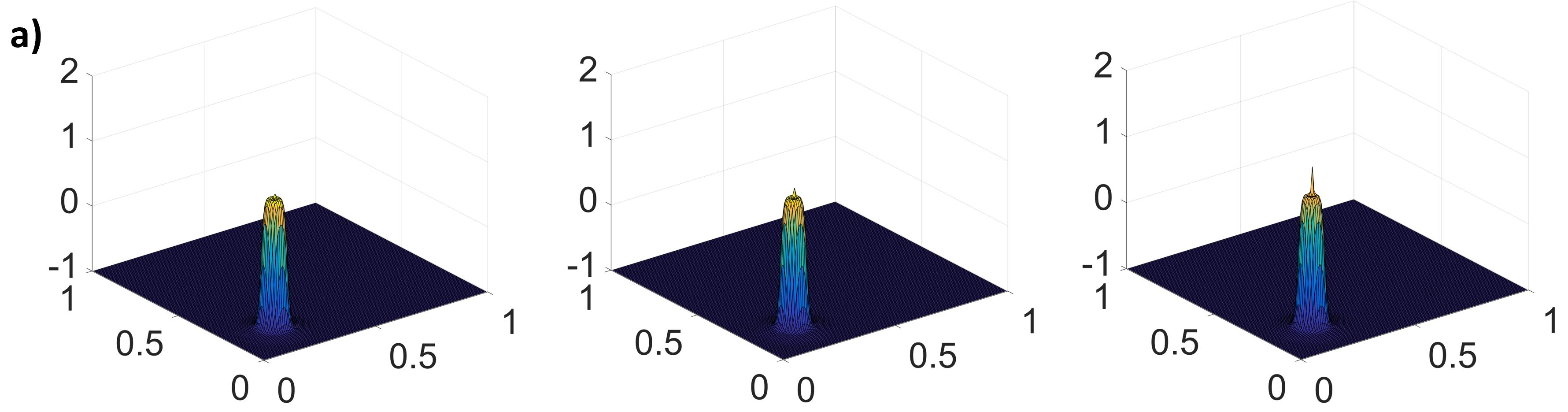}
	\includegraphics[scale=.4]{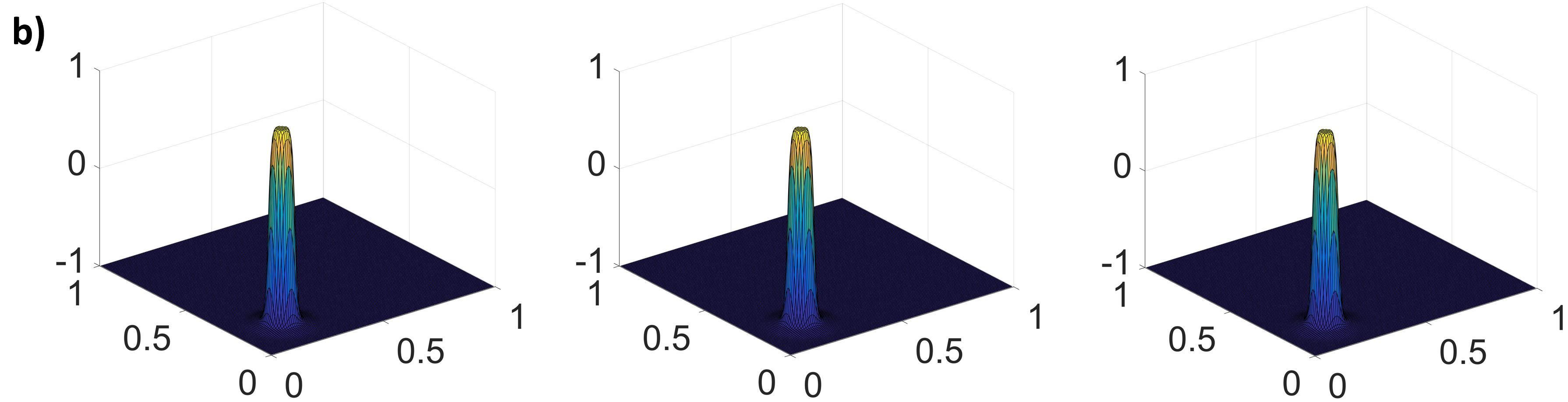}
	\includegraphics[scale=.4]{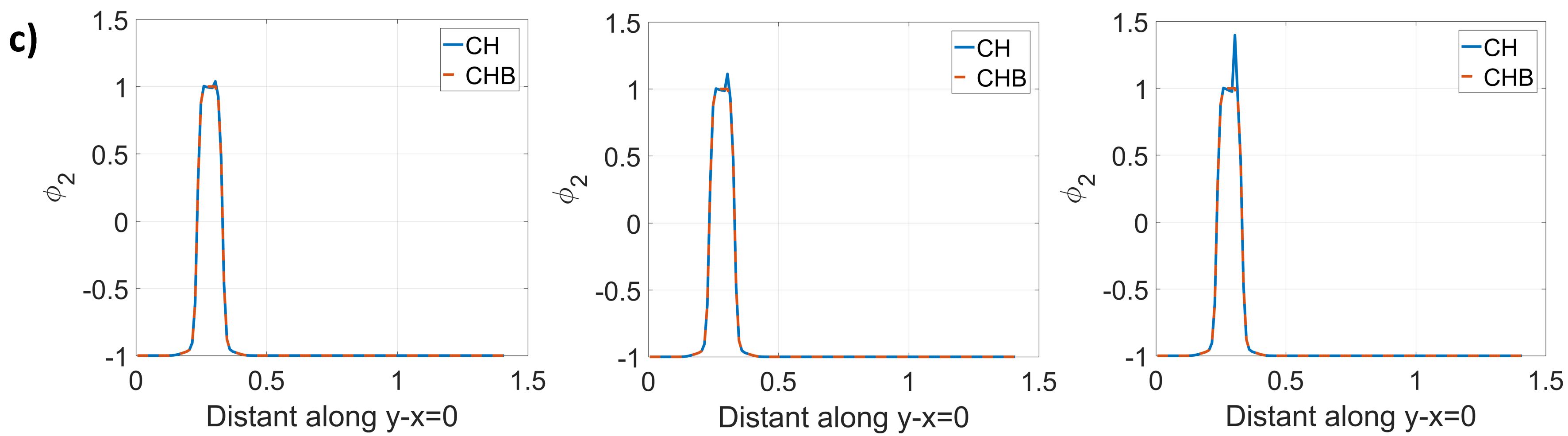}
	\caption{Profiles of $\phi_2$ at selected moments of the under-resolved structures from a) CH and b) CHB. c) $\phi_2$ along $y-x=0$ from CH and CHB. The first column: $t=781.2500$. The second column: $t=820.3125$. The third column: $t=832.0313$.\label{Fig SmallFigure-Profile}}
\end{figure}

Without the boundedness mapping, the result using CH becomes unstable after the smallest circle disappears. We can observe in the first column of Fig.\ref{Fig SmallFigure-Sup} \textbf{a)} that the maximum of $\phi_2$ from CH is larger than $1$ and increases with a dramatic rate. Due to the summation constraint for the order parameters, the minimum of $\phi_4$ decreases beyond $-1$ with a similar behavior, as shown in the first column of Fig.\ref{Fig SmallFigure-Sup} \textbf{b)}. We can infer that the out-of-bound error appears at the interfacial region between Phases 2 and 4.
Fig.\ref{Fig SmallFigure-Profile} \textbf{a)} shows the profiles of $\phi_2$ at selected moments from CH. At the interfacial region of Phases 2 and 4, an out-of-bound error initializes, which is consistent with the remark in Section \ref{Sec Apply}. As time goes on, the out-of-bound error keeps growing and becomes a spike. As a result, the profile of $\phi_2$ is significantly contaminated and becomes unphysical. Eventually, numerical instability is triggered due to the large out-of-bound error. We supplement the profiles of $\phi_2$ using CHB at the same moments in Fig.\ref{Fig SmallFigure-Profile} \textbf{b)}. Thanks to the boundedness mapping, the out-of-bound error is eliminated. Consequently, the physical profile of $\phi_2$ is preserved and the computation is stable. More clearly, the profile of $\phi_2$ from CH along $y-x=0$, which crosses the centers of the spike and the circle of Phase 2, is shown in Fig.\ref{Fig SmallFigure-Profile} \textbf{c)}, and compared to the one from CHB.

In summary, CAC has a better ability to preserve under-resolved structures than CH. The boundedness mapping is beneficial to improve the robustness of the scheme and to provide a physical solution. 

\subsection{Large-Density-Ratio advection}\label{Sec Advection}
An appropriate coupling between the Phase-Field models and the momentum equation Eq.(\ref{Eq Momentum}) should satisfy the \textit{consistency of mass conservation} and the \textit{consistency of mass and momentum transport}. The large-density-ratio advection is performed to demonstrate that the consistency conditions are achieved at the discrete level.

The domain considered is $[1\times1]$ and its boundaries are all periodic. $[128\times128]$ cells are used to discretize the domain and the time step is $\Delta t=0.1h=\frac{0.1}{128}$. Neither the viscosity of the fluids nor the surface force is considered such that $\{\mu_p\}_{p=1}^N$ and $\mathbf{f}_s$ are set to be zero. The densities of Phases 1, 2, and 3 are $10^6$, $10^3$, and $1$, respectively. We set $\eta=3h$, $M_0=10^{-7}$, and the off-diagonal elements of $\lambda_{p,q}$ are $\frac{0.03}{2\sqrt{2}}\eta$. Phase 1 is enclosed by a circle at $(0.65,0.65)$ with a radius $0.15$. Phase 2 is enclosed by a circle at $(0.3,0.4)$ with a radius $0.1$. Phase 3 occupies the rest of the domain. Initially, the velocity is homogeneous, i.e., $u|_{t=0}=v|_{t=0}=1$.

\begin{figure}[!t]
	\centering
	\includegraphics[scale=.4]{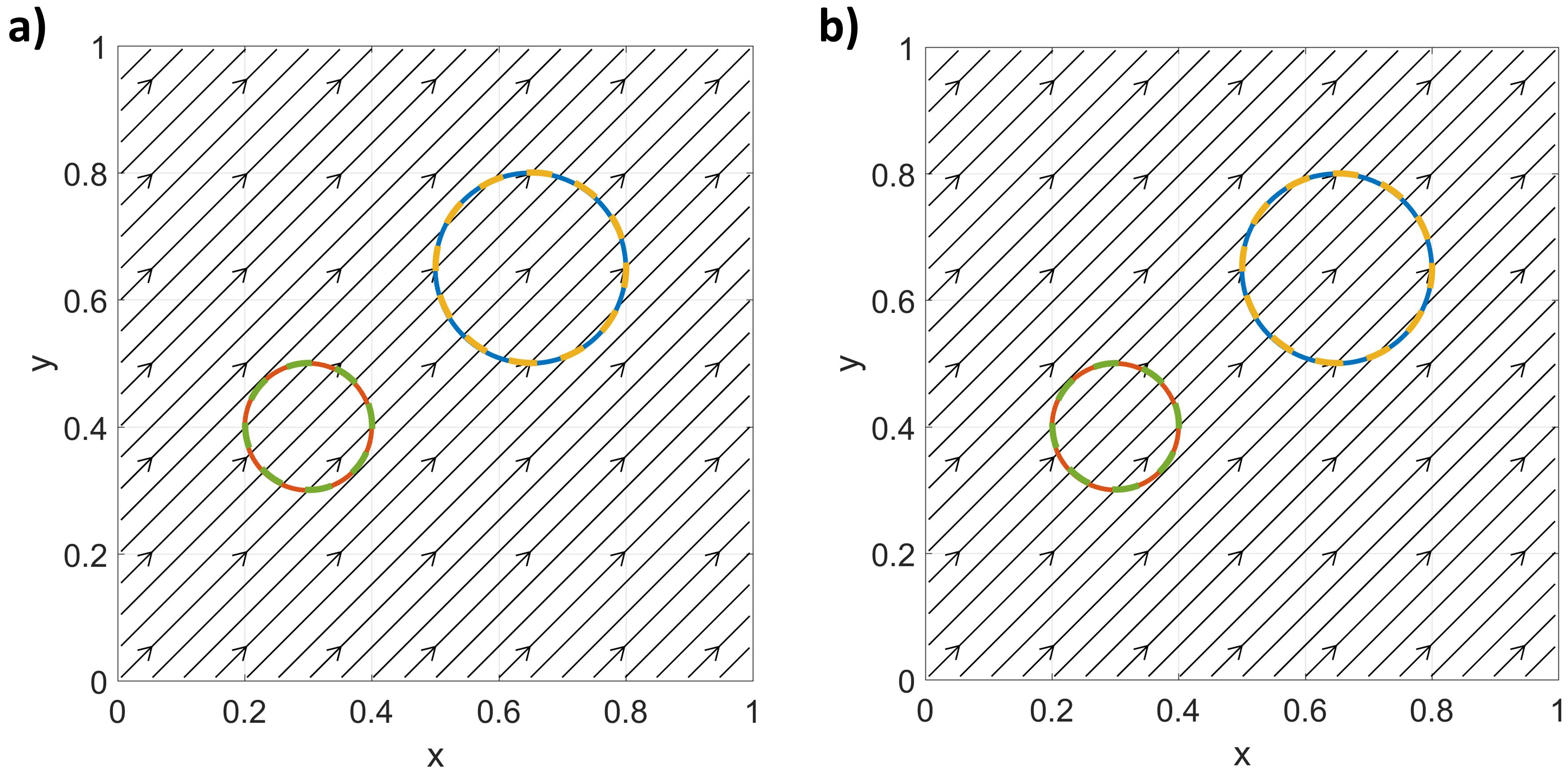}
	\caption{Results of the large-density-ratio advection from a) CHB and b) CACB. Blue solid line: Interface of phase 1 at $t=0$. Red solid line: Interface of phase 2 at $t=0$. Yellow dashed line: Interface of phase 1 at $t=1$. Green dashed line: Interface of phase 2 at $t=1$. Black arrow: Stream lines at $t=1$. \label{Fig LDA}}
\end{figure}

The interfaces should be translated by the homogeneous velocity, without any deformation. At the same time, the translation of the interfaces should not change the velocity. These should be true, independent of the density ratio. Therefore, at $t=1$, the interfaces should return to their original locations.  
Fig.\ref{Fig LDA} shows the results from CHB and CACB. At $t=1$, the interfaces return to their original locations without any deformation and the velocity preserves its initial value, even though the maximum density ratio is $10^6$. 

\begin{figure}[!t]
	\centering
	\includegraphics[scale=.45]{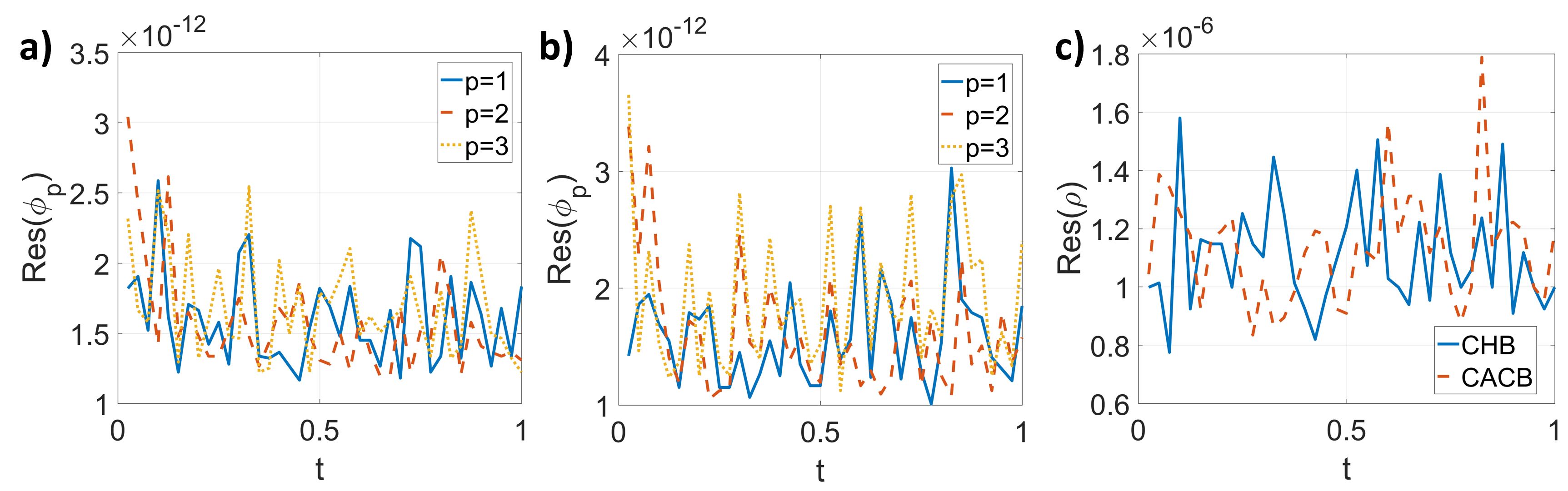}
	\caption{Residues of Eq.(\ref{Eq Phase-Field conservation discrete}) and Eq.(\ref{Eq Mass discrete}) of the large-density-ratio advection from CHB and CACB. a) Residue of Eq.(\ref{Eq Phase-Field conservation discrete}) from CHB. b) Residue of Eq.(\ref{Eq Phase-Field conservation discrete}) from CACB. c) Residue of Eq.(\ref{Eq Mass discrete}) from CHB ant CACB. \label{Fig LDA-Res}}
\end{figure}

The critical factor in the problem is to satisfy the \textit{consistency of mass conservation} and the \textit{consistency of mass and momentum transport}. Thanks to the consistent formulation Eq.(\ref{Eq Consistent formulation discrete}), the fully-discretized Phase-Field model including the boundedness mapping is able to be written in a conservative form, i.e., Eq.(\ref{Eq Phase-Field conservation discrete}). As a result, the discrete mass conservation equation Eq.(\ref{Eq Mass discrete}) is satisfied as well by using the discrete consistent mass flux Eq.(\ref{Eq Mass flux discrete}) which is applied in the momentum equation.  
Fig.\ref{Fig LDA-Res} shows the residues of Eq.(\ref{Eq Phase-Field conservation discrete}) and Eq.(\ref{Eq Mass discrete}) from CHB and CACB. The residue of Eq.(\ref{Eq Phase-Field conservation discrete}) from either CHB or CACB is in the order of the round-off error. The residue of Eq.(\ref{Eq Mass discrete}) is in the order of the maximum density ratio of the problem times the residue of Eq.(\ref{Eq Phase-Field conservation discrete}).
Violating these consistency conditions can introduce unphysical velocity fluctuations and interface deformation, which trigger the instability in large-density-ratio problems. This has been demonstrated in previous studies, e.g., \cite{Huangetal2020,Huangetal2020N,Huangetal2020CAC}.

\subsection{Horizontal shear layer}\label{Sec Horizontal shear layer}
\begin{figure}[!t]
	\centering
	\includegraphics[scale=.45]{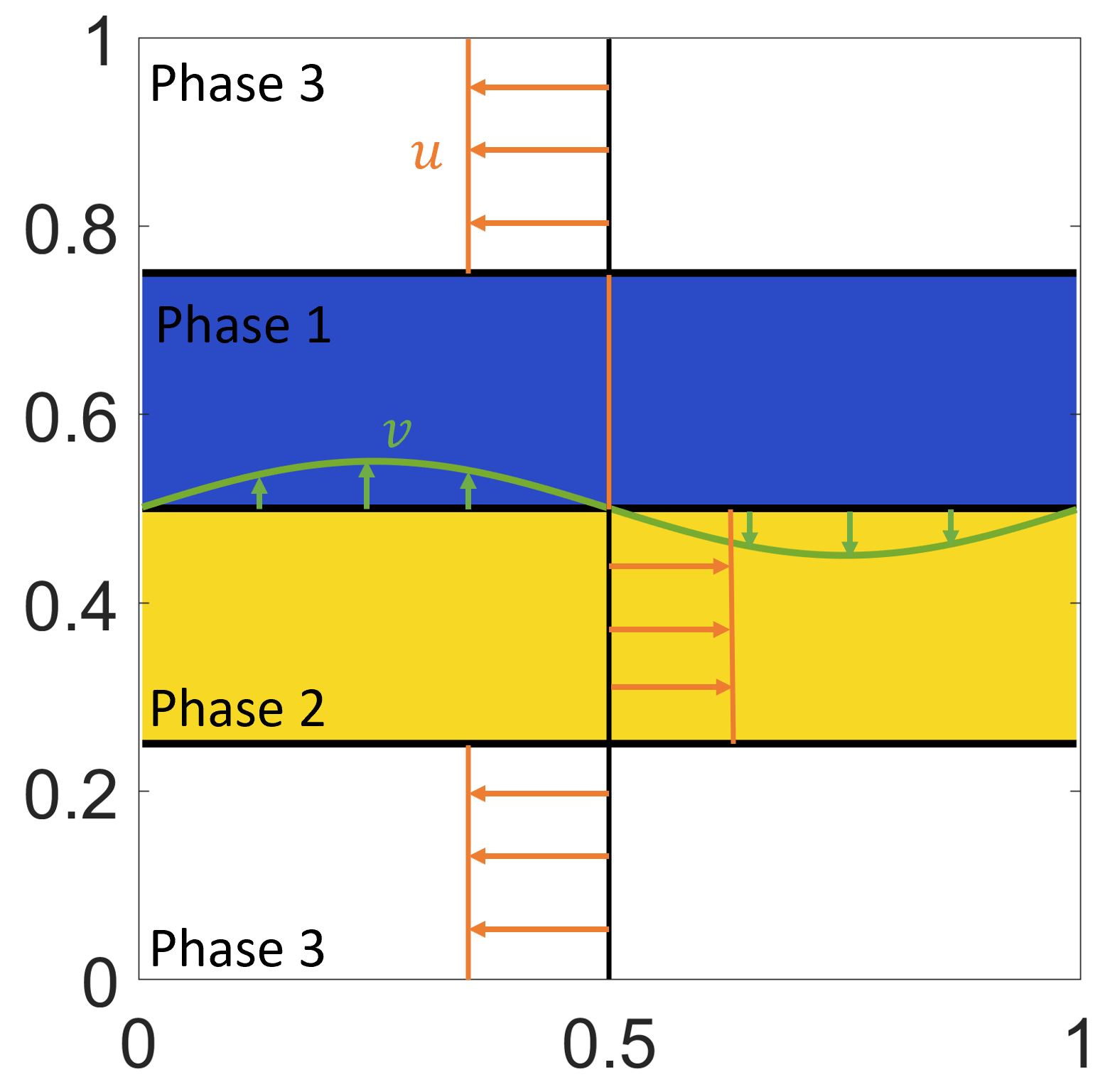}
	\caption{Schematic of the setup of the horizontal shear layer.  \label{Fig HSL-Schematic}}
\end{figure}

To demonstrate the properties of the Phase-Field models when they are coupled to the flow dynamics, the three-phase horizontal shear layer is performed.
The domain considered is $[1\times1]$ and its boundaries are all periodic. The domain is discretized by $[128\times128]$ cells and the time step is $\Delta t=0.1 h=\frac{0.1}{128}$. The density of Phase 1 is $50$ and its viscosity is $0.01$. Phase 2 has a density $10$ and a viscosity $0.1$. The density and viscosity of Phase 3 are $1$ and $0.05$, respectively. The surface tensions are $\sigma_{1,2}=0.05$, $\sigma_{1,3}=0.01$, and $\sigma_{2,3}=0.1$. We set $\eta=h$ and $M_0=10^{-7}$. Initially, Phase 1 is at $y_0<y<y_2$ and is stationary. Phase 2 is at $y_1<y<y_0$ and is moving to the right with a unity speed. Phase 3 is at the rest of the domain and is moving to the left with a unity speed. We set $y_0 = 0.5$, $y_1 = 0.25$, and $y_2 = 0.75$. A sinusoidal vertical velocity is added, whose amplitude is $0.05$ and wavelength is $2\pi$. The schematic of the setup is shown in Fig.\ref{Fig HSL-Schematic}.

\begin{figure}[!t]
	\centering
	\includegraphics[scale=.45]{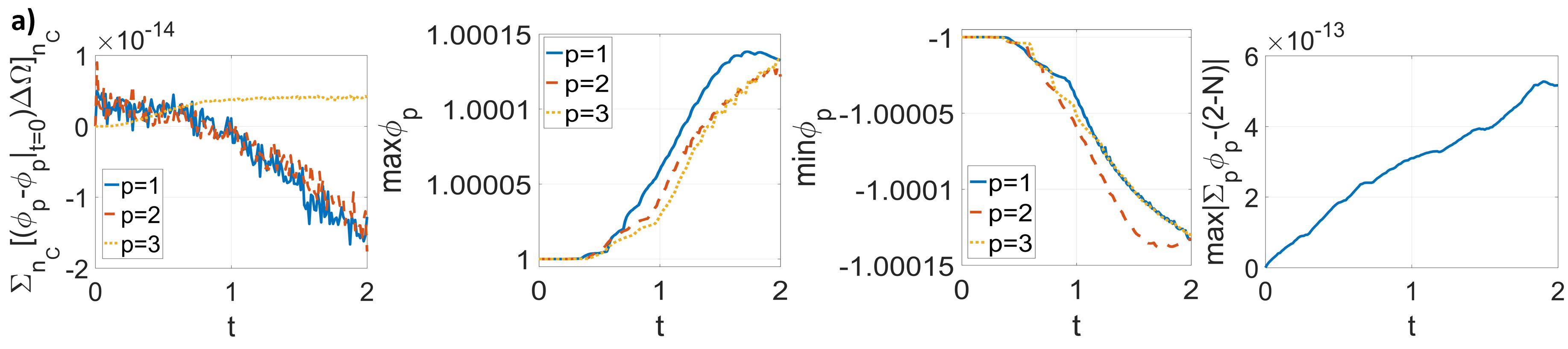}
	\includegraphics[scale=.45]{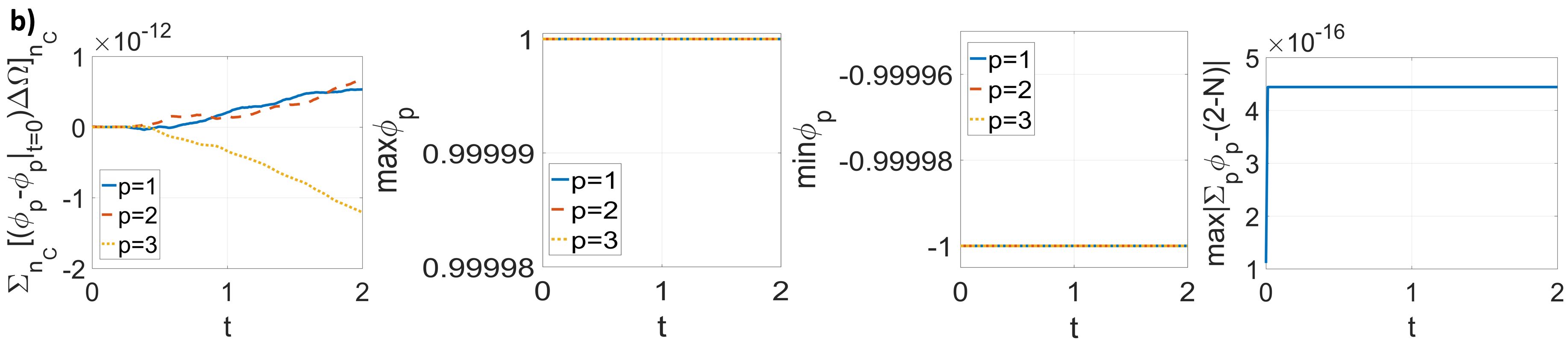}
	\includegraphics[scale=.45]{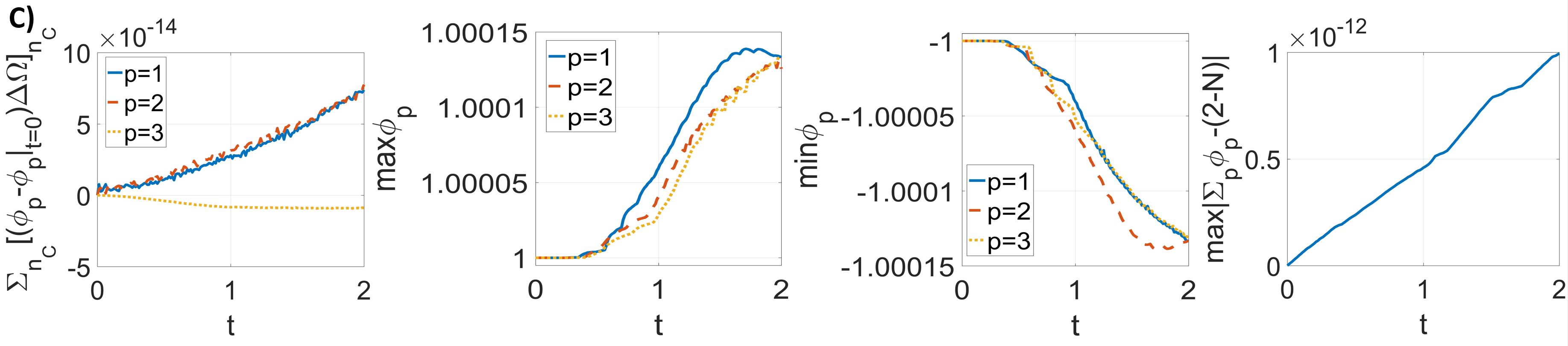}
	\includegraphics[scale=.45]{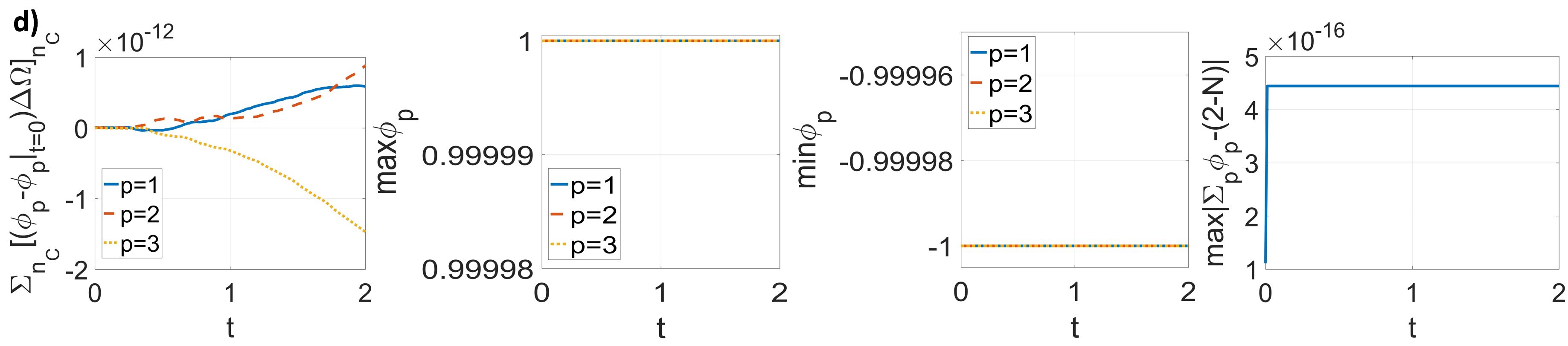}
	\caption{Results of the three-phase horizontal shear layer from a) CH, b) CHB, c) CAC, and d) CACB. The first column: Errors of the mass conservation of individual phases. The second column: Maximum of the order parameters. The third column: Minimum of the order parameters. The fourth column: Error of the summation constraint for the order parameters. \label{Fig HSL-3Phases}}
\end{figure}

We first investigate the mass conservation of individual phases Eq.(\ref{Eq Conservation phi}), the maximum and minimum of the order parameters, and the summation constraint for the order parameters Eq.(\ref{Eq Summation constraint phi}). The results from CH, CHB, CAC, and CACB are shown in Fig.\ref{Fig HSL-3Phases}. It is clear that all the results satisfy the mass conservation and the summation constraint for the order parameters, no matter whether the boundedness mapping is included. However, without the boundedness mapping, i.e., from CH and CAC, the order parameters do not stay in the physical interval $[-1,1]$. The out-of-bound error appears at about $t=0.4$. It grows with time and finally reaches $O(10^{-4})$. On the other hand, with the help of the boundedness mapping, none of the order parameters go beyond the physical interval $[-1,1]$. 

\begin{figure}[!t]
	\centering
	\includegraphics[scale=.45]{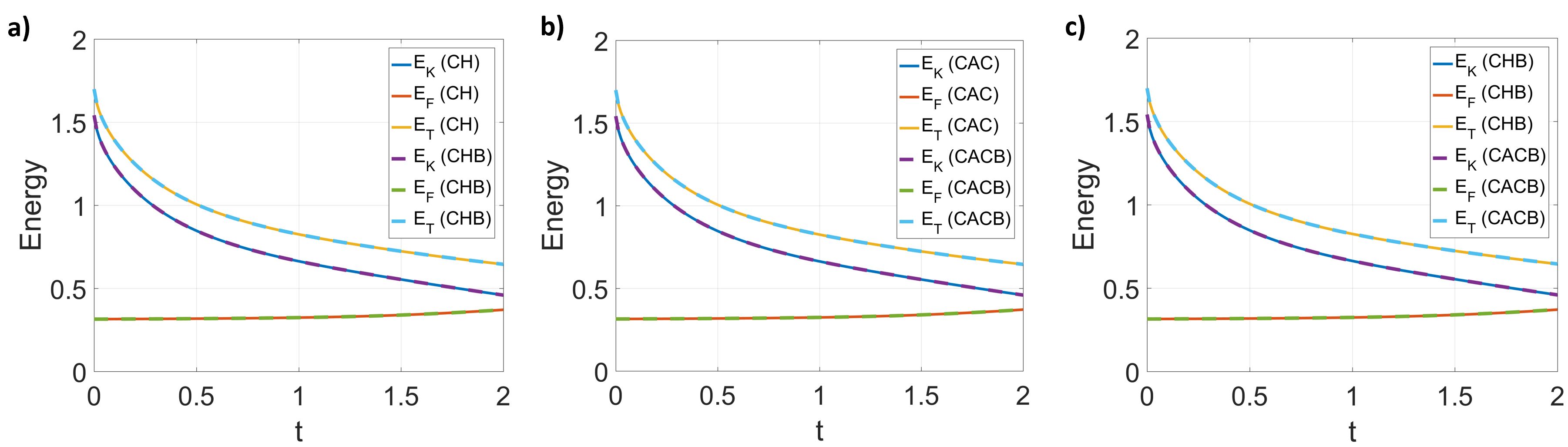}
	\caption{Time histories of the energies from a) CH and CHB, b) CAC and CACB, and c) CHB and CACB. \label{Fig HSL-3Phases-Energy}}
\end{figure}

Next, we quantify the problem by the time histories of the kinetic energy, 
\[
E_K= \int_{\Omega} \frac{1}{2} \rho (u^2+v^2) d\Omega,
\]
the free energy, 
\[
E_F= \int_{\Omega} \sum_{p,q=1}^N \frac{\lambda_{p,q}}{2} \left( \frac{1}{\eta^2} 
\left( g_1(\phi_p)+g_1(\phi_q)-g_2(\phi_p+\phi_q) \right)-\nabla \phi_p \cdot \nabla \phi_q \right) d\Omega,
\]
and the total energy,
\[
E_T=E_K+\frac{1}{2}E_F,
\]
and they are shown in Fig.\ref{Fig HSL-3Phases-Energy}. The results with/without the boundedness mapping are on top of each other since the out-of-bound error is small. In addition, the difference between CHB and CACB is unobservable in this case. The decay of the total energy is consistent with the energy law in \cite{Huangetal2020N}.

\begin{figure}[!t]
	\centering
	\includegraphics[scale=.45]{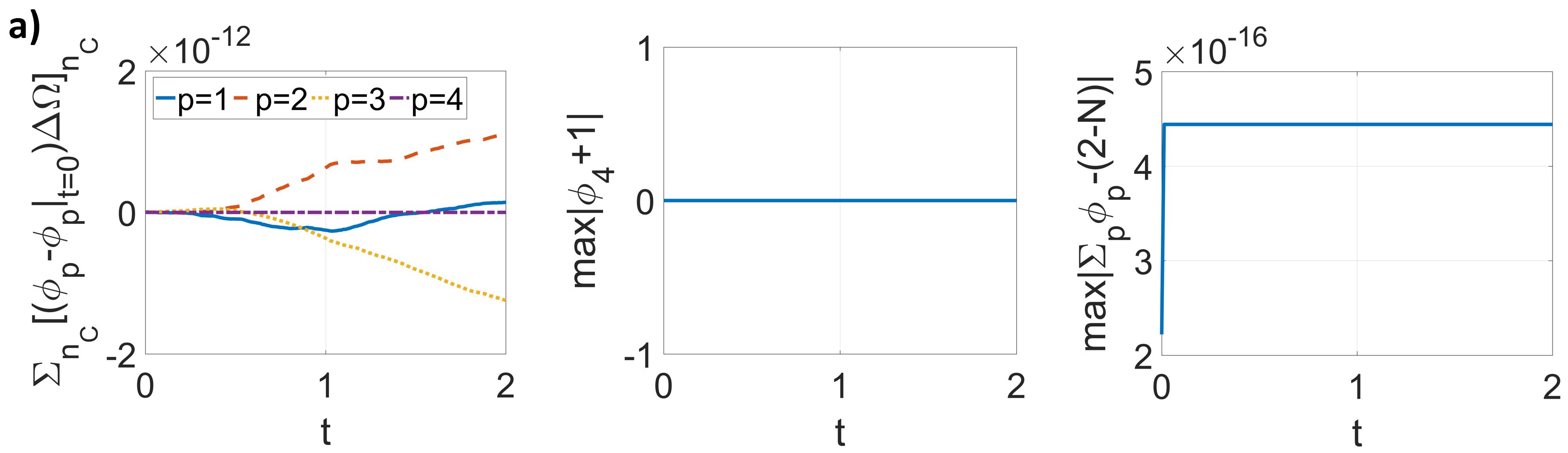}
	\includegraphics[scale=.45]{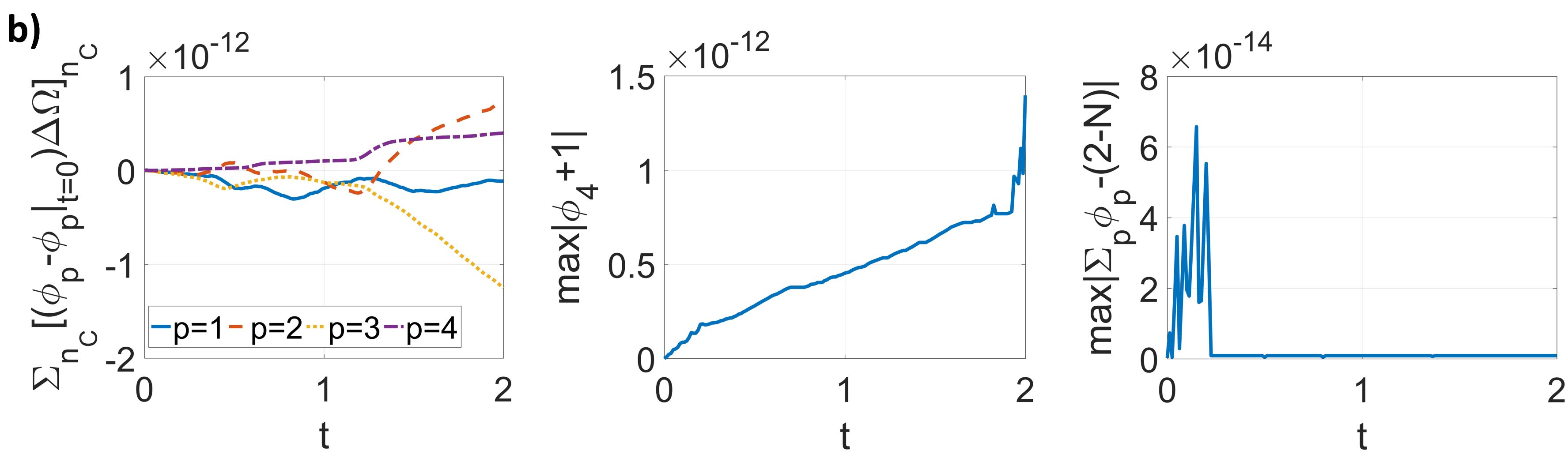}
	\caption{Results of the four-phase horizontal shear layer from a) CHB and b) CACB. The first column: Errors of the mass conservation of individual phases. The second column: Error of $\phi_4$. The third column: Error of the summation constraint for the order parameters.\label{Fig HSL-4Phases}}
\end{figure}

\begin{figure}[!t]
	\centering
	\includegraphics[scale=.45]{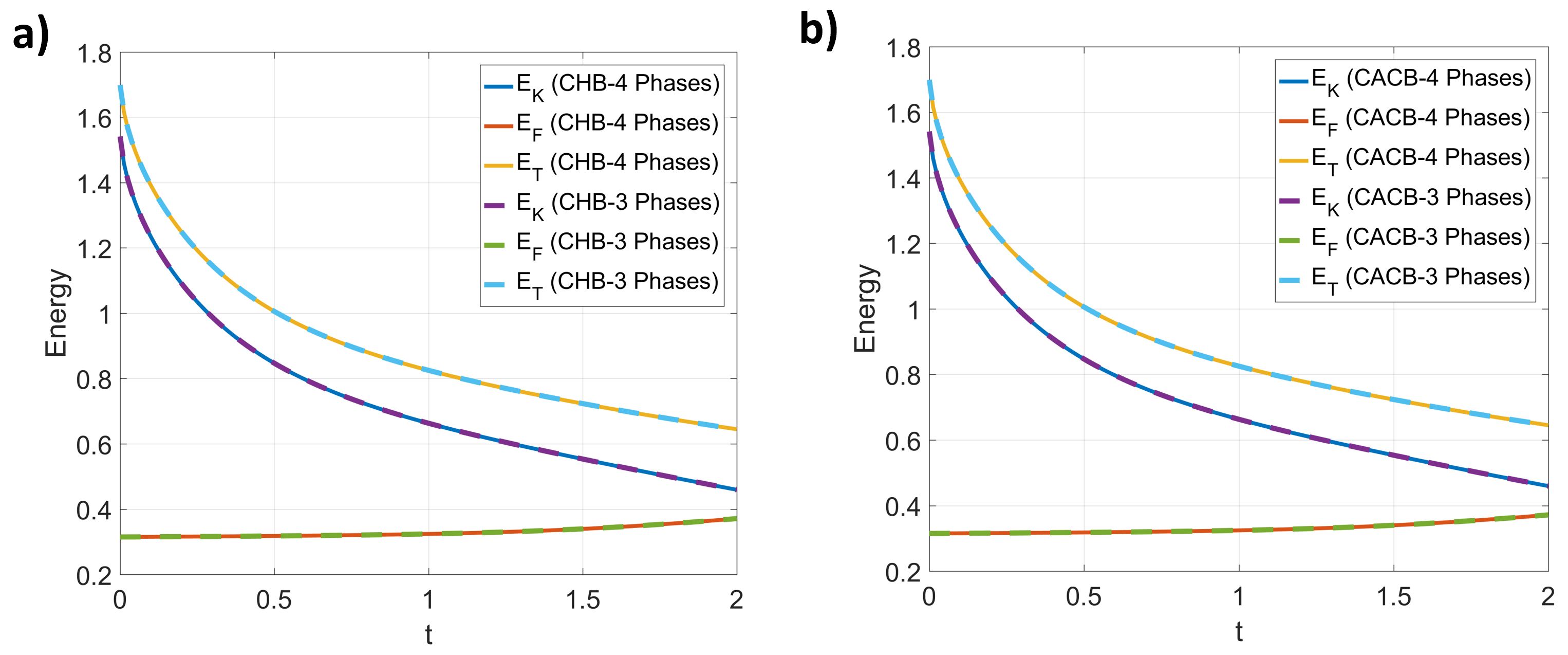}
	\caption{Time histories of the energies from a) CHB and b) CACB. \label{Fig HSL-4Phases-Energy}}
\end{figure}

We finally validate the \textit{consistency of reduction} by adding the 4th phase, whose density is $0.5$, viscosity is $0.08$, and surface tensions are $\sigma_{1,4}=0.08$, $\sigma_{2,4}=0.02$, and $\sigma_{3,4}=0.2$. However, Phase 4 is absent at the beginning, i.e., $\phi_4|_{t=0}=-1$. Therefore, it should not appear during the computation. The results from CHB and CACB are shown in Fig.\ref{Fig HSL-4Phases}. The mass conservation and the summation constraint for the order parameters are again satisfied. $\phi_4$ from CHB is exact $-1$ at every cell and every time step. However, the difference between $\phi_4$ from CACB and $-1$ is the round-off error. Therefore, both the Phase-Field models along with the boundedness mapping do not generate any fictitious phases. Then we compare the energies from the three-phase case to those from the 4-phase case in Fig.\ref{Fig HSL-4Phases-Energy}, and the difference between them is unobservable. The three-phase dynamics is reproduced by the four-phase setup with a phase absent. Therefore the \textit{consistency of reduction} is satisfied.

In summary, with the proposed schemes, the order parameters satisfy their summation constraint, the mass conservation, and the \textit{consistency of reduction}. They, in addition, are bounded in their physical interval if the boundedness mapping is supplemented.

\subsection{Convergence tests}\label{Sec Convergence}
The convergence behaviors of the models and schemes are investigated. In the first test, we fix $\eta$ and $M_0$ in order to study the convergence of the numerical solutions to the exact solutions of the Phase-Field models. This test is related to the truncation error and should be able to indicate the formal order of accuracy of the schemes. In the second test, $\eta$ and $M_0$ are correlated to the grid/cell size $h$. The interface thickness $\eta$ is reduced when the cell is refined, and the convergence of the numerical solutions of the Phase-Field models to the sharp-interface solution can be studied. 

The domain considered is $[1\times2]$ with free-slip boundaries at the left and right, and with no-slip boundaries at the top and bottom. Initially, the flow is stationary, and a circular bubble of Phase 1 is at $(0.5,0.5)$ with a radius of $0.25$. Phase 2 occupies the rest of the domain. The number of cells on a unit length is ranging from $16$ to $256$, and the time step is proportional to the cell size, i.e., $\Delta t=0.128h$. The computation is stopped at $t=1$. The density and viscosity of Phase 1 are $1$ and $0.1$, while the density and viscosity of Phase 2 are $1000$ and $10$. The surface tension is $1.96$, and the gravity is $0.98$ pointing downward. The same setup is considered in \cite{Hysingetal2009}, and the sharp-interface solutions using either the Level-Set or Arbitrary Lagrangian-Eulerian (ALE) method are available. The results are well-validated, and the differences of the results from the different sharp-interface methods are negligible in the time period considered. The circularity
\[
\psi_c=\frac{P_a}{P_b}=\frac{2\sqrt{\int_{\phi_1>0}\pi d\Omega}}{P_b},
\]
where $P_b$ and $P_a$ are the perimeters of the rising bubble and of the circle whose area is identical to the bubble,
the center of mass
\[
y_c=\frac{\int_{\Omega} y \frac{1+\phi_1}{2} d\Omega}{\int_{\Omega} \frac{1+\phi_1}{2} d\Omega},
\]
and the rising velocity
\[
v_c=\frac{\int_{\Omega} v \frac{1+\phi_1}{2} d\Omega}{\int_{\Omega} \frac{1+\phi_1}{2} d\Omega},
\]
are defined as the benchmark quantities of the problem. 

\begin{figure}[!t]
	\centering
	\includegraphics[scale=.51]{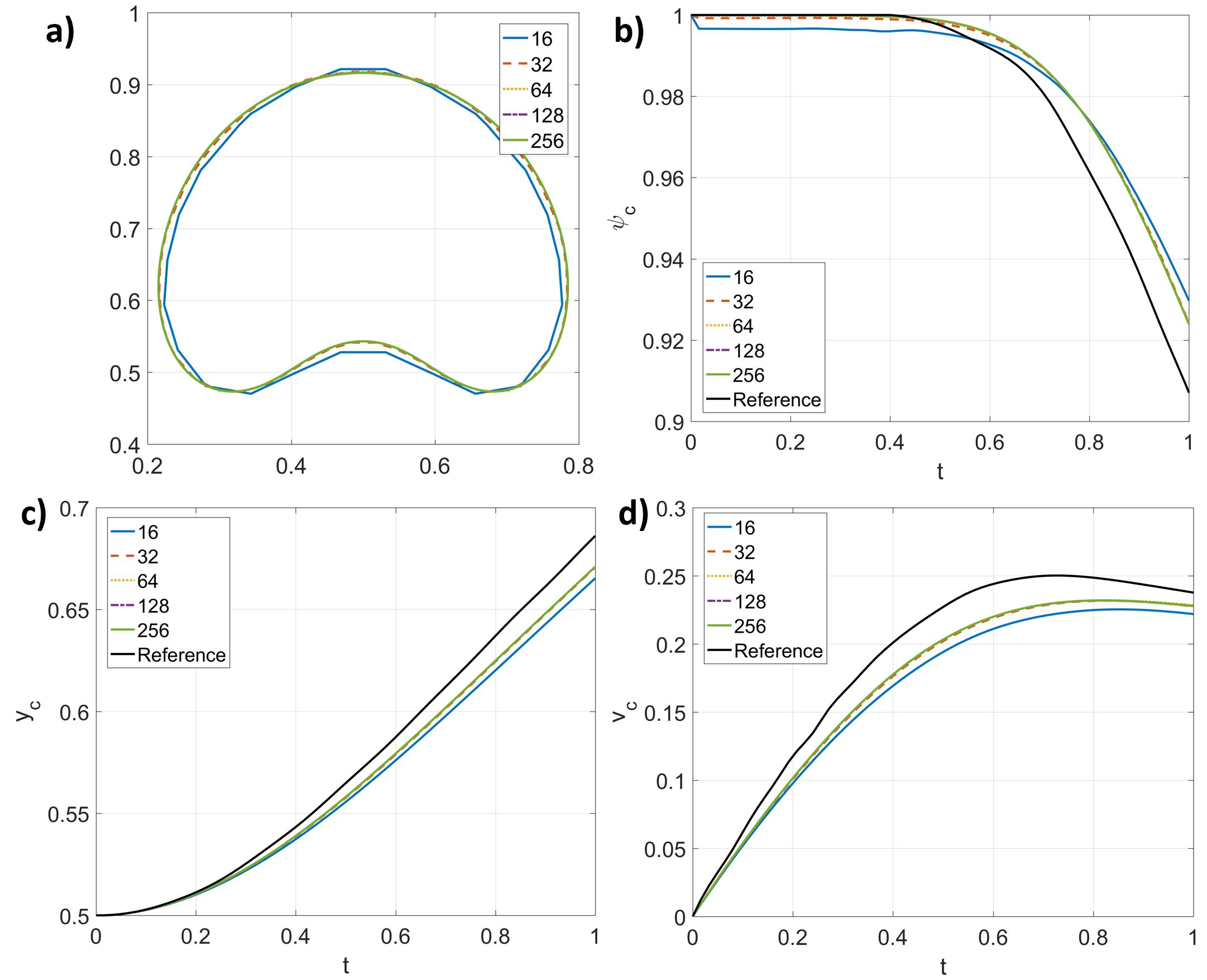}
	\caption{Results of the convergence test with $\eta=\eta_0$ from CHB. a) Shape of the bubble at $t=1$. b) Circularity. c) Center of mass. d) Rising velocity. ``16'', ``32'', ``64'', ``128'', and ``256'' in the legend are the numbers of cells on a unite length of the domain. ``Reference'' in the legend is the sharp-interface solution from \citep{Hysingetal2009}. \label{Fig Convergence-Eta0-CHB}}
\end{figure}

\begin{figure}[!t]
	\centering
	\includegraphics[scale=.51]{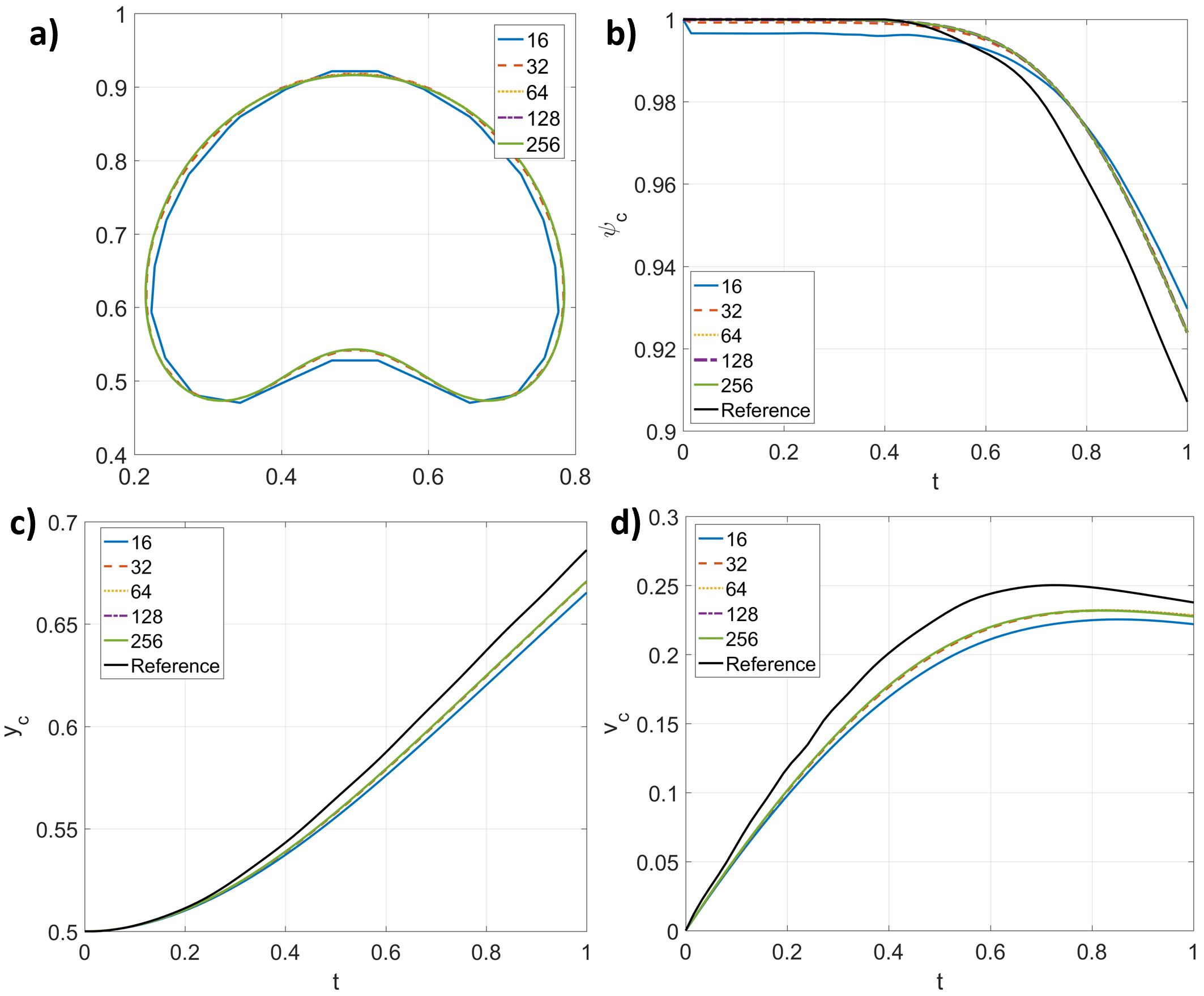}
	\caption{Results of the convergence test with $\eta=\eta_0$ from CACB. a) Shape of the bubble at $t=1$. b) Circularity. c) Center of mass. d) Rising velocity. ``16'', ``32'', ``64'', ``128'', and ``256'' in the legend are the numbers of cells on a unite length of the domain. ``Reference'' in the legend is the sharp-interface solution from \citep{Hysingetal2009}. \label{Fig Convergence-Eta0-CACB}}
\end{figure}

Fig.\ref{Fig Convergence-Eta0-CHB} and Fig.\ref{Fig Convergence-Eta0-CACB} show the results from CHB and CACB, respectively, using fixed $\eta=\eta_0=\frac{1}{32}$ and $M_0=10^{-7}$. As the cell size becomes smaller, the numerical solutions gradually approach the exact solutions of the Phase-Field models, which is different from the sharp-interface solution, since the interface thickness is not reducing. To quantify the convergence behavior, we compute the $L_2$ errors, i.e., the root-mean-square errors, of the three benchmark quantities. We consider the solutions from the finest grid as the exact solutions of the Phase-File models. The errors are summarized in Table \ref{Table Convergence-Eta0}, and the convergence rates are 2nd-order. Therefore, the schemes for CHB and CACB are formally 2nd-order accurate.
\begin{table}[!t]
	\caption{$L_2$ errors of the convergence test with $\eta=\eta_0$\label{Table Convergence-Eta0}}
	\centering
	\includegraphics[scale=.4]{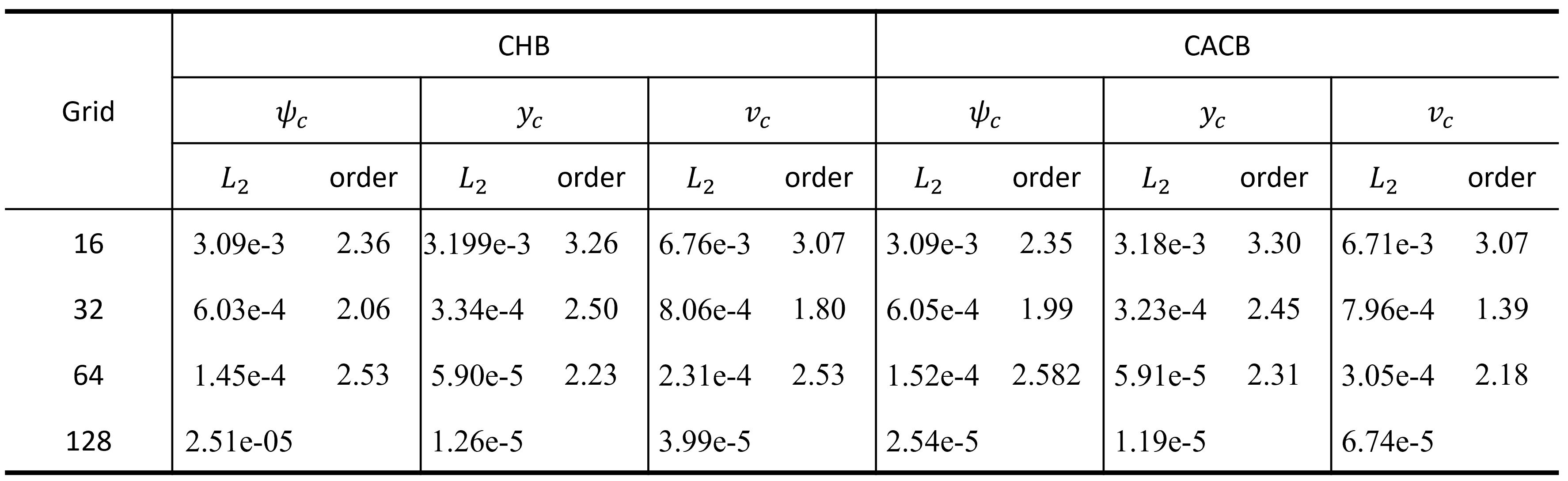}
\end{table}

\begin{figure}[!t]
	\centering
	\includegraphics[scale=.51]{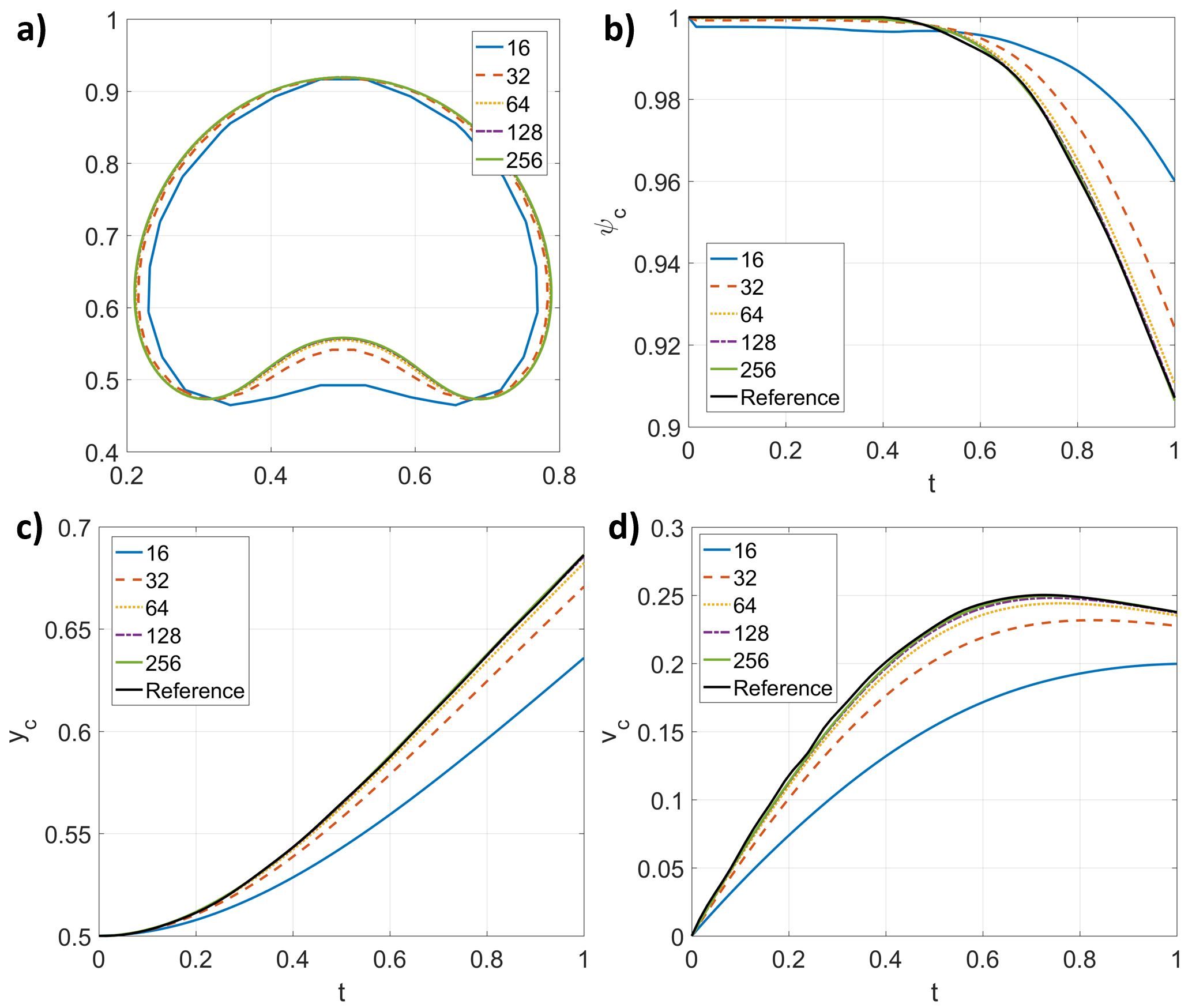}
	\caption{Results of the convergence test with $\eta=h$ from CHB. a) Shape of the bubble at $t=1$. b) Circularity. c) Center of mass. d) Rising velocity. ``16'', ``32'', ``64'', ``128'', and ``256'' in the legend are the numbers of cells on a unite length of the domain. ``Reference'' in the legend is the sharp-interface solution from \citep{Hysingetal2009}. \label{Fig Convergence-CHB}}
\end{figure}

\begin{figure}[!t]
	\centering
	\includegraphics[scale=.51]{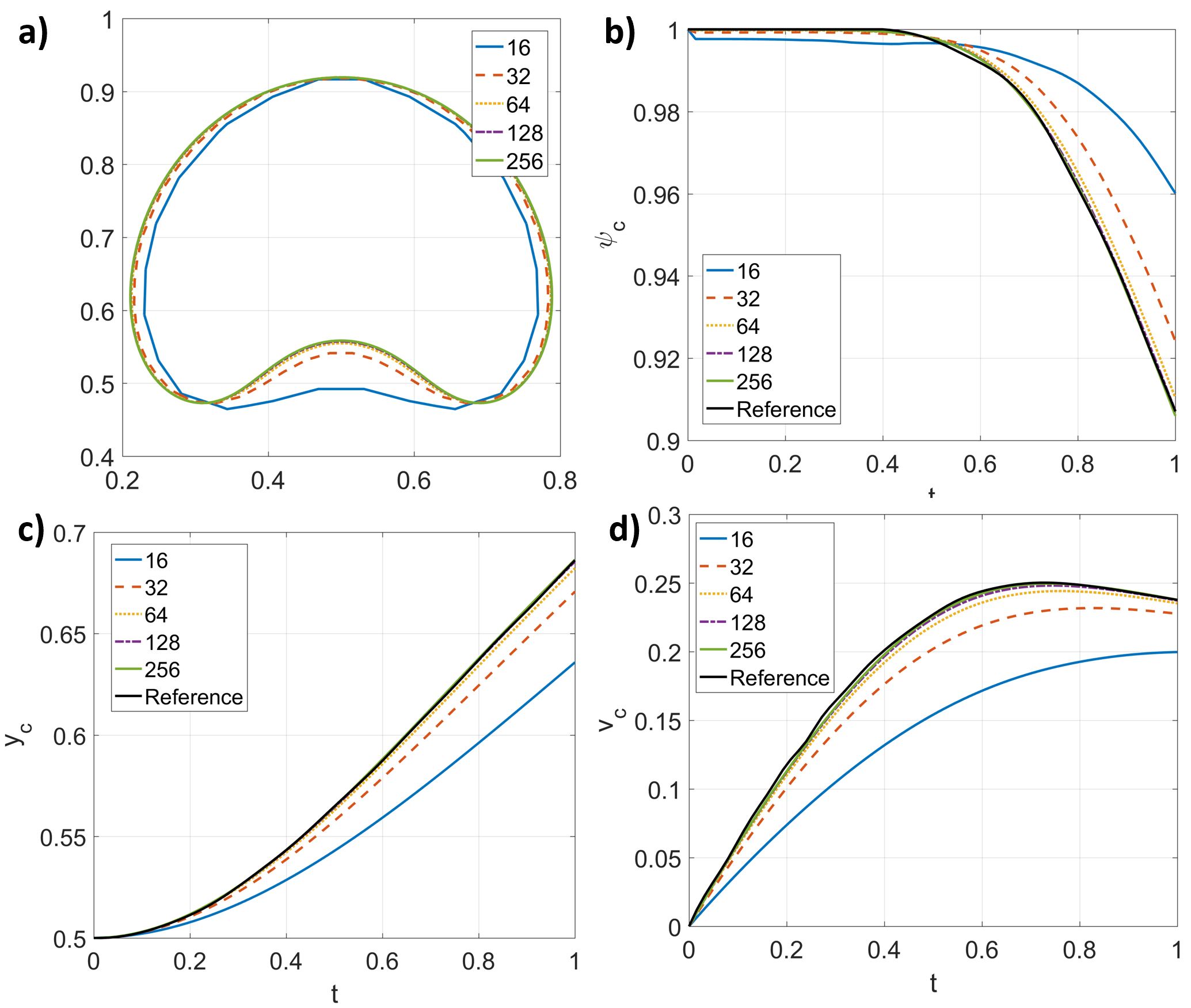}
	\caption{Results of the convergence test with $\eta=h$ from CACB. a) Shape of the bubble at $t=1$. b) Circularity. c) Center of mass. d) Rising velocity. ``16'', ``32'', ``64'', ``128'', and ``256'' in the legend are the numbers of cells on a unite length of the domain. ``Reference'' in the legend is the sharp-interface solution from \citep{Hysingetal2009}. \label{Fig Convergence-CACB}}
\end{figure}

Fig.\ref{Fig Convergence-CHB} and Fig.\ref{Fig Convergence-CACB} show the results from CHB and CACB, respectively, using $\eta=h$ and $M_0=10^{-7}(\eta/\eta_0)$. The convergence of the numerical solutions from the Phase-Field models to the sharp-interface solution is observed, as the cell size, as well as the interface thickness, is refined. The $L_2$ errors of the three benchmark quantities are computed using the sharp-interface solution as the reference. The errors are summarized in Table \ref{Table Convergence}, and both CHB and CACB share a similar convergence behavior. The circularity $\psi_c$, which quantifies the shape of the bubble, converges to the sharp-interface solution at a rate close to 2nd-order. The convergence rate is around 1.5th-order for the dynamics of the bubble, quantified by both $y_c$ and $v_c$. Therefore, the numerical solutions of the Phase-Field models converge to the sharp-interface solution.   
\begin{table}[!t]
	\caption{$L_2$ errors of the convergence test with $\eta=h$\label{Table Convergence}}
	\centering
	\includegraphics[scale=.4]{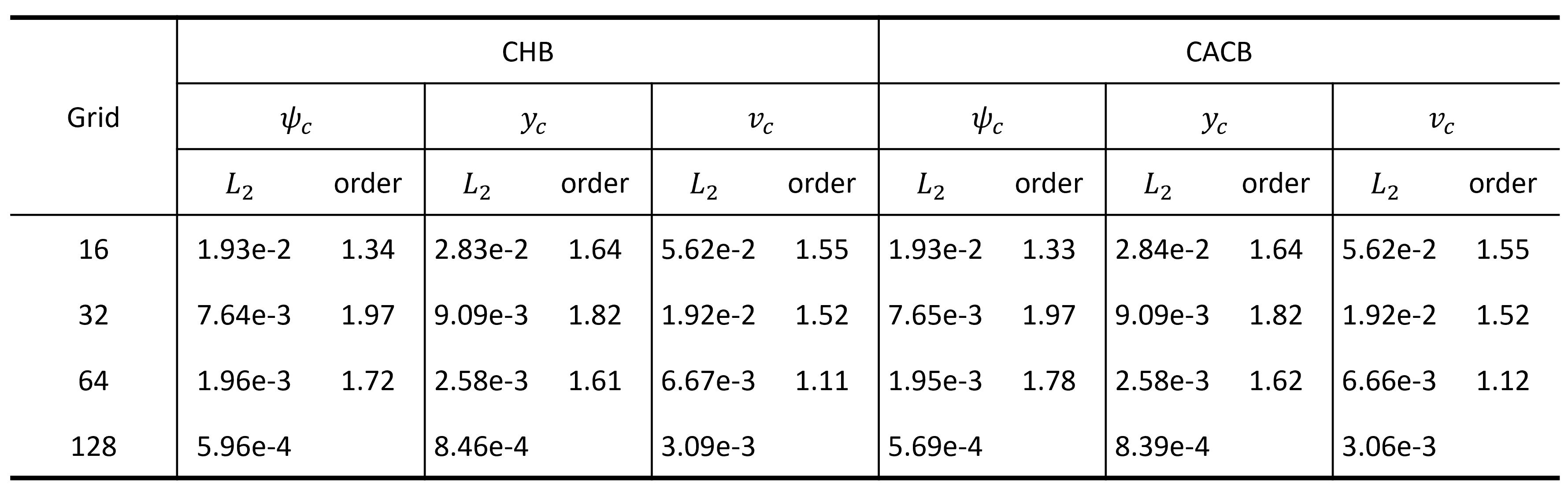}
\end{table}

The same studies have been performed in \cite{Huangetal2020N} for CH and a similar convergence behavior is observed. Therefore, the boundedness mapping has little effect on either the formal order of accuracy or the convergence rate to the sharp-interface solution.

\subsection{Three-Phase dam break}\label{Sec Dam break}
To demonstrate the capability of the proposed CHB and CACB models for complicated multiphase problems, the three-phase dam-break problem is performed. 

The three phases considered are water (Phase 1), whose density is $998.207 \mathrm{kg/m^3}$ and viscosity is $1.002\times10^{-3}\mathrm{Pa\cdot s}$, oil (Phase 2), whose density is $557 \mathrm{kg/m^3}$ and viscosity is $9.15 \times 10^{-2}\mathrm{Pa\cdot s}$, and air (Phase 3), whose density is $1.2041 \mathrm{kg/m^3}$ and viscosity is $1.78 \times 10^{-5} \times 10^{-2}\mathrm{Pa\cdot s}$. The surface tensions between them are $\sigma_{1,2}=0.04 \mathrm{N/m}$, $\sigma_{1,3}=0.0728 \mathrm{N/m}$, and $\sigma_{2,3}=0.055 \mathrm{N/m}$. The gravity is pointing downward with a magnitude $9.8 \mathrm{m/s^2}$. The governing equations are non-dimensionalized by a density scale $1.204\mathrm{kg/m^3}$ (or the air density), a length scale $5.715\mathrm{cm}$ (or the initial height of the water or oil column), and an acceleration scale $9.8\mathrm{m/s^2}$ (or the magnitude of the gravity). The rest of the setup and the results of the problem are reported in their dimensionless forms using the scales mentioned. The domain considered is $[8 \times 2]$ and all the boundaries are no-slip. The domain is discretized by $[512\times128]$ cells. $\eta$ and $M_0$ are $0.01$ and $10^{-7}$, respectively. The time step is $\Delta t=5\times10^{-4}$. Initially, the flow is stationary, a square water column with a width $1$ is at the left of the domain, and an oil column with the same size is at the right of the domain. 

\begin{figure}[!t]
	\centering
	\includegraphics[scale=.45]{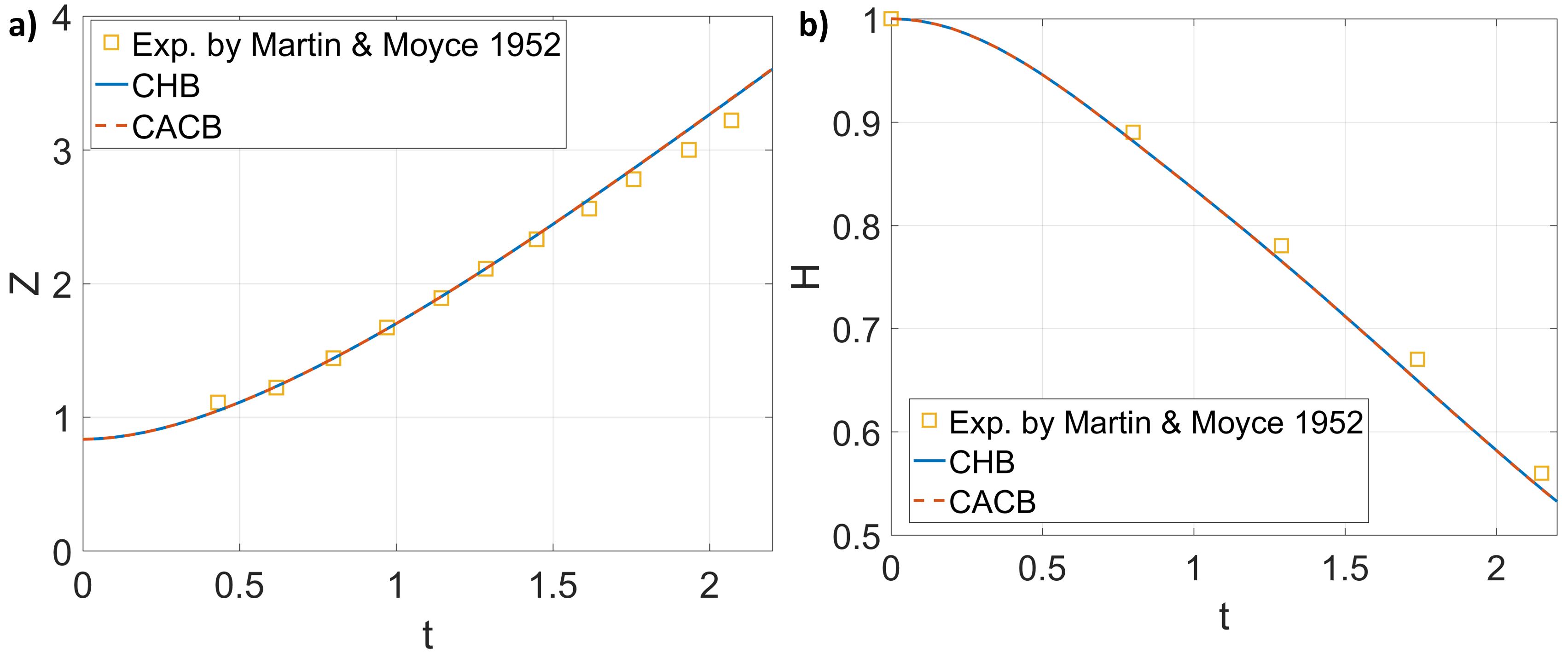}
	\caption{Front and height of the water column vs. $t$. a) Front of the water column. b) Height of the water column. \label{Fig DB-Location}}
\end{figure}

The initial dynamics is quantified by measuring the front $Z$ and height $H$ of the water column. Since the water and oil are far away separated at the beginning, there is no interaction between them and we can compare the numerical solutions from both CHB and CACB to the experimental data from Martin and Moyce \cite{MartinMoyce1952}. We calibrate the numerical results by setting $Z$ equal to $1.44$, when $t$ is $0.8$, and $H$ equal to $1$, when $t$ is $0$, as those in \cite{MartinMoyce1952}. The results are shown in Fig.\ref{Fig DB-Location}. The difference between CHB and CACB is unobservable and both results agree well with the experimental data. 

\begin{figure}[!t]
	\centering
	\includegraphics[scale=.2]{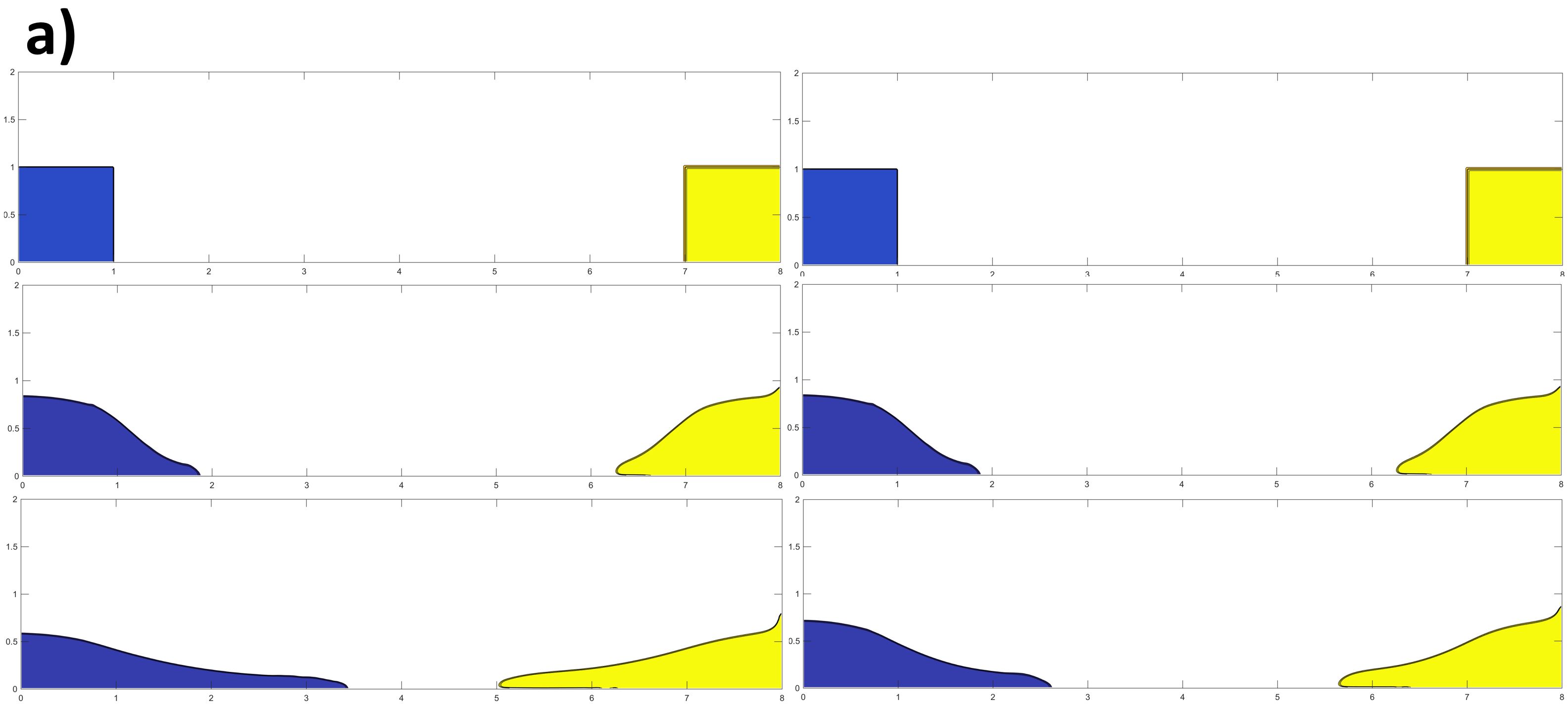}
	\includegraphics[scale=.2]{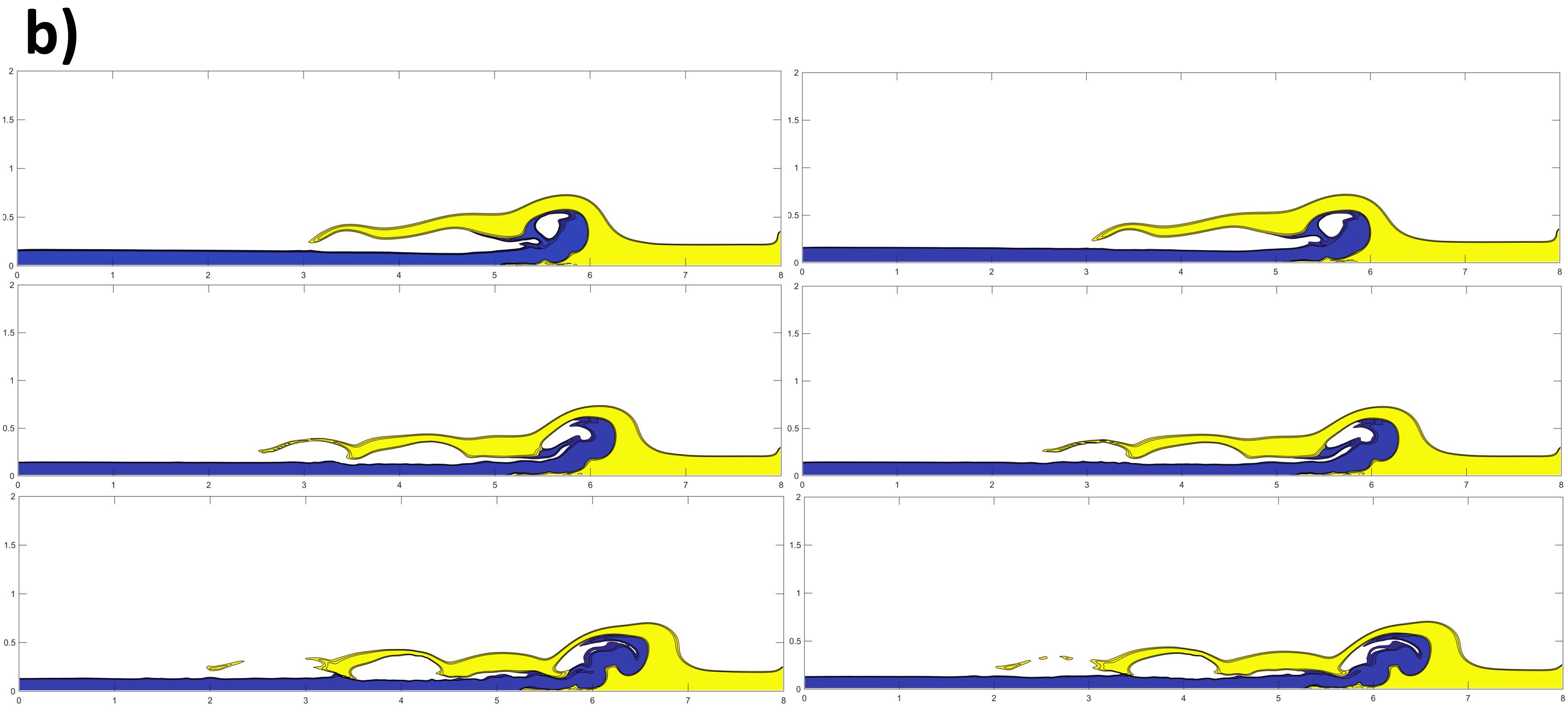}\\
	\includegraphics[scale=.2]{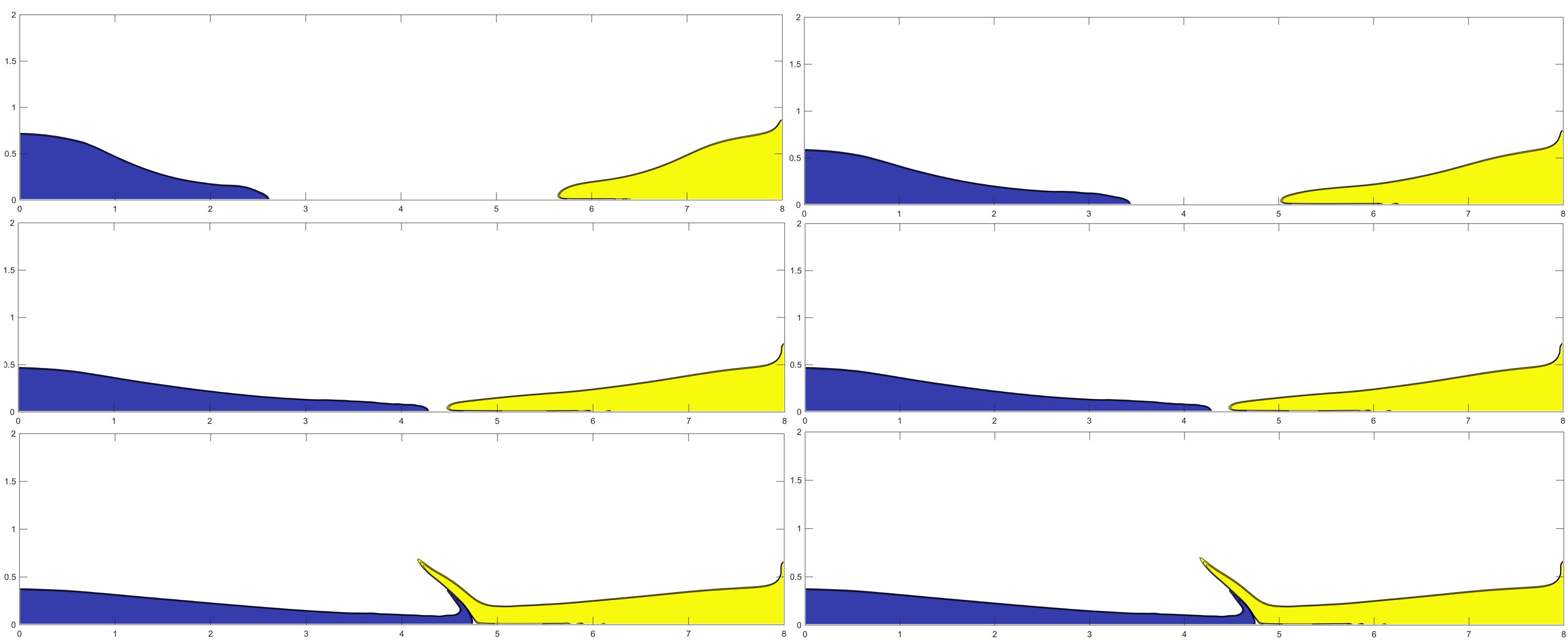}
	\includegraphics[scale=.2]{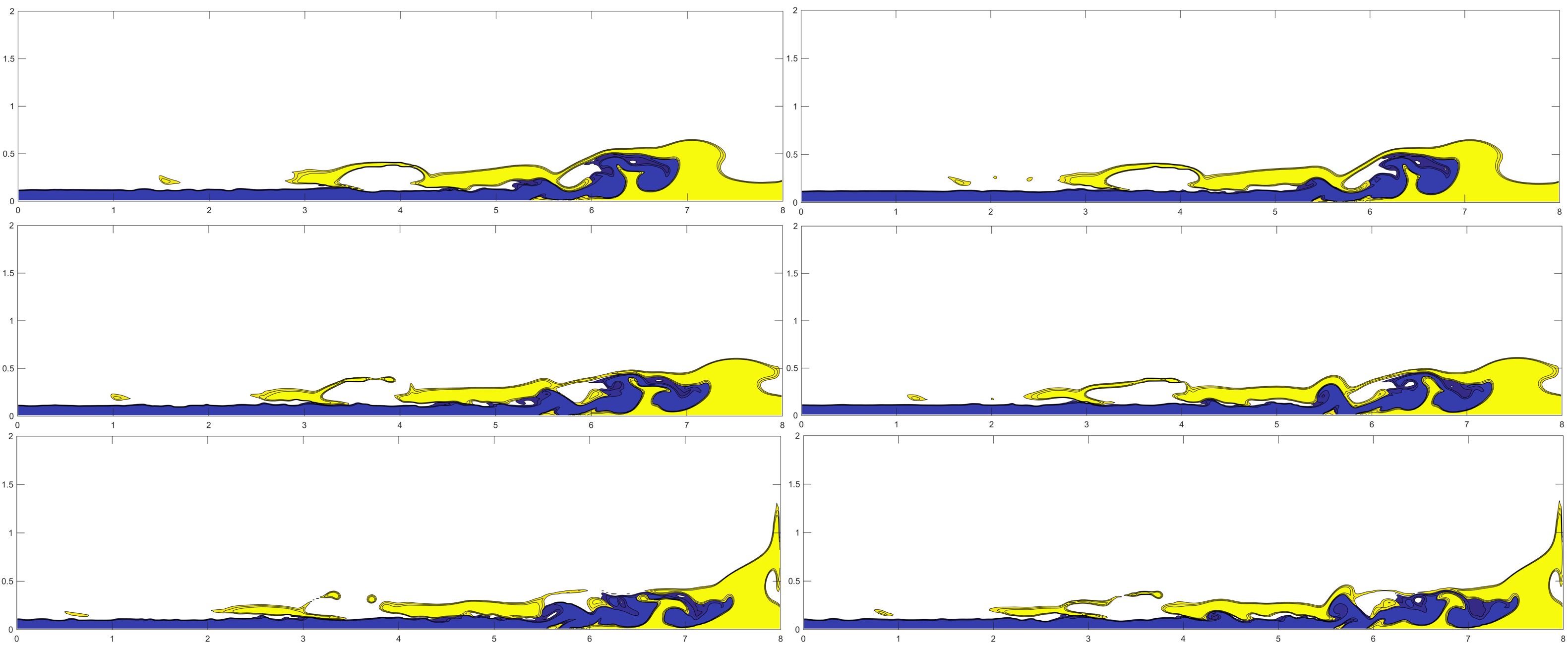}\\
	\includegraphics[scale=.2]{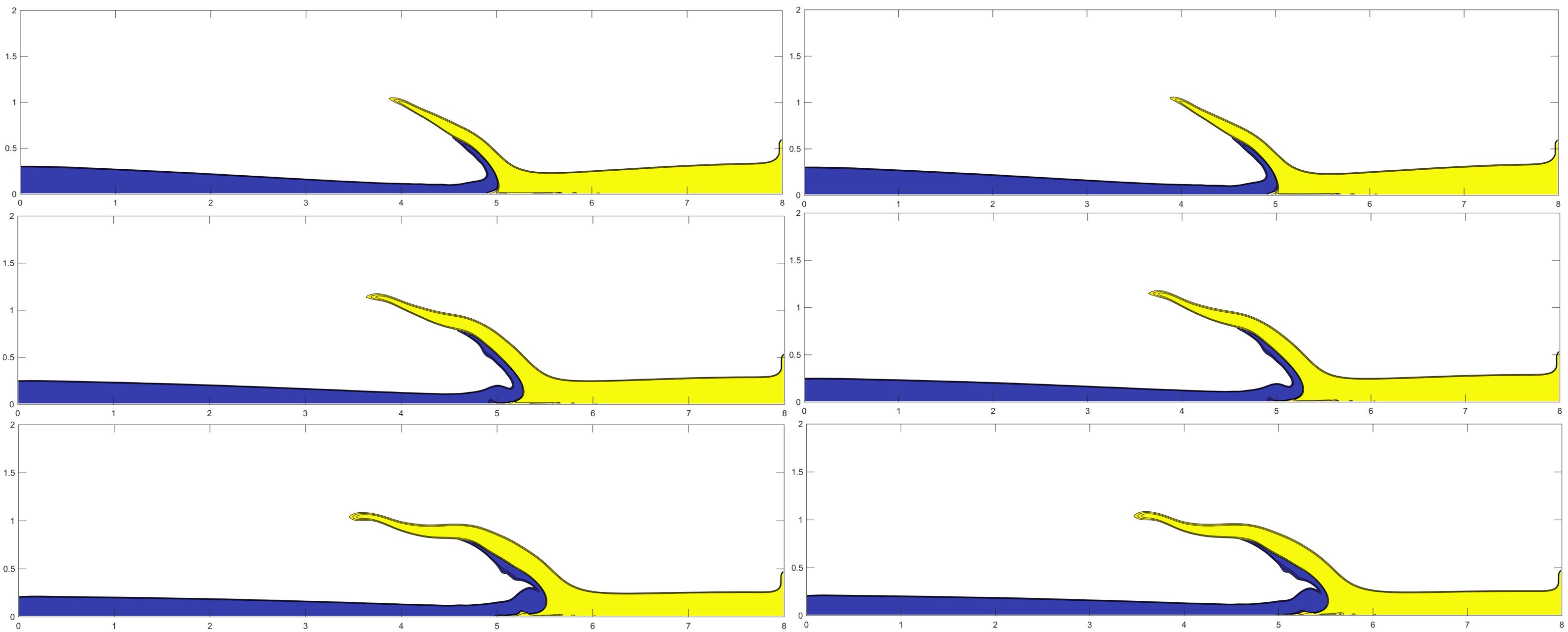}
	\includegraphics[scale=.2]{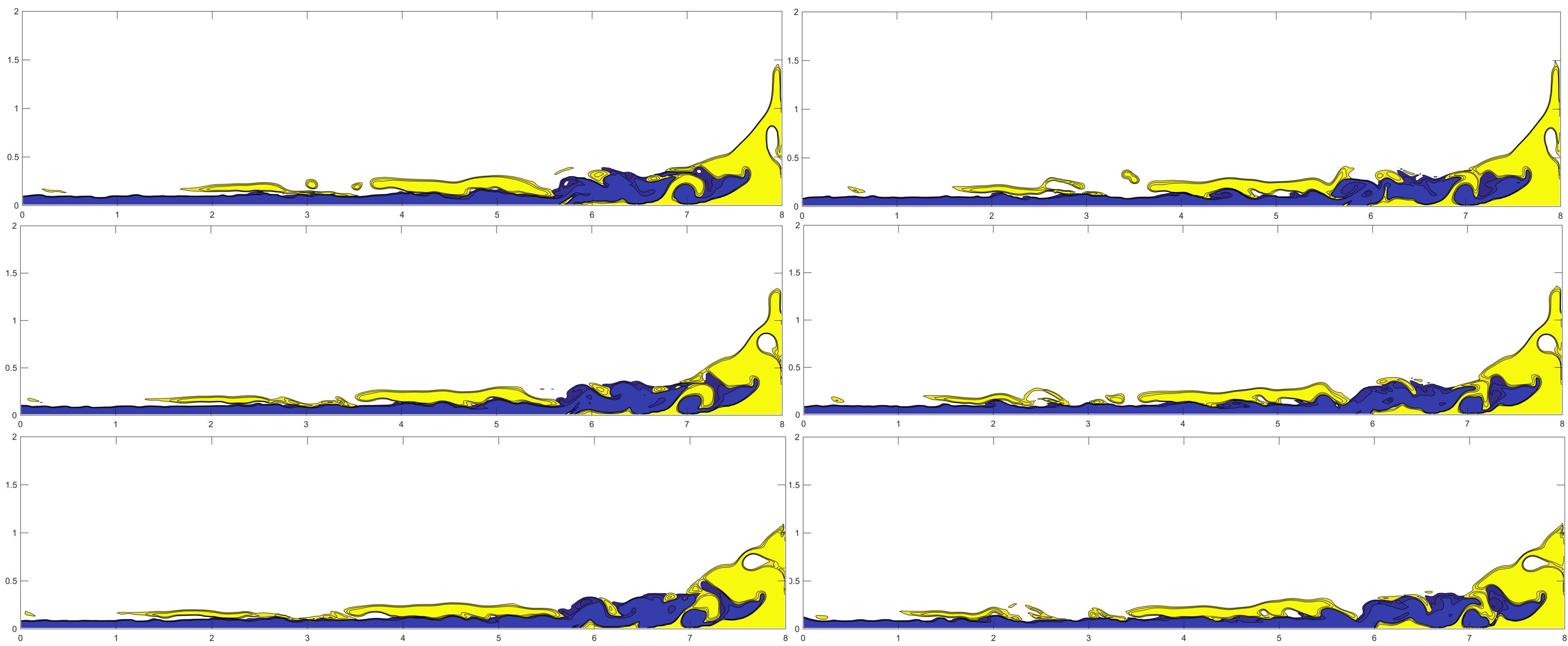}\\
	\includegraphics[scale=.2]{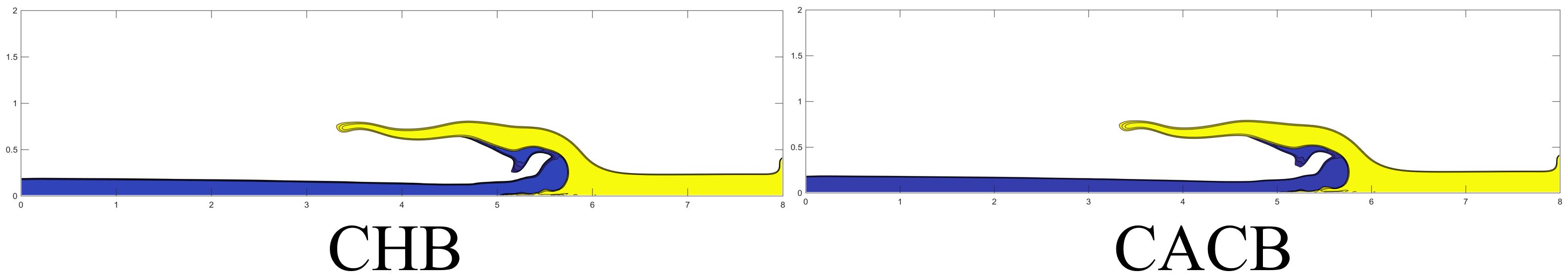}
	\includegraphics[scale=.2]{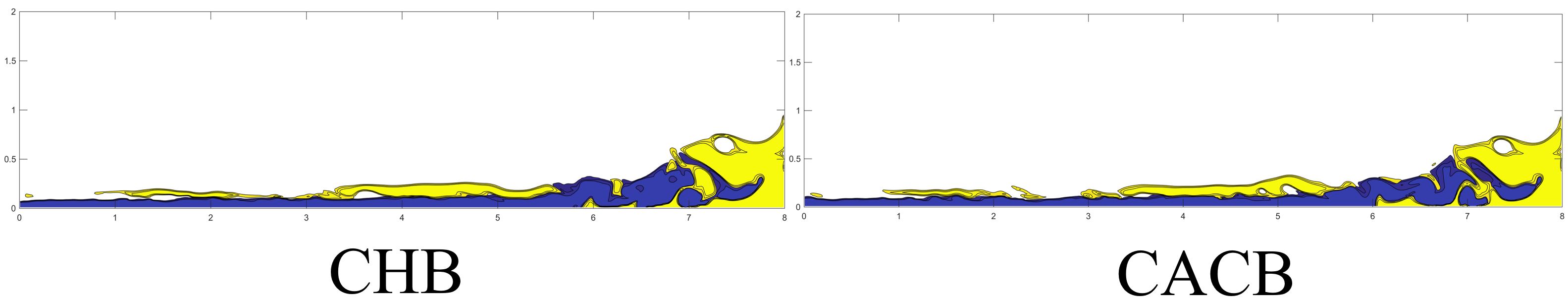}
	\caption{Configurations of the three-phase dam break from CHB and CACB. a) from top to bottom, $t=0.0, 1.0,1.5,2.0,2.5,3.0,3.5,4.0,4.5,5.0$. b) from top to bottom, $t=5.5, 6.0,6.5,7.0,7.5,8.0,8.5,9.0,9.5,10.0$. Blue: Water (phase 1). Yellow: Oir (phase 2). White: Air (phase 3).\label{Fig DB-Interface-CHB}}
\end{figure}

\begin{figure}[!t]
	\centering
	\includegraphics[scale=.43]{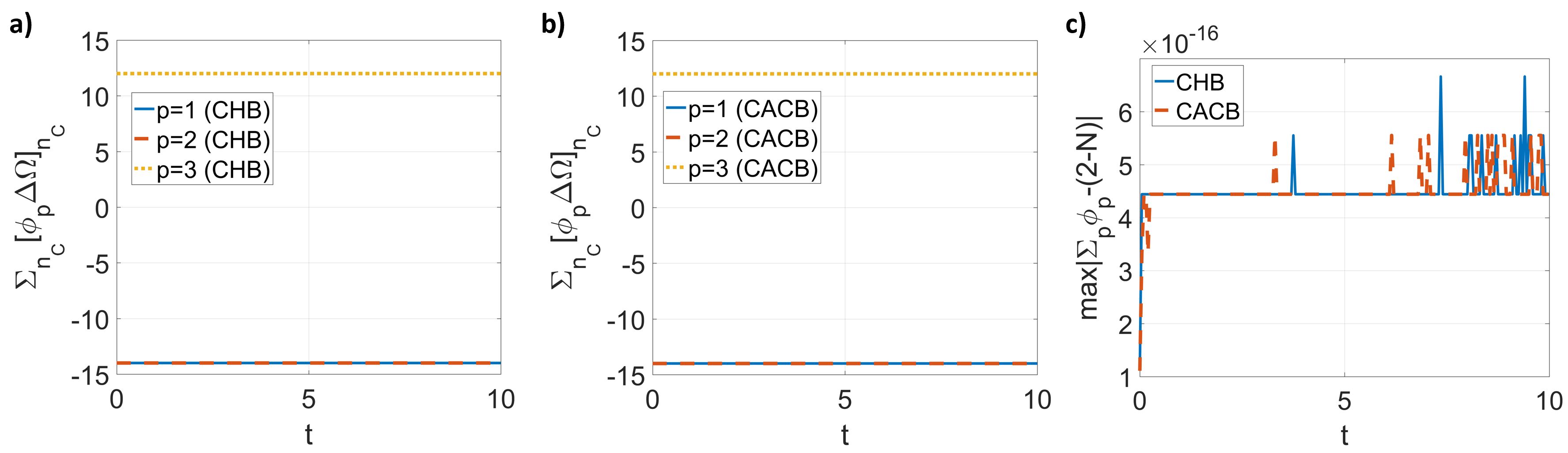}
	\caption{Results of the three-phase dam break. a) Mass conservation of individual phases from CHB. b) Mass conservation of individual phases from CACB. c) Error of the summation constraint for the order parameters from CHB and CACB. \label{Fig DB-Mass}}
\end{figure}

Fig.\ref{Fig DB-Interface-CHB} show the configurations of the interfaces from CHB and CACB, up to $t=10$. Both models give a similar picture of the problem. The water and oil columns collapse at the beginning and start sliding along the bottom wall. The water is moving faster than the oil since the oil is about $100$ times more viscous than the water. Due to the high viscosity, the oil close to the lateral wall is falling down more slowly than other parts of it. When the fronts of the water and oil meet together, the oil, which is lighter and moving slower, is squeezed upward by the water, and the water climbs along the bottom of the oil. The front of the oil, squeezed by the water, collapses again and lays above the water, along with breaking up into small droplets and filaments. At the same time, the water keeps moving toward the right and pushing the oil at the bottom moving backward. It can be observed that the interactions among different phases are very complicated. Even though the problem is challenging, the mass conservation and the summation constraint for the order parameters are always satisfied, as shown in Fig.\ref{Fig DB-Mass}.

\begin{figure}[!t]
	\centering
	\includegraphics[scale=.45]{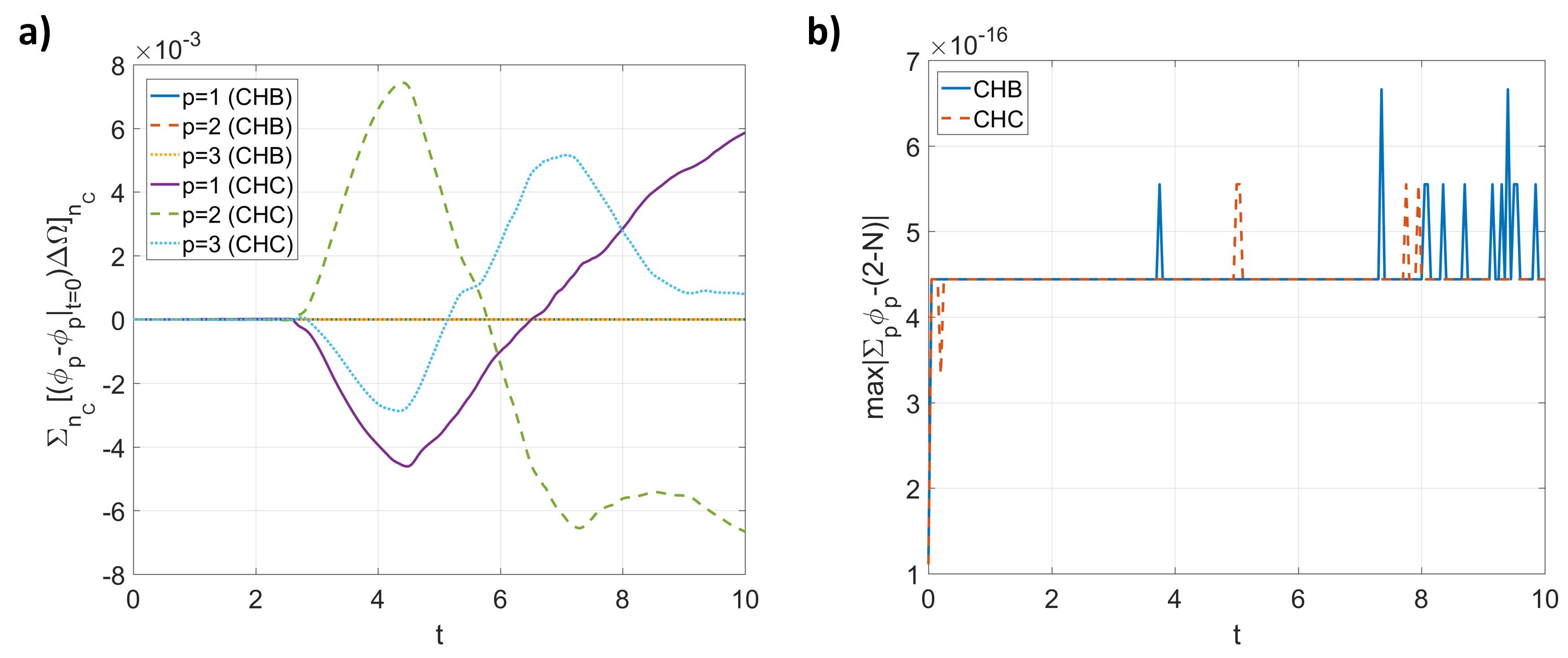}
	\caption{Results of the three-phase dam break from CHB and CHC (which is CH but including the clipping and rescaling steps in the boundedness mapping). a) Errors of mass conservation of individual phases. b) Errors of the summation constraint for the order parameters. \label{Fig DB-Compare}}
\end{figure}

Fig.\ref{Fig DB-Compare} shows the results from CHC, which is CH but including only the clipping and rescaling steps in the boundedness mapping in Section \ref{Sec Boundedness mapping}. The results are compared to those from CHB. Thanks to adding the rescaling step, the summation constraint for the order parameters are satisfied from CHC. The mass change due to simply clipping and rescaling the order parameters is significant, although the out-of-bound error is small, in the order of $10^{-5}$, in one time step. 
The problem is sensitive to out-of-bound errors since it has a maximum density ratio of about $1000$ and a maximum viscosity ratio of about $5000$. Both the clipping operation in previous studies and the boundedness mapping proposed in the present work improve the robustness of the scheme, which is important for the success of the simulation. Moreover, the boundedness mapping additionally enforces the mass conservation of each phase, which is also important for long-time simulation.  

\section{Conclusion}\label{Sec Conclusions}
In the present work, the general multiphase volume distribution problem, which is an important component of developing multiphase Phase-Field models, is addressed consistently and conservatively by the proposed algorithm in Section \ref{Sec Volume distribution}. The algorithm honors the summation constraint, conservation constraint, and \textit{consistency of reduction}, so that no fictitious phases, voids, or overfilling are generated after the volume distribution.
It is challenging to satisfy all the constraints in a general multiphase case, and we discover that the phase-wise formula, which works for two-phase problems, is not feasible. 
Then the problem is turned into a linear system representing the interactions among different phases. A weight function for volume distribution is carefully selected, so that the aforementioned constraints are satisfied and the coefficient matrix of the linear system is not only symmetric but also diagonally dominant.
A scaling argument is supplemented so that the solution of the linear system is admissible.
To the best of our knowledge, this is the first volume distribution algorithm that is general for an arbitrary number of phases and satisfies all the physical constraints mentioned. 

The proposed volume distribution algorithm is successfully applied to determine the Lagrange multipliers that enforce the mass conservation for general multiphase Phase-Field models. As an example of this application, a multiphase conservative Allen-Cahn model Eq.(\ref{Eq Allen-Cahn}) that honors the summation constraint for the order parameters, the mass conservation, and the \textit{consistency of reduction}, simultaneously, is developed in Section \ref{Sec L_p^c}. Such a kind of model is first reported in the present study. In addition, the multiphase conservative Allen-Cahn model exactly reduces to the one proposed in \cite{BrasselBretin2011} for two-phase problems. 
A corresponding consistent and conservative numerical scheme is developed in Section \ref{Sec Scheme CAC} for the model, and we show that the scheme preserves the physical properties of the model on the discrete level. 
The \textit{consistency of reduction} is an important property of a multiphase model and its scheme, since it eliminates any generation of fictitious phases. Our numerical studies in Section \ref{Sec Results} show that fictitious phases are unphysically generated at interfacial regions by the multiphase model in \citep{KimLee2017} that violates the \textit{consistency of reduction}. On the other hand, there are no fictitious phases generated by either the multiphase conservative Allen-Cahn model, proposed in the present work, or the multiphase Cahn-Hilliard model, proposed and studied in \cite{Dong2018,Huangetal2020N}, since both the models and their schemes are reduction consistent. 
A comparison study is also performed and it shows that the conservative Allen-Cahn model has a better ability than the Cahn-Hilliard model to preserve under-resolved structures.

Another application of the proposed volume distribution algorithm is the development of the boundedness mapping in Section \ref{Sec Boundedness mapping}, which is a numerical procedure to address the out-of-bound order parameters resulting from either the defect of a multiphase model or the numerical error. The out-of-bound order parameters can lead to a negative density or viscosity of the fluid mixture, especially when the density or viscosity ratio of a problem is large.
The boundedness mapping maps the out-of-bound order parameters into their physical interval but does not violate the physical properties of the order parameters, i.e., their summation constraint, mass conservation, and \textit{consistency of reduction}.
In the present study, the boundedness mapping is applied to both the multiphase Cahn-Hilliard and conservative Allen-Cahn models. Along with the consistent and conservative schemes for those models, the order parameters are therefore reduction consistent, mass conservative, and bounded, which has been carefully analyzed and numerically validated. It is observed in the numerical tests in Section \ref{Sec Results} that the out-of-bound error can grow as the computation goes on, contaminate the physical solution, and in the end, trigger numerical instability, although it is small in one time step. The boundedness mapping removes the out-of-bound error while preserves the physical properties of the order parameters, which is beneficial to improve the robustness of the scheme and to provide a physical solution. 

In addition to the \textit{consistency of reduction}, the \textit{consistency of mass conservation} and the \textit{consistency of mass and momentum transport} \citep{Huangetal2020,Huangetal2020CAC,Huangetal2020N} need to be satisfied, when coupling the Phase-Field model to the hydrodynamics. Begin with a generic form of the Phase-Field model for multiphase flows and with the help of the consistent formulation proposed in \citep{Huangetal2020CAC}, the governing equations in Section \ref{Sec Governing equations} are presented in a generally valid way, and different parts of them are connected physically.
The boundedness mapping is modeled as a set of discrete Lagrange multipliers and is included in the implementation of the consistent formulation discretely in Section \ref{Sec Apply}. As a result, the discrete consistent mass flux that satisfies the \textit{consistency of mass conservation} is obtained. Finally, the momentum conservative scheme that satisfies the \textit{the consistency of mass and momentum transport} in \cite{Huangetal2020} is applied to solve the momentum equation. An advection problem having a density ratio $10^6$ is successfully performed, indicating that those two consistency conditions are satisfied on the discrete level.
The convergence tests show that the proposed schemes are formally 2nd-order accurate, and that the numerical Phase-Field solutions from both the Cahn-Hilliard and conservative Allen-Cahn models converge to the sharp-interface solution with a similar behavior. In addition, the boundedness mapping does not influence either the order of accuracy of the scheme or the convergence behavior to the sharp-interface solution. 
A complicated three-phase dam-break problem, which has large density and viscosity ratios, is performed to demonstrate the capability of the Phase-Field models in multiphase flows. The numerical solutions agree well with the experimental data, and strong interactions among different phases are captured. The mass conservation and the summation constraint for the order parameters are always satisfied even though the problem considered is highly dynamical. The results also show that simply clipping and rescaling the order parameters, instead of performing the boundedness mapping, leads to significant mass changes.

In summary, the multiphase volume distribution problem is appropriately addressed in the present work and two applications of it are performed. The first one is on the continuous level, developing the physical Lagrange multipliers that enforce the mass conservation for general Phase-Field models. The proposed multiphase conservative Allen-Cahn model that satisfies the \textit{consistency of reduction} is an example of this application. The second one is on the discrete level, developing the boundedness mapping that maps the out-of-bound order parameters into their physical interval. Therefore, the consistent and conservative volume distribution algorithm is an important tool for modeling and simulating multiphase flows.

\section*{Acknowledgments}
A.M. Ardekani would like to acknowledge the financial support from the National Science Foundation (CBET-1705371). This work used the Extreme Science and Engineering Discovery Environment (XSEDE) \cite{Townsetal2014}, which is supported by the National Science Foundation grant number ACI-1548562 through allocation TG-CTS180066  and TG-CTS190041.
G. Lin gratefully acknowledges the support from the National Science Foundation (DMS-1555072, DMS-1736364, CMMI-1634832, CMMI-1560834, and DMS-2053746), and Brookhaven National Laboratory Subcontract 382247, ARO/MURI grant W911NF-15-1-0562, and U.S. Department of Energy (DOE) Office of Science Advanced Scientific Computing Research program DE-SC0021142.

\section*{Appendix}
Proof of Theorem \ref{Theorem Volume consistency of reduction}:
\begin{proof}\label{Proof Volume consistency of reduction}
Without loss of generality and for a clear presentation, we consider that the last phase of a $N$-phase $(N \geqslant 2)$ system is absent at the location where $\phi_N=-1$. Then from Eq.(\ref{Eq L_p}), we have
\begin{equation}\nonumber
L_N=\sum_{q=1}^N W_{N,q} B_q=0,
\quad
L_p=\sum_{q=1}^N W_{p,q} B_q = \sum_{q=1}^{N-1} W_{p,q} B_q,
\quad
1 \leqslant p \leqslant N-1,
\end{equation}
due to $W_{p,q}|_{\phi_p=-1}=W_{q,p}|_{\phi_p=-1}=0$ from Eq.(\ref{Eq W_{p,q}}). Therefore, the $N$-phase formulation of $\{L_p\}_{p=1}^N$ reduces to the corresponding $(N-1)$-phase formulation for the present phases and its value for the absent phase is zero. Notice that one can choose any other phases as the absent phase and reach the same conclusion. By induction, the \textit{consistency of reduction} is true. For $N=1$, the trivial solution $L_1=0$ is obtained due to $\phi_1 \equiv 1$, corresponding to that the volume of the domain does not change after the volume distribution.
\end{proof}

Proof of Theorem \ref{Theorem CAC summation}:
\begin{proof}\label{Proof CAC summation}
Summing Eq.(\ref{Eq Allen-Cahn}) over $p$, both sides of the summed equation are zero, given $\sum_{p=1}^N \phi_p = (2-N)$, and Eq.(\ref{Eq Divergence-free}), Eq.(\ref{Eq Allen-Cahn parts}), and Eq.(\ref{Eq Summation constraint L_p^c}) are used.
\end{proof}

Proof of Theorem \ref{Theorem CAC conservation}:
\begin{proof}\label{Proof CAC conservation}
Integrating Eq.(\ref{Eq Allen-Cahn}) over domain $\Omega$ and applying the divergence theorem, the boundary integrals vanish due to the boundary condition, and $\frac{d}{dt} \int_{\Omega} \phi_p d\Omega=0$ $(1 \leqslant p \leqslant N)$ is obtained with the help of Eq.(\ref{Eq Conservation L_p^c}).
\end{proof}

Proof of Theorem \ref{Theorem CAC reduction}:
\begin{proof}\label{Proof CAC reduction}
Without loss of generality and for a clear presentation, we consider that the last phase of a $N$-phase $(N \geqslant 2)$ system is absent at the location where $\phi_N=-1$ and $|\nabla \phi_N|=|\nabla^2 \phi_N|=0$. From Eq.(\ref{Eq Allen-Cahn}), we have
\begin{equation}\nonumber
\frac{\partial \phi_N}{\partial t}=0
\end{equation}
\begin{equation}\nonumber
\frac{\partial \phi_p}{\partial t}
+
\nabla \cdot ( \mathbf{u} \phi_p )
=
M_0 \lambda_0 \nabla^2 \phi_p
-\frac{M_0 \lambda_0}{\eta^2} \left(  g'_1(\phi_p)  -  \frac{1+\phi_p}{2} L^s \right)
+L_p^c,
\quad
1 \leqslant p \leqslant N-1,
\end{equation}
where
\begin{equation}\nonumber
L_p^c=\sum_{q=1}^{N-1} W_{p,q} B_q,
\quad
L^s=\sum_{p=1}^{N} g'_1(\phi_p)=\sum_{p=1}^{N-1} g'_1(\phi_p),
\end{equation}
from Theorem \ref{Theorem Volume consistency of reduction}
and due to $g'_1(\phi_N)=g'_1(-1)=0$, respectively.
Therefore, the absent phase remains absent, i.e., $\phi_N=-1$ at that location, while Eq.(\ref{Eq Allen-Cahn}) for the other phases reduces to the corresponding $(N-1)$-phase one. By induction, Eq.(\ref{Eq Allen-Cahn}) satisfies the \textit{consistency of reduction}. 
\end{proof}

Proof of Theorem \ref{Theorem Boundedness reduction}:
\begin{proof}\label{Proof Boundedness reduction}
Without loss of generality and for a clear presentation, we consider that the last phase of a $N$-phase $(N \geqslant 2)$ system is absent at the location where $\phi_N=-1$. From the clipping step Eq.(\ref{Eq Boundedness Clipping}), we have
\begin{equation}\nonumber
\phi_N^{b*}=-1,
\quad
\phi_p^{b*}=\left\{
\begin{array}{ll}
1, \phi_p \geqslant 1,\\
-1,\phi_p \leqslant -1,\\
\phi_p, \mathrm{else},
\end{array}
\right.
\quad
1 \leqslant p \leqslant N-1.
\end{equation}
From the rescaling step Eq.(\ref{Eq Boundedness rescaling}), we have
\begin{eqnarray}\nonumber
C_N^{b*}=\frac{1+\phi_N^{b*}}{2}=0,
\quad
C_p^{b*}=\frac{1+\phi_p^{b*}}{2},
\quad
1 \leqslant p \leqslant N-1,\\
\nonumber
C_N^{b**}=\frac{C_N^{b*}}{\sum_{q=1}^N C_q^{b*}}=0,
\quad
C_p^{b**}=\frac{C_p^{b*}}{\sum_{q=1}^N C_q^{b*}}=\frac{C_p^{b*}}{\sum_{q=1}^{N-1} C_q^{b*}},
\quad
1 \leqslant p \leqslant N-1,\\
\nonumber
\phi_N^{b**}=2C_N^{b**}-1=-1,
\quad
\phi_p^{b**}=2C_p^{b**}-1,
\quad
1 \leqslant p \leqslant N-1.
\end{eqnarray}
From the conservation step Eq.(\ref{Eq Boundedness conservation}) along with the constraints in Eq.(\ref{Eq Boundedness Constraints}), we have
\begin{equation}\nonumber
\phi_N^b=\phi_N^{b**}=-1,
\quad
\phi_p^b=\phi_p^{b**}+\mathcal{L}_p^b,
\quad 1 \leqslant p \leqslant N-1.
\end{equation}
Noticing that $\{\mathcal{L}_p^b\}_{p=1}^N$ is determined by the consistent and conservative volume distribution algorithm in Section \ref{Sec Volume distribution}, which has been shown to be reduction consistent, see Theorem \ref{Theorem Volume consistency of reduction}. 

Therefore, the absent phase remains absent while the $N$-phase formulations for the other phases reduce to the corresponding $(N-1)$-phase ones. Iterating the above three steps by letting the output $\{\phi_p^b\}_{p=1}^N$ become the new input does not change the conclusion. By induction, we show that the boundedness mapping satisfies the \textit{consistency of reduction}. 
\end{proof}

The fully-discretized equation of the multiphase conservative Allen-Cahn model Eq.(\ref{Eq Allen-Cahn}) from the proposed scheme is 
\begin{equation}\label{Eq CAC discrete}
\frac{\gamma_t \phi_p^{n+1}-\hat{\phi_p}}{\Delta t}
+
\tilde{\nabla} \cdot (\mathbf{u}^{*,n+1} \tilde{\phi}_p^{*,n+1})
=
M_0 \lambda_0 \tilde{\nabla} \cdot \tilde{\nabla} \phi_p^{*}
-
\frac{M_0 \lambda_0}{\eta^2} 
\left( \tilde{g}'_1(\phi_p^{*})
-
\frac{1+\phi_p^n}{2} \tilde{L}^s 
\right)
+ 
\tilde{L}_p^c,
\quad 1 \leqslant p \leqslant N.
\end{equation}

Proof of Theorem \ref{Theorem CAC summation discrete}:
\begin{proof}\label{Proof CAC summation discrete}
Given $\sum_{p=1}^N \phi_p^n=\sum_{p=1}^N \phi_p^{n-1}=\cdots=\sum_{p=1}^N \phi_p^0=(2-N)$ at every discrete cell in the domain, and summing Eq.(\ref{Eq CAC discrete}) over all $p$, 
we have 
$\sum_{p=1}^N \hat{\phi}_p=\widehat{\sum_{p=1}^N \phi_p}=\gamma_t (2-N)$, 
$\sum_{p=1}^N \tilde{\nabla} \cdot (\mathbf{u}^{*,n+1} \tilde{\phi}_p^{*,n+1})=0$, and $\sum_{p=1}^N \frac{1+\phi_p^n}{2}=1$. 
From Eq.(\ref{Eq L^s discrete}) and the first property of $\{\tilde{L}_p^c\}_{p=1}^N$ in Eq.(\ref{Eq Constraints L_p^c discrete}), the right-hand side of the summed Eq.(\ref{Eq CAC discrete}) over $p$ becomes zero. 
At the end, we reach
\begin{equation}\nonumber
\sum_{p=1}^N \phi_p^{n+1}=\frac{1}{\gamma_t} \sum_{p=1}^N \hat{\phi}_p=(2-N),
\end{equation}
at every discrete location. By induction, $\sum_{p=1}^N \phi_p=(2-N)$ is true at every time level as well.
\end{proof}

Proof of Theorem \ref{Theorem CAC conservation discrete}:
\begin{proof}\label{Proof CAC conservation discrete}
Given $\sum_{n_C} [\phi_p^n \Delta \Omega]_{n_C}=\sum_{n_C} [\phi_p^{n-1} \Delta \Omega]_{n_C}=\cdots=\sum_{n_C} [\phi_p^0 \Delta \Omega]_{n_C}$, and summing Eq.(\ref{Eq CAC discrete}) over all the cells after multiplying it to $\Delta \Omega$, 
we have $\sum_{n_C} [\hat{\phi_p}\Delta \Omega]_{n_C}=\widehat{\sum_{n_C}[\phi_p \Delta \Omega]_{n_C}}=\gamma_t \sum_{n_C} [\phi_p^0 \Delta \Omega]_{n_C}$. 
All the terms having the discrete divergence operator in Eq.(\ref{Eq CAC discrete}) vanish after the summation, as mentioned at the beginning of Section \ref{Sec Discretizations}. 
From Eq.(\ref{Eq Conservation L_p^c discrete}) and the second property of $\{\tilde{L}_p^c\}_{p=1}^N$ in Eq.(\ref{Eq Constraints L_p^c discrete}), the right-hand side of the summed Eq.(\ref{Eq CAC discrete}) over $n_C$ becomes zero. 
At the end, we reach
\begin{equation}\nonumber
\sum_{n_C} [\phi_p^{n+1} \Delta \Omega]_{n_C}=\frac{1}{\gamma_t} \sum_{n_C} [\hat{\phi}_p \Delta \Omega]_{n_C}=\sum_{n_C} [\phi_p^0 \Delta \Omega]_{n_C},
\quad
1 \leqslant p \leqslant N.
\end{equation}
By induction, the conservation constraint is enforced at every time step.
\end{proof}

Proof of Theorem \ref{Theorem CAC reduction discrete}:
\begin{proof}\label{Proof CAC reduction discrete}
Without loss of generality and for a clear presentation, we consider that the last phase of a $N$-phase $(N \geqslant 2)$ system is absent globally, i.e., $\phi_N = -1$ in the entire domain.

Consider the absent phase, i.e., Phase $N$, first, and we have $\phi_N^n=\phi_N^{*,n+1}=-1$, $\hat{\phi}_N=-\gamma_t$, $\tilde{\nabla} \cdot (\mathbf{u}^{*,n+1} \tilde{\phi}_N^{*,n+1})=-\tilde{\nabla} \cdot \mathbf{u}^{*,n+1}=0$, and $\tilde{g}'_1(\phi_N^*)=2(\phi_N^*+1)$. Then $\phi_N^*=-1$ is the solution of \textbf{Step 1} Eq.(\ref{Eq CAC discrete step1}). We don't need to consider \textbf{Step 2} at this moment. From \textbf{Step 3}, i.e., the last property of $\{\tilde{L}_p^c\}_{p=1}^N$ in Eq.(\ref{Eq Constraints L_p^c discrete}), we have $\tilde{L}_N^c=0$, and finally from \textbf{Step 4} Eq.(\ref{Eq CAC discrete step2}) we obtain $\phi_N^{n+1}=\phi_N^*=-1$.

For the rest of the phases, \textbf{Setp 1} Eq.(\ref{Eq CAC discrete step1}) does not include any coupling among the phases so it automatically satisfies the \textit{consistency of reduction}. The contribution of Phase $N$ to $\tilde{L}^s$ in \textbf{Step 2} disappears due to $\tilde{\nabla}\cdot \tilde{\nabla} \phi_N^*=\tilde{g}'_1(\phi_N^*)=0$, and, as a result, the summation in Eq.(\ref{Eq L^s discrete}) is only from $p=1$ to $p=N-1$. $\{\tilde{L}_p^c\}_{p=1}^N$ in \textbf{Step 3} is determined by the consistent and conservative volume distribution algorithm in Section \ref{Sec Volume distribution}, and therefore it is reduction consistent, see Theorem \ref{Theorem Volume consistency of reduction}. Consequently, all the terms in \textbf{Step 4} Eq.(\ref{Eq CAC discrete step2}) for the rest of the phases reduce to the corresponding $(N-1)$-phase ones.

In summary, the absent phase, i.e., Phase $N$, remains absent, and the rest of the phases are updated from the formulations that reduce to the corresponding $(N-1)$-phase ones, without any influences from the absent phase. By induction, the \textit{consistency of reduction} is satisfied by the proposed scheme on the discrete level. 
\end{proof}
\textit{\textbf{Remark:}
\begin{itemize}
    \item
    When discussing the \textit{consistency of reduction} on the discrete level, we consider the absence of Phase $N$ globally for convenience. It becomes more involved when considering the local absence because there is a matrix inversion, which couples all the information in the domain, in \textbf{Step 1} Eq.(\ref{Eq CAC discrete step1}). If a fully explicit scheme is used in \textbf{Step 1} Eq.(\ref{Eq CAC discrete step1}), then at the location where $\phi_N^{n}=\phi_N^{*,n+1} = -1$ and $|\tilde{\nabla} \tilde{\phi}_N^{*,n+1}|=\tilde{\nabla} \cdot \tilde{\nabla} \phi_N^n=0$, the consistency of reduction can be proof in the same manner. The difference between the solutions of the present scheme in \textbf{Step 1} Eq.(\ref{Eq CAC discrete step1}) and the fully explicit scheme is of the order of $\frac{\partial \phi_N^*}{\partial t} \Delta t$, where $\frac{\partial \phi_N^*}{\partial t}$ is the time derivative of the Allen-Cahn model (without $L_s$ and $\{L_p^c\}_{p=1}^N$). As a result, the highest spatial derivative in the difference of the schemes is $\tilde{\nabla} \cdot \tilde{\nabla} (\tilde{\nabla} \cdot \tilde{\nabla} \phi_N^n)$. It is reasonable to expect that Theorem \ref{Theorem CAC reduction discrete} is still valid at the location where $\phi_N^{n}=\phi_N^{*,n+1} = -1$ and $|\tilde{\nabla} \tilde{\phi}_N^{*,n+1}|=\tilde{\nabla} \cdot \tilde{\nabla} \phi_N^n=|\tilde{\nabla}\cdot\tilde{\nabla}(\tilde{\nabla} \tilde{\phi}_N^{*,n+1})|=\tilde{\nabla} \cdot \tilde{\nabla} (\tilde{\nabla} \cdot \tilde{\nabla} \phi_N^n)=0$. Our numerical implementation in Section \ref{Sec Fictitious phases} demonstrates that the \textit{consistency of reduction} on the discrete level holds locally and the results are shown in Fig.\ref{Fig Fictitious}.
    \item 
    The convection term $\{\tilde{\nabla} \cdot (\mathbf{u}^{*,n+1} \tilde{\phi}_p^{*,n+1})\}_{p=1}^N$ in Eq.(\ref{Eq CAC discrete step1}) is corrected by the gradient-based phase selection procedure, proposed in \citep{Huangetal2020N}, so that $\sum_{q=1}^N \tilde{\nabla} \cdot (\mathbf{u}^{*,n+1} \tilde{\phi}_q^{*,n+1})=(2-N) \tilde{\nabla} \cdot \mathbf{u}^{*,n+1}=0$, and $\tilde{\nabla} \cdot (\mathbf{u}^{*,n+1} \tilde{\phi}_p^{*,n+1})=-\tilde{\nabla} \cdot \mathbf{u}^{*,n+1}=0$ given Phase $p$ absent $(1 \leqslant p \leqslant N)$, even though the non-linear WENO scheme is used. These properties of $\{\tilde{\nabla} \cdot (\mathbf{u}^{*,n+1} \tilde{\phi}_p^{*,n+1})\}_{p=1}^N$ are critical in the proofs of Theorem \ref{Theorem CAC summation discrete} and Theorem \ref{Theorem CAC reduction discrete}.
    \item 
    In previous studies, e.g., \cite{Dong2014,Dong2015,Dong2017,Dong2018,Kim2009,LeeKim2015,KimLee2017}, only the first $(N-1)$ order parameters are solved from the Phase-Field model numerically and the $N$th one is computed algebraically from the summation constraint, i.e., Eq.(\ref{Eq Summation constraint phi}). Such a strategy can easily violate the consistency of reduction and, as a result, produce fictitious phases. Consider the following three-phase example where Phase 3 is absent, and therefore, the exact solution of the order parameters have the following properties, i.e., $\phi_1^E+\phi_2^E+\phi_3^E=-1$ and $\phi_3^E=-1$, everywhere. If only the first two order parameters are solved from the Phase-Field model, there is no explicit restriction to enforce $\phi_1+\phi_2=0$, especially at interfacial regions where the gradients of $\phi_1$ and $\phi_2$ are large. Then, $\phi_3=-1-(\phi_1+\phi_2)$ is possibly not $-1$ everywhere. In other words, Phase 3 can be numerically generated following the above strategy. On the other hand, the proposed scheme solves all the order parameters from the Phase-Field model. The consistency of reduction of the scheme, i.e., Theorem \ref{Theorem CAC reduction discrete}, ensures that $\phi_3=-1$ after solving the equation for Phase 3, and at the same time $\phi_1+\phi_2+\phi_3=0$ is ensured, see Theorem \ref{Theorem CAC summation discrete}. Consequently, no Phase 3 is being numerically generated and $\phi_1+\phi_2=0$ is valid everywhere by the proposed schemes.
\end{itemize}
}

\bibliographystyle{plain}
\bibliography{refs.bib}

\end{document}